\documentclass[
  aps,
  prd,
  reprint,
  superscriptaddress,
  amsmath,
  amssymb,
  longbibliography,
  nofootinbib,
  floatfix
]{revtex4-2}

\usepackage[english]{babel}
\usepackage{amsthm}
\usepackage{simplewick}

\usepackage{graphicx}
\usepackage[export]{adjustbox}
\usepackage{array}
\usepackage{xcolor}
\usepackage[framemethod=default]{mdframed}

\usepackage{tikz}
\usetikzlibrary{
  external,
  shapes.misc,
  positioning,
  arrows,
  arrows.meta,
  bending,
  matrix,
  shapes,
  fit,
  tikzmark,
  calc,
  patterns,
  topaths,
  decorations.pathmorphing,
  decorations.markings,
  decorations.pathreplacing
}

\tikzset{>=latex}

\usepackage{hyperref}

\definecolor{darkblue}{rgb}{0.15,0.35,0.55}
\definecolor{reddish}{rgb}{0.65,0.2,0.2}

\hypersetup{
  colorlinks=true,
  citecolor=darkblue,
  linkcolor=reddish,
  urlcolor=darkblue,
  pdfauthor={Ji-Yuan Ke and Ping He},
  pdftitle={Splitting Rules from Kinematic Flow: Local Evolution and the Emergence of Time}
}

\definecolor{blue3}{RGB}{31,119,180}
\definecolor{red3}{RGB}{214,39,40}
\definecolor{orange3}{RGB}{255,127,14}
\definecolor{green3}{RGB}{44,160,44}

\tikzset{
  cross line/.style={
    decoration={
      markings,
      mark=at position 0.5 with {
        \draw[solid,thick,-] (-3pt,-3pt) -- (3pt,3pt);
        \draw[solid,thick,-] (-3pt,3pt) -- (3pt,-3pt);
      }
    },
    postaction={decorate}
  }
}

\newcommand{\functionfig}[1]{
  \includegraphics[
    scale=0.3,
    valign=c
  ]{kinematic_flow/Figures/three-site_case/functions/#1.pdf}
}
\newcommand{\x}{\boldsymbol{x}}
\newcommand{\bk}{\boldsymbol{k}}
\newcommand{\dd}{\mathrm{d}}
\newcommand{\Comment}[1]{}

\def\eps{\varepsilon}

\makeatletter
\newlength{\apb@width}
\newcommand{\autoparbox}[2][c]{%
  \settowidth{\apb@width}{#2}%
  \parbox[#1]{\apb@width}{#2}%
}

\makeatother

\allowdisplaybreaks[1]

\begin{document}

\title{Splitting Rules from Kinematic Flow: Local Evolution and Emergent Time}

\author{Ji-Yuan Ke}
\email{kejy22@mails.jlu.edu.cn}
\affiliation{
  State Key Laboratory of High Pressure and Superhard Materials,
  College of Physics, Jilin University,
  Changchun 130012, China
}
\affiliation{
  Center for Theoretical Physics and College of Physics,
  Jilin University,
  Changchun 130012, China
}

\author{Ping He}
\email{hep@jlu.edu.cn}
\affiliation{
  Center for Theoretical Physics and College of Physics,
  Jilin University,
  Changchun 130012, China
}
\affiliation{
  Center for High Energy Physics,
  Peking University,
  Beijing 100871, China
}

\date{\today}

\begin{abstract}
The differential equations satisfied by the wavefunction coefficients of conformally coupled scalars in a power-law cosmology can be recast as an iterative differential system of basis functions.
These functions can be encoded within graph tubings and are
governed by a set of rules describing how they flow in kinematic space.
In this paper, we formulate a set of splitting rules equivalent to the kinematic flow at tree level by reversing the flow direction of graph tubings.
Specifically, for any basis function, these rules identify all other basis functions whose total differentials contain it.
From the splitting perspective, we uncover deeper physical information captured by graph tubings, such as the formation of singularity structures and the realization of local evolution.
In an alternative basis based on time ordering, the splitting processes correspond to the emergence of time integrals.
These rules can also be generalized naturally to the
$\mathrm{tr}\,\phi^3$ theory beyond individual graphs, providing a
physical interpretation of the structure of the associahedron.
This suggests that graph tubings and the kinematic flow may be more
fundamental objects than the differential equations, and may have a
life of their own.
\end{abstract}

\maketitle

\section{Introduction}
Cosmology is one of the most compelling frontiers motivating the exploration of more fundamental physics. A profound realization within this context is that the concept of ``cosmological time" should be discarded, and replaced by a description relying purely on spatial kinematic variables \cite{Arkani-Hamed:2018kmz,Arkani-Hamed:2023kig,Arkani-Hamed:2023bsv,Lee:2023kno}. This requires the wavefunction coefficients to satisfy the corresponding differential equations, which determine how they change as the external kinematics are varied. For example, by using the bulk spacetime isometries on the future boundary of de Sitter space,\footnote{On the future boundary, the isometries of the bulk spacetime will become conformal symmetries, enabling the application of the momentum-space conformal field theory \cite{Bzowski:2013sza, Bzowski:2015pba, Bzowski:2017poo, Bzowski:2019kwd, Bzowski:2020kfw, Dymarsky:2014zja, Gillioz:2018mto, Gillioz:2019lgs, Gillioz:2020mdd}.} together with specific boundary conditions analogous to those for scattering amplitudes \cite{Baumann:2021fxj, Goodhew:2020hob, Goodhew:2021oqg, Melville:2021lst, Raju:2012zr, Maldacena:2011nz}, one can determine the tree-level four-point function for a scalar particle exchange \cite{Arkani-Hamed:2018kmz}. Furthermore, correlators of operators differing by integer weights and spins can be directly related through some weight-shifting and spin-raising operators \cite{Costa:2011dw,Costa:2018mcg,Karateev:2017jgd,Baumann:2019oyu,Baumann:2020dch}, which significantly expands the range of tractable models. This approach, known as the cosmological bootstrap \cite{Arkani-Hamed:2015bza,Arkani-Hamed:2018kmz, Baumann:2019oyu, Baumann:2020dch, Baumann:2022jpr, Sleight:2019mgd, Sleight:2019hfp, Pajer:2020wxk, Albayrak:2020fyp, Salcedo:2022aal, DuasoPueyo:2023kyh, Meltzer:2021zin, DiPietro:2021sjt, Heckelbacher:2022hbq, Hogervorst:2021uvp, Pimentel:2022fsc, Jazayeri:2021fvk, Jazayeri:2022kjy, Loparco:2023rug, SalehiVaziri:2024joi, Armstrong:2022vgl, Benincasa:2020aoj, Albayrak:2023hie, Lee:2023kno, Aoki:2024uyi, Liu:2024xyi, Wang:2025qww}, has seen rapid development and widespread applications in recent years.

However, this description based on spatial kinematics is not even constrained by the symmetries of bulk spacetime. Recent studies have shown that it remains possible to derive appropriate differential equations for the correlators of conformally coupled scalars in a power-law cosmology \cite{Arkani-Hamed:2023kig,Baumann:2025qjx,De:2024zic,De:2023xue,Fevola:2024nzj,He:2024olr}. Specifically, starting from the time-integral representation, the derivatives of the wavefunction coefficients with respect to kinematic variables can be expressed as a linear combination of basis functions. By iteratively differentiating the newly appearing basis functions until they close under differentiation, we obtain a system of differential equations. Although this approach is systematic, as the Feynman diagrams become more complicated, the number of basis functions grows rapidly, making both the derivation and the resulting differential equations intricate.

Fortunately, a set of rules is naturally encoded in these differential equations, revealing how the basis functions flow in kinematic space \cite{Arkani-Hamed:2023kig,Arkani-Hamed:2023bsv}. To construct this, we need to organize the basis functions and the possible singularities in the differential equations (letters) into graph tubings. Then, the kinematic flow describes how tubes are activated and gradually enlarge through growth and absorption. 
As some rules depend only on the boundary data, the kinematic flow not only simplifies the derivation of these differential equations, but also captures the physics of bulk time evolution \cite{Arkani-Hamed:2023bsv}. More recently, the kinematic flow for an alternative basis based on time ordering has been identified in \cite{Baumann:2025qjx}. In this case, these rules are not only simpler to implement, but also carry richer physical interpretations: The merger of two tubes corresponds to the collapse of a time-ordering in the cosmological integrands.

We propose that rather than serving merely as static labels for the basis functions, the graph tubings might encode a richer set of information. Therefore, it is natural to ask whether these graph tubings are governed by other internal rules. In this paper, we propose a new perspective that relates the complete tubings by reversing the evolution direction of the kinematic flow. Specifically, this approach starts from the simplest class of functions in the systems of differential equations, whose total differentials depend only on themselves and the total-energy letters. Then, by iteratively tracing which functions receive a given basis function as a source, i.e. identifying in whose total differentials it appears, we can establish a series of splitting rules among the tubings. These rules reproduce the same system of differential equations for the wavefunction coefficient as the kinematic flow. However, the wavefunction coefficients arise as the final result of the splitting processes. In an alternative basis based on time-ordering \cite{Baumann:2025qjx}, these splitting rules also become considerably simpler.

The key advantage of these splitting rules is that they make manifest the physical content encoded in graph tubings, which is otherwise obscure in the kinematic flow formulation. For example, these rules reveal how the singularity structure associated with each basis function is built up through the system of differential equations. In contrast, while we can deduce the distribution of singularities for the wavefunction coefficients, it is still challenging to extract such structure of any given basis function from the kinematic flow.  Furthermore, this perspective allows us to clearly understand how local evolution is captured by graph tubings. We can view the vertices residing in the same tube as mutually non-local, while the splitting processes will restore  this locality by separating vertices and introducing the corresponding internal energies.

The kinematic flow provides a way to bootstrap the wavefunction coefficients purely from spatial kinematics. Therefore, understanding how time evolution emerges from  kinematic space becomes a crucial task \cite{Arkani-Hamed:2023bsv}. Surprisingly, this can be naturally achieved through our proposed framework of tubing splitting. This is best illustrated in the new basis introduced in \cite{Baumann:2025qjx}, where tubes  enlarge exclusively via merging, and each merger eliminates one time integral. Consequently, each allowed splitting in this basis corresponds to the emergence of a time integral, with different splitting channels arising directly from all possible time orderings. This conclusion is even independent of the choice of function basis, and can be viewed as a consequence of implementing locality or introducing internal energies.

Similar to the case of scattering amplitudes, individual Feynman diagrams are typically not physical observables \cite{Parke:1986gb,Benincasa:2007xk,Hodges:2012ym,Cheung:2014dqa}. It is therefore necessary to generalize the splitting rules to the sum over different channels contributing to a given process, where they now act on the kinematic sub-polygons. We illustrate that, with only minor modifications, the splitting rules remain applicable for analyzing the relations among the basis functions in this context. After summing over graphs, since a basis function appears only within different triangulations rather than levels, the splitting of the kinematic sub-polygons exhibits identical features to those of the tubings: The introduction of each internal energy leads to the emergence of a time integral. This produces a new viewpoint on the deep connections between the tubing kinematics and bulk physics. The perspective of the inverse flow may also endow the relations among the elements of the corresponding associahedron \cite{Arkani-Hamed:2017mur} with a physical interpretation.

\textbf{Outline.} The outline of this paper is as follows: In section~\ref{sec:background}, we review the differential equations and the kinematic flow for the toy model of conformally coupled scalars in a power-law cosmology. Readers familiar with the kinematic flow can skip this section. In section~\ref{sec:alternative}, we construct the splitting rules to analyze the relations among the basis functions, and in section~\ref{sec:examples}, we illustrate with several examples that these rules reproduce the correct differential equations. We also discuss how these rules reveal further structures underlying the differential equations. In section~\ref{sec:another-basis}, we generalize the splitting rules to an alternative basis based on time ordering, and show that each splitting in this context corresponds to the emergence of a time integral. In section~\ref{sec:beyond}, we discuss the case of summing over graphs and construct the appropriate splitting rules. Finally, we draw our conclusions in section~\ref{sec:conclusion}. 

Two appendices are provided to supplement the main text. In appendix~\ref{app:diffeq3site}, we provide the complete set of differential equations in the three-site chain case, which motivated our discovery of the splitting rules. Appendix~\ref{app:more-example} is devoted to more explicit examples to illustrate how the splitting rules are implemented in more complicated cases.
\section{Differential Equations and Kinematic Flow}
We begin by showing how to construct the differential equations and the kinematic flow for the toy model of conformally coupled scalars in a power-law cosmology. While the viability of this construction is rigorously guaranteed by twisted cohomology in mathematics \cite{Mizera:2019ose,Mastrolia:2018uzb,Weinzierl:2022eaz}, we wish to present its derivation from a more physical perspective.
\label{sec:background}
\subsection{Power-Law Cosmology}
\label{sec:power-law}
The present work is concerned with the theory of a conformally-coupled scalar in a power-law Friedmann-Robertson-Walker (FRW) cosmology, with polynomial interactions
\begin{equation}
    S  = \int \dd^4x \, \sqrt{-g} \left[-\frac{1}{2}(\partial \phi)^2 -\frac{1}{12}R\phi^2 - \sum_{p=3} ^\infty\frac{\lambda_p}{p!}\phi^p\right] \,, 
\end{equation}
where $R$ is the Ricci scalar, and the spacetime metric takes a power-law form
\begin{equation}
    \dd s^2 = a^2(\eta)(-\dd \eta^2 + \dd \x^2 )\,,  \; \text{where} \; a(\eta) =\left(\frac{\eta}{\eta_0}\right)^{-(1+\eps)} \,.
\end{equation}
Here, $\eta_0$ is a reference time which is often normalized to unity $\eta_0=1$. The constant parameter $\eps$ characterizes different cosmological scenarios, such as $\eps =0$ (de Sitter), $\eps \approx 0$ (inflation), $\eps =-1$ (Minkowski), $\eps =-2$ (radiation) and $\eps=-3$ (matter). The correlations will be evaluated at a fixed time $\eta_* = 0$, which for an accelerating universe is the future boundary of the spacetime.

While this is only a toy model for studying cosmological correlations, it captures invaluable clues about their underlying structure. Under a Weyl transformation $g_{\mu \nu} \mapsto a^2 g_{\mu \nu}$, $\phi \mapsto a^{-1} \phi$, this action can be transformed to that of a massless field in flat space 
\begin{equation}
    S = \int \dd^4 x\, \left[ -\frac{1}{2}(\partial \phi)^2 - \sum_{p=3}^\infty \frac{\lambda_p(\eta)}{p!}\phi^p\right]\,,
\end{equation}
and all time dependence of the scale factor is encoded in the coupling constant $\lambda_p(\eta) \equiv \lambda_p a(\eta)^{4-p} \propto \eta^{(p-4)(1+\eps)}$. The equivalence between these two actions implies that the mode functions are now identical to those in flat space, $\phi_k^{\rm flat} (k) = e^{ik\eta}/\sqrt{2k} $, which greatly simplifies the computation of the correlation functions.

Our analysis will be focused on the wavefunction of the universe $\Psi[\phi]$, which is the overlap between the Bunch-Davies vacuum and the basis of field eigenstates $|\varphi(\x)\rangle$ at the future boundary 
\begin{equation}
    \Psi[\varphi(\x)] = \langle \varphi(\x) |0 \rangle  = \int^{\phi(0) = \varphi }_{\phi(-\infty) = 0 } {\cal D} \phi \, e^{iS[\phi]} \, .
\end{equation}
Analogous to quantum mechanics, the modulus squared of the wavefunction can be interpreted as a probability distribution in field space, which directly determines the cosmological correlators \cite{Baumann:2026lecture}. In perturbation theory, the wavefunction admits an expansion in powers of field fluctuations (in Fourier space)
\begin{equation}
\begin{aligned}
    \Psi&[\varphi] = \exp \Bigg[-\sum_{n=2}^{\infty} \int \frac{\dd^3 k_1}{(2\pi)^3} \cdots \frac{\dd^3 k_n}{(2\pi)^3} \\ &\times (2\pi)^3   \delta(\bk_1 +\cdots +\bk_n)\, \psi_n(\bk_1,\cdots,\bk_n)\, \varphi_{\bk_1} \cdots \varphi_{\bk_n} \Bigg]  \,, 
\end{aligned}
\end{equation}
where the kernel functions $\psi_n$ are called \textit{wavefunction coefficients}. The wavefunction coefficients serve as the primary objects of our study and
exhibit many features similar to those of scattering amplitudes, such as factorization near singularities \cite{Baumann:2021fxj}. A systematic diagrammatic interpretation of these wavefunction coefficients can be established by constructing appropriate Feynman rules. These rules can be found, for example, in \cite{Arkani-Hamed:2023kig,Baumann:2024mvm} and we only present a
few elements here: the bulk-to-boundary propagator is the solution to the free equation of motion under suitable boundary conditions, which takes the form
\begin{equation}
    K (k,\eta) = e^{ik\eta} \,.
\end{equation}
This corresponds to the external lines in the Feynman diagram. The bulk-to-bulk propagator is the Green's function corresponding to the Klein-Gordon operator $\Box^2-m^2$, satisfying
\begin{equation}
\begin{aligned}
    G(k;\eta,\eta') &= \frac{1}{2k} \Bigg[ e^{-ik(\eta-\eta')} \theta (\eta-\eta')  \\  &+e^{ik(\eta-\eta')}\theta(\eta'-\eta) -e^{ik(\eta+\eta')} \Bigg] \,, \label{bulk-to-bulk-propagator}
\end{aligned}
\end{equation}
which is associated to the internal lines. For each bulk vertex, we should introduce a time-dependent coupling constant $\lambda_p(\eta)$ and integrate over the time $\eta$ at which the interaction occurs. Finally, if the Feynman diagrams contain loops, we should integrate over the corresponding loop momenta.

These simple rules are sufficient to derive the expressions for the wavefunction coefficients in full generality. To illustrate this, we consider the simplest nontrivial case of a two-site chain with $p=3$, representing a single-particle exchange contribution to the four-point function. The wavefunction coefficient is given by 
\begin{align}
    \psi_{(2)} &= \includegraphics[scale=0.5,valign=c]{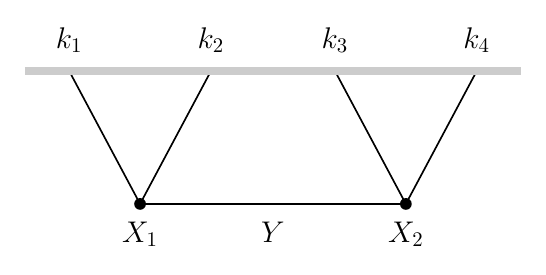} \nonumber \\ &= -\int_{-\infty}^0 \dd \eta \, \dd \eta' e^{iX_1\eta}  \, \lambda_3(\eta) \, G(Y;\eta,\eta') \, \lambda_3(\eta') \, e^{iX_2\eta'} \,,  \label{psi-twosite}
\end{align}
which has been expressed in terms of energy flowing into each vertex, $X_1 \equiv k_1 +k_2$, $X_2 \equiv k_3+k_4$, and the energy of the exchanging particle $Y \equiv |\bk_1 +\bk_2|$. By expanding the time-dependent coupling in frequency space, this expression can be recast into a more compact representation as a twisted integral, whose kernel is linked to the flat-space wavefunction coefficient \cite{Arkani-Hamed:2017fdk,Arkani-Hamed:2023kig}
\begin{equation}
\begin{aligned}
    \psi_{(2)}  & \propto  -\int_0^\infty \dd x_1 \dd x_2 \, (x_1x_2)^\eps \\  &\times\frac{2Y}{(X_1+X_2+x_1+x_2)(X_1+x_1+Y)(X_2 +x_2+Y)} 
    \\ & \equiv  -\int_0^\infty \dd x_1 \dd x_2 \, (x_1x_2)^\eps \frac{2Y}{B_1 B_2B_3}\,, \label{eq:two-site coefficient}
\end{aligned}
\end{equation}
where the linear factors are given by
\begin{equation}
\begin{aligned}
    B_1 &= X_1 +x_1 +Y \,, \qquad B_3 = X_1 +X_2+x_1+x_2 \,, \\ B_2 &= X_1 +x_2 +Y \, .
\end{aligned}
\end{equation}
This formulation transforms the FRW wavefunction coefficient $\psi_{(2)}$ into a two-dimensional rational integrand twisted by powers of integration variables. 
In fact, this structure is shared by all wavefunction coefficients in this model, and is analogous to the loop amplitudes in dimensional regularization.
Based on this feature, pioneering works developed a systematic approach based on differential equations \cite{Arkani-Hamed:2023kig}\footnote{Recently, the method of differential equations has witnessed rapid development in the study of cosmological correlators, with concrete applications ranging from two-site graphs and cosmological loop integrands \cite{De:2023xue,Benincasa:2024ptf,Westerdijk:2025ywh,He:2024olr,Fan:2024iek,Hang:2024xas}, massive particles \cite{Liu:2024str,Gasparotto:2024bku,Baumann:2026atn} and ``unparticles" (bulk CFT primaries) \cite{Pimentel:2025rds,Westerdijk:2026msm}.} to uncover their analytic properties. 
Before introducing this method in detail, we provide the wavefunction coefficient in the case of the three-site chain, which we will study most extensively in this paper
\begin{equation}
\begin{aligned}
     \psi_{(3)} &= \includegraphics[scale=0.5,valign=c]{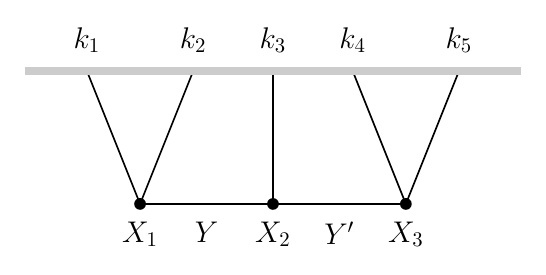} \\ & \propto \int_{0}^\infty \dd x_1 \dd x_2 \dd x_3 (x_1 x_2 x_3)^\eps \frac{4Y Y'}{B_1 B_2 B_3 B_4} \left(\frac{1}{B_5}+\frac{1}{B_6}\right) , \label{eq:three-site coefficient}
\end{aligned}
\end{equation}
with the linear factors defined by 
\begin{equation}
\begin{aligned}
B_1 &\equiv X_1+x_1+Y\,,\\
B_2 &\equiv X_2+x_2+Y+Y'\,,\\
B_3 &\equiv X_3+x_3+Y'\,,\\[2pt]
B_4 &\equiv X_1+X_2+X_3+x_1+x_2+x_3\,,\\
B_5 &\equiv X_1+X_2+x_1+x_2+Y'\,,\\
B_6 &\equiv X_2+X_3+x_2+x_3+Y\,.
\end{aligned}
\label{three-site factor}
\end{equation}
Since all propagators in this model coincide with those in flat space, the kernels of the twist integrals correspond directly to the flat-space wavefunction coefficients $\psi_{\rm flat}$, with the external energies shifted by the integration variables $x_i$ \cite{Arkani-Hamed:2017fdk}.
We are now ready to discuss the differential equations satisfied by these wavefunction coefficients.
\subsection{Differential Equations}
\label{sec:diffeq}
As proposed in \cite{Arkani-Hamed:2023kig}, the twisted integrals in eqs.~\eqref{eq:two-site coefficient} and \eqref{eq:three-site coefficient} are members of a finite-dimensional family of master integrals. These integrals share the same singularity structure and are mutually connected via integration by parts and partial fraction identities. A finite basis of independent master integrals spans a vector space, whose dimension is equal to the number of bounded regions defined by the singularities of the integrand (including the twisted plane). \footnote{The number of independent master integrals is also equal to the dimension of the twisted cohomology group defined by the connection matrix $\tilde A$, as these integrals correspond to different differential forms in twisted space.} Choosing $\psi$ as one of the basis functions, i.e. $\vec{I} \equiv [\psi,I_2,\cdots,I_n]$, and differentiating this basis with respect to the kinematic variables $Z_I \equiv (X_v,Y_e)$ leads to a closed system of differential equations due to the completeness of the basis
\begin{equation}
    \dd \vec{I} = \eps \tilde{A} \,  \vec{I} \,, 
\end{equation}
where $\tilde A$ is an $N \times N$ connection matrix that shares the same characteristics as an Abelian flat connection \cite{Fevola:2024nzj,Capuano:2025ehm}. This feature implies that $\tilde A$ can be expressed as a sum of dlog forms
\begin{equation}
    \tilde A = \sum_i \alpha_i \, \dd \log \Phi_i(Z) \, , \label{connetcion-matrix}
\end{equation}
where $\alpha_i$ are constant matrices and $\Phi_i$ are referred to as \textit{letters}. The complete alphabet of letters captures all possible singularities of the basis functions appearing in the differential equations.

We now briefly illustrate how this method is implemented through the case of two-site chain. In this case, the five lines representing the singularities intersect to form four bounded regions \cite{Arkani-Hamed:2023kig,De:2023xue}. We then define four basis functions via the canonical forms \cite{Arkani-Hamed:2017tmz} associated with each region, one of which is precisely the wavefunction coefficient
\begin{align}
   \begin{bmatrix}
       \psi \\ F \\\tilde F \\Z 
   \end{bmatrix} = \int(x_1 x_2)^\eps \begin{bmatrix}
       \Omega_\psi \\ \Omega_F \\ \Omega_{\tilde F}  \\ \Omega_Z
   \end{bmatrix} \,.
\end{align}
The explicit expressions for the four canonical forms can be found in \cite{Arkani-Hamed:2023kig}. Next, we differentiate the wavefunction coefficient with respect to the kinematic variables $Z_I$ and express the results as linear combinations of different basis functions. The newly appearing functions are referred to as the source functions of $\psi$. By iteratively taking the total differentials of these source functions until the system of differential equations closes, i.e. the total differential depends only on itself, we obtain a complete system of differential equations. In the case of the two-site chain, the differential equations take the form \cite{Arkani-Hamed:2023kig}
\begin{equation}
\begin{aligned}
\dd\psi
&=\eps\Bigl[
 (\psi-F)\,\dd\log(X_1+Y)
 +F\,\dd\log(X_1-Y)
\\[-1pt]
&\quad{}
 +(\psi-\tilde F)\,\dd\log(X_2+Y)
 +\tilde F\,\dd\log(X_2-Y)
 \Bigr]\,,\\[3pt]
\dd F
&=\eps\Bigl[
 F\,\dd\log(X_1-Y)
 +(F-Z)\,\dd\log(X_2+Y)
\\[-1pt]
&\quad{}
 +Z\,\dd\log(X_1+X_2)
 \Bigr]\,,\\[3pt]
\dd\tilde F
&=\eps\Bigl[
 \tilde F\,\dd\log(X_2-Y)
 +(\tilde F-Z)\,\dd\log(X_2+Y)
\\[-1pt]
&\quad{}
 +Z\,\dd\log(X_1+X_2)
 \Bigr]\,,\\[3pt]
\dd Z
&=2\eps\,Z\,\dd\log(X_1+X_2)\,.
\end{aligned}
\label{eq:two-site-differential-system}
\end{equation}
We note that in addition to $\psi$, the system of differential equations  contains another distinguished basis function $Z$, whose total differential takes the simplest form and exhibits only a total energy singularity \cite{Maldacena:2011nz,Raju:2012zr}. When solving these differential equations, we typically employ the solution of this function as a source term, proceeding iteratively to determine the wavefunction coefficient. As the iteration proceeds, higher-level functions acquire an increasing number of singularities (although some of the singularities will be eliminated by imposing appropriate boundary conditions \cite{Arkani-Hamed:2018kmz}), finally yielding $\psi$ with the most complicated singularity structure. We will return to this pattern in detail later.

For the three-site chain and more complicated cases, manipulating these integrals directly becomes highly nontrivial. Fortunately, the authors in \cite{Arkani-Hamed:2023kig} established a convenient basis by introducing the projective simplices associated with all possible combinations of hyperplanes.
This approach significantly simplifies the differentiation process, allowing for a systematic derivation of these differential equations.
All basis functions for the three-site chain case, together with the explicit formulas for the differential equations are presented in appendix~\ref{app:diffeq3site}.
Remarkably, although the singularity hyperplanes define $25$ bounded regions in this case, only $16$ master integrals are required to produce a closed system of differential equations for $\psi$. In this work, we will attempt to explain this phenomenon from a new perspective.
\subsection{Kinematic Flow}
\label{sec:kinematicflow}
While the explicit expressions for these differential equations become intricate for higher-point functions, it is striking that they can be derived straightforwardly by following a series of rules known as ``kinematic flow". These concise and elegant rules unveil hidden structures underlying the differential equations, and are not even confined by the physical interpretation of the cosmological wavefunction \cite{Arkani-Hamed:2023bsv}. To illustrate this, we first introduce the concept of \textit{graph tubings} as in \cite{Arkani-Hamed:2023kig}. Since the wavefunction coefficient depends only on the total external energy flowing into each vertex and the internal energies, we remove all the external lines of the Feynman diagram and assign a cross to each internal line; we refer to such diagrams as marked graphs. A tube is defined as a connected subgraph of the marked graph, represented by circling the enclosed vertices (crosses) and edges. Note that a tube enclosing only a cross but no vertices is forbidden. Two tubes are said to be compatible if their intersections are empty. Furthermore, we define a complete tubing as a maximal set of compatible (non-overlapping) tubes on a graph. Under this formulation, every basis function appearing in the differential equations is in one-to-one correspondence with a complete tubing of the marked graph. Taking the two-site chain as an example, the four basis functions can be represented as follows
\begin{equation}
\label{two-site function}
    \begin{tikzpicture}[baseline=(current  bounding  box.center)]
    \node at (-0.8,0)  {$\psi$};
\node[inner sep=0pt] at (0.1,0)
   {\includegraphics[scale=0.3]{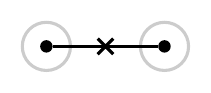}};
   \node at (1.5,0)  {$F$};
 \node[inner sep=0pt] at (2.4,0)
   {\includegraphics[scale=0.3]{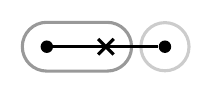}}; 
 \node at (1.5,-0.6)  {$\tilde F$};
    \node[inner sep=0pt] at (2.4,-0.6)
   {\includegraphics[scale=0.3]{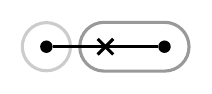}};
    \node at (3.5,0)  {$Z$};
 \node[inner sep=0pt] at (4.4,0)
   {\includegraphics[scale=0.3]{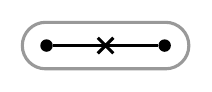}}; 
    \end{tikzpicture}
\end{equation}
The introduction of graph tubings also provides a convenient diagrammatic representation for the letters appearing in the differential equations, such as the letters in eq.~\eqref{connetcion-matrix}. Specifically, the letter associated with a tubing is defined as the dlog of the sum of the energies flowing into the enclosed vertices and the energies of the internal lines crossing the boundary of the corresponding tube. If a cross is enclosed on one side of this tube, the sign of the corresponding internal energy should be reversed. For the two-site chain graph, there are five such letters in the differential equations, which can be graphically represented by
\begin{equation}
\begin{aligned}
\includegraphics[scale=0.3,valign=c]{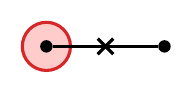} &\equiv \dd \log (X_1 +Y) \, , \qquad \includegraphics[scale=0.3,valign=c]{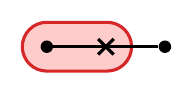} \equiv \dd \log (X_1 - Y) \,, \\
\includegraphics[scale=0.3,valign=c]{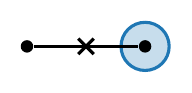} &\equiv \dd \log(X_2+Y)\,, \qquad 
\includegraphics[scale=0.3,valign=c]{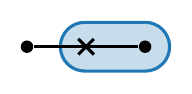} \equiv \dd \log (X_2-Y)\,, \\
\includegraphics[scale=0.3,valign=c]{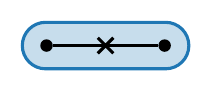} &\equiv \dd \log (X_1+X_2) \,.
\end{aligned}
\end{equation}
These letters connect the total differential of a given basis function to other basis functions and represent the possible singularities of the master integrals.

The kinematic flow states that the differential equation associated with a given parent function defined by a complete tubing is determined by the following three steps (see \cite{Arkani-Hamed:2023bsv,Arkani-Hamed:2023kig} and \cite{Baumann:2024mvm} for cosmological loop integrands):
\begin{itemize}
    \item[1.] \textbf{Activation:} The first step is to identify all tubes in the complete tubing and mark them with distinct colors. These colored tubes are called ``activated", and each activation yields a corresponding branch of the function tree.
    \item[2.] \textbf{Growth and merger:} An activated tube containing no cross can grow by absorbing adjacent crosses. Should any of these crosses be part of another tube, the two tubes merge, with the resulting union becoming activated. New functions generated by this procedure are called the descendants along this branch.
    \item[3.] \textbf{Absorption:} If the crossed side of an activated tube is adjacent to another tube containing a cross, the former can absorb the latter to yield a larger descendant. Note that for every absorption occurring in the path, the resulting descendant functions acquire an additional factor of $-1$.
\end{itemize}
These rules generate a function tree that consists of all basis functions and letters appearing in the total differential of the parent function. With the function tree in hand, the corresponding differential equation can be read off directly: For each graph in the tree, the activated tube corresponds to a letter in the differential equation, weighted by the number of vertices it contains. Then we multiply this letter by the difference between the function associated with the graph and the sum of its immediate descendants, with an overall constant factor $\eps$.

As a concrete example, consider the wavefunction coefficient $\psi$ for the two-site chain case. By applying the rules defined above, one obtains the following function tree
\begin{equation}
\includegraphics[scale=0.7,valign=c]{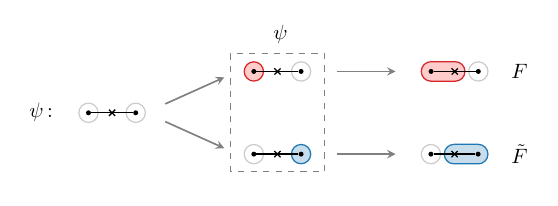}
\end{equation}
In this case, the tubing of the parent function $\psi$ contains two tubes, which give rise to two distinct activation channels. For each channel, a further growth step can be performed, and no absorption process is involved. From this, we can read off the differential equation as follows
\begin{equation}
\begin{aligned}
    \dd \psi &= \eps \Big[(\psi -F) \includegraphics[scale=0.3,valign=c]{kinematic_flow/Figures/two-site_chain/letters/X1+red.pdf}  + F \includegraphics[scale=0.3,valign=c]{kinematic_flow/Figures/two-site_chain/letters/X1-red.pdf}\\ & + (\psi - \tilde F) \includegraphics[scale=0.3,valign=c]{kinematic_flow/Figures/two-site_chain/letters/X2+blue.pdf} +
    \tilde F \includegraphics[scale=0.3,valign=c]{kinematic_flow/Figures/two-site_chain/letters/X2-blue.pdf}\Big] \,,
\end{aligned}
\end{equation}
and there are two new source functions, $F$ and $\tilde F$. The differential equations satisfied by these two functions can be obtained by constructing the function tree via the same procedure. For the case of $F$, this leads to
\begin{equation}
\label{kinematic-flow-F}
\includegraphics[scale=0.7,valign=c]{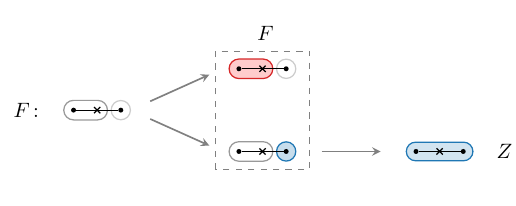}
\end{equation}
Note that at this stage, the lower active tube undergoes a merger, eventually yielding the function $Z$. The function tree for $\tilde F$ can be obtained directly through the symmetry between $X_1$ and $X_2$. The differential equations now are
\begin{align}
    \dd F &= \eps \left[F \includegraphics[scale=0.3,valign=c]{kinematic_flow/Figures/two-site_chain/letters/X1-red.pdf} +(F-Z) \includegraphics[scale=0.3,valign=c]{kinematic_flow/Figures/two-site_chain/letters/X2+blue.pdf} +Z \includegraphics[scale=0.3,valign=c]{kinematic_flow/Figures/two-site_chain/colored/ZZcolored.pdf}\right] \,, \\
    \dd \tilde F &= \eps \left[\tilde F \includegraphics[scale=0.3,valign=c]{kinematic_flow/Figures/two-site_chain/letters/X2-blue.pdf} + (\tilde F -Z) \includegraphics[scale=0.3,valign=c]{kinematic_flow/Figures/two-site_chain/letters/X1+red.pdf} + Z \includegraphics[scale=0.3,valign=c]{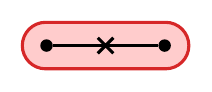}\right]\,.
\end{align}
Finally, there is also a special basis function $Z$. Since the associated tubing leaves no room for growth and absorption, the resulting differential equation takes a particularly simple form
\begin{equation}
    \dd Z = 2 \eps \, Z \includegraphics[scale=0.3,valign=c]{kinematic_flow/Figures/two-site_chain/colored/ZZcolored.pdf} \, .
\end{equation}
We now turn to a more detailed discussion of the kinematic flow. 
Within this framework, the basis functions and letters of the system of differential equations are encoded into distinct graph tubings. Consequently, the relations among these tubings directly mirror the structure of physical observables (correlation functions). The kinematic flow precisely captures these connections through the growth kinematics of tubings.
This naturally induces a hierarchical structure:
As one iteratively differentiates these basis functions, the letters evolve via growth (merger) and absorption, leading to a progressive expansion of the associated tubes. In other words, the evolution originates from the wavefunction coefficient $\psi$, follows different evolutionary paths (growing into different functions), and finally terminates at another special function $Z$, 
\footnote{In more complicated examples discussed in \cite{Arkani-Hamed:2023kig}, the last level of the basis function tree often involves multiple functions --- for example, $\{g,\tilde g,Z \}$ for the three-site chain case. However, it is easy to observe from eqs.~\eqref{gdiffeq} and \eqref{tildeg1} that, the total differentials of both $g$ and $\tilde g$ involve $Z$. The same pattern also holds for other examples.
Consequently, we conclude that all growth channels eventually culminate in $Z$.}
which incorporates all vertices and crosses in the diagram. 
In \cite{Arkani-Hamed:2023kig}, the evolution of these tubings is interpreted as the replacement of different denominator factors by twisted planes. 

This naturally motivates us to explore whether an alternative perspective exists for understanding the relations between these basis functions (complete tubings). More specifically, is there an alternative formulation for these function trees based on graph tubings? 
Guided by these hints, it is natural to consider commencing from the opposite side of the function tree, such as the function $Z$ in the two-site chain case.
Under the kinematic flow rules, such a function corresponds to the terminal stage of  tubing growth.
Thus, if we invert the flow of evolution, then the wavefunction coefficient will emerge as the final result. While such an inverse prescription may be viewed as a mere technical reformulation, the underlying physics is nevertheless highly nontrivial: Starting from the simplest function which depends only on the graph geometry (total energy), we iteratively introduce internal energies by splitting different tubings, and ultimately arrive at the same result as the integrands associated with bulk time evolution.

Given a basis function, the kinematic flow tells us which source functions will appear in its total differential.
Conversely, can we construct an equivalent rule by iteratively asking \textit{for which basis functions a given function serves as a source}?
This question forms the central focus of this work, and we now proceed to elaborate it in detail.

\section{Splitting Rules}
\label{sec:alternative}
We wish to construct a series of rules to systematically recover the wavefunction coefficient from the final descendant of the kinematic flow via intermediate basis functions. 
To achieve this, we define the function associated to a complete tubing enclosing all the sites and vertices as the generating function.  
Such functions exhibit only the total energy singularity, and their differential equations close automatically. 
We now provide some explicit examples for several cases:
\begin{equation}
\label{generating-functions}
\begin{tikzpicture}
\node at (-1.2,0)  {two-site chain\,:};  
\node[inner sep=0pt] at (2,0)
{\includegraphics[scale=0.3,valign=c]{kinematic_flow/Figures/two-site_chain/functions/ZZ.pdf}}; 
\node at (3.5,0)  {$Z$};  
\node at (-1.2,-1)  {three-site chain\,:};
\node[inner sep=0pt] at (2,-1)
{\includegraphics[scale=0.3,valign=c]{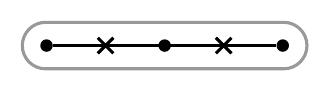}};
\node at (4,-1)  {$Z$};  
\node at (-1.2,-2)  {four-site chain\,:};
\node[inner sep=0pt] at (2,-2)
{\includegraphics[scale=0.3,valign=c]{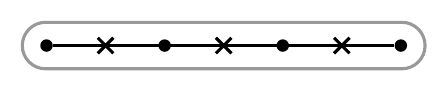}};
\node at (4.3,-2)  {$F_{12_2 3_2 4}$}; 
\node at (-1.2,-3)  {four-site star\,:};
\node[inner sep=0pt] at (2,-3)
{\includegraphics[scale=0.3,valign=c]{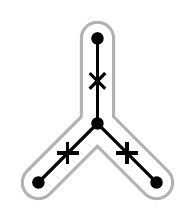}};
\node at (4.3,-3)  {$F_{12 3 4_7}$}; 
\end{tikzpicture} \nonumber
\end{equation}
Here, our notations for the generating functions follow the conventions established in \cite{Arkani-Hamed:2023kig}. 
In this section, we will show how the relations among the complete tubings can be revisited through the lens of tubing splitting. Furthermore, we shall see that this perspective also permits a systematic approach equivalent to the kinematic flow, which produces the correct differential equations.

\subsection{Splitting the Complete Tubings}
\label{sec:splitting-rules}
Since the differential equations for the two-site chain case are relatively trivial, we begin our analysis with the three-site chain, which contains richer information.
In this case, the system of differential equations for $\psi$ contains 16 basis functions, each of them corresponding to a distinct complete tubing
\begin{equation}
{
\setlength{\arraycolsep}{0pt}
\begin{array}{
  @{}
  r@{\hspace{0.35em}}l
  @{\hspace{1.4em}}
  r@{\hspace{0.35em}}l
  @{\hspace{1.4em}}
  r@{\hspace{0.35em}}l
  @{}
}
\psi
& \functionfig{psi}
&
f
& \functionfig{f}
&
g
& \functionfig{g}
\\[3pt]
{}
& {}
&
q_1
& \functionfig{q1}
&
\tilde g
& \functionfig{tildeg}
\\[3pt]
F
& \functionfig{F}
&
q_2
& \functionfig{q2}
&
Z
& \functionfig{Z}
\\[3pt]
\widetilde F
& \functionfig{tildeF}
&
q_3
& \functionfig{q3}
&
{}
& {}
\\[3pt]
Q_1
& \functionfig{Q1}
&
\tilde q_1
& \functionfig{tildeq1}
&
{}
& {}
\\[3pt]
Q_2
& \functionfig{Q2}
&
\tilde q_2
& \functionfig{tildeq2}
&
{}
& {}
\\[3pt]
Q_3
& \functionfig{Q3}
&
\tilde q_3
& \functionfig{tildeq3}
&
{}
& {}
\end{array}
}
\label{three-site source}
\end{equation}

We first note that the differential equation satisfied by the generating function $Z$ takes a particularly simple form 
\begin{equation}
    \dd Z =  3 \eps \, Z  \,\dd \log (X_1 +X_2 +X_3) \, , 
\end{equation}
and the differential equations satisfied by other basis functions are collected in appendix~\ref{app:diffeq3site}. 
The distinctive feature of this function stems from the replacement of all linear factors (except $B_4$) in eq.~\eqref{three-site factor} by the twisted planes. From the perspective of twisted cohomology, this function corresponds to the canonical form of the region formed by all twisted planes and the ``total energy plane" $\sum_i (X_i+x_i)$ \cite{Arkani-Hamed:2023kig}.
More subtly, this function also serves as the generating function in an alternative basis proposed in \cite{Baumann:2025qjx}, 
\footnote{Recently, this basis has also been referred to as the dual basis. The invariant behavior of the generating function in both bases arises from their association with the same bounded region. This pattern holds for all Feynman diagrams within this model.} 
where it carries profound physical implications --- corresponding to the collapse of all time orderings. We will discuss this in detail in section~\ref{sec:another-basis}.

We now seek the basis functions that are connected to $Z$ by differential equations. By inspection, we identify four basis functions $\{  q_2,g,\tilde g, \tilde q_2\}$, whose total differentials involve only themselves and $Z$, and share a similar differential structure
\begin{align}
\dd q_2
&=\eps\Bigl[
 2q_2\,\dd\log(X_1+X_2-Y')
 +(q_2-Z)\,\dd\log(X_3+Y)
 \notag\\[-1pt]
&\quad{}
 +Z\,\dd\log(X_1+X_2+X_3)
 \Bigr]\,,
\label{q2diffeq}\\[3pt]
\dd g
&=\eps\Bigl[
 2g\,\dd\log(X_1+X_2+Y')
 +(g+Z)\,\dd\log(X_3-Y)
 \notag\\[-1pt]
&\quad{}
 -Z\,\dd\log(X_1+X_2+X_3)
 \Bigr]\,,
\label{gdiffeq2}\\[3pt]
\dd\tilde g
&=\eps\Bigl[
 2\tilde g\,\dd\log(X_2+X_3+Y)
 +(\tilde g+Z)\,\dd\log(X_1-Y)
 \notag\\[-1pt]
&\quad{}
 -Z\,\dd\log(X_1+X_2+X_3)
 \Bigr]\,,
\label{tildegdiffeq}\\[3pt]
\dd\tilde q_2
&=\eps\Bigl[
 2\tilde q_2\,\dd\log(X_2+X_3-Y)
 +(\tilde q_2-Z)\,\dd\log(X_1+Y)
 \notag\\[-1pt]
&\quad{}
 +Z\,\dd\log(X_1+X_2+X_3)
 \Bigr]\,.
\label{tilderq2diffeq}
\end{align}
Additionally, we notice that the differentials of three other basis functions, $\{q_1,f,\tilde q_1\}$, also depend on $Z$, but their connections are mediated by the four aforementioned functions (see eqs.~\eqref{diffeqq1} to \eqref{diffeqtildeq1}). We refer to this configuration as indirectly connected to $Z$.

Although these features remain somewhat opaque through the simplex notation in \cite{Arkani-Hamed:2023kig}, they become transparent from the perspective of graph tubings shown in eq.~\eqref{three-site source}. Specifically, under the kinematic flow rules, the tubings associated with $\{q_2,g,\tilde g, \tilde q_2\}$ require only a single growth (merger) or absorption step to reach $Z$. In contrast, the tubings for $\{q_1,f,\tilde q_1\}$ require an intermediate procedure to reach those of $\{q_2,g,\tilde g, \tilde q_2\}$, before connecting to $Z$ via an additional absorption.
Given that the processes of growth (merger) and absorption essentially amount to fusing two tubes into a larger one, the inverse processes can be naturally interpreted as tubing splittings. 
This is easily verified from the tubing representation eq.~\eqref{three-site source}: the tubings of the functions $\{q_2,g,\tilde g, \tilde q_2\}$ correspond to a single splitting of those of  $Z$, while $\{q_1,f,\tilde q_1\}$ emerge as the result of two successive splittings.
Accordingly, we can classify the basis functions into different levels based on the number of splittings required to derive them from the generating function. 
\footnote{In fact, this classification coincides with that employed in constructing the Pfaffian chain for the cosmological correlators \cite{Grimm:2024mbw}. In that context, functions derived from different numbers of splittings are characterized by a corresponding number of cuts.}

This motivates us to formulate a set of rules for organizing these basis functions from the perspective of splitting.
In constructing these rules, it is crucial to note that the growth (merger) and absorption operations in the kinematic flow exhibit distinct properties. For instance, according to the rules detailed in section~\ref{sec:kinematicflow}, a basis function generated via absorption inherently acquires an overall minus sign.
Therefore, during the splitting procedure, we must carefully distinguish whether it corresponds to the inverse of a growth or absorption.

\noindent \textbf{Splitting Rules.} With these considerations in mind, we summarize the splitting rules for the complete tubings as follows:
\begin{itemize}
    \item [1.]  A tube in the complete tubing containing more than one site can split into two smaller ones, with the splitting occurring anywhere between a site and a cross.
    \item [2.] If one of the resulting tubes contains only a single site, we denote this splitting with a solid line. The inverse of this process corresponds to a growth (merger). Otherwise, the splitting is represented by a dashed line, with its inverse corresponding to an absorption.
    \item [3.] If a splitting yields a tube containing only a single cross, this process is forbidden, unless the parent tube 
    \footnote{Since the tubing evolution under these splitting rules is reversed compared to that in \cite{Arkani-Hamed:2023kig,Arkani-Hamed:2023bsv}, the parent functions hereafter should be interpreted as the basis functions that yield the given function through splitting. Similarly, we also refer to the parents of a parent function as ancestor functions.}
    consists of only one site and one cross. Such forbidden processes are denoted by crossed dashed lines, indicating that the corresponding functions have no direct relation.
\end{itemize}
We now illustrate how these rules work with a few concrete examples.
First, an inspection of the tubing of $Z$ reveals four possible locations where a splitting can occur,
\begin{equation}\nonumber
\begin{tikzpicture}
\node[inner sep=0pt] at (2,-1)
{\includegraphics[scale=0.3,valign=c]{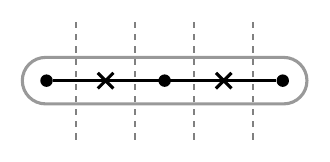}};
\end{tikzpicture}
\end{equation}
and performing these splittings yields precisely
\begin{equation}\nonumber
\includegraphics[
  scale=0.3,
  valign=c
]{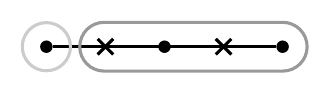}
\hspace{0.8em}
\includegraphics[
  scale=0.3,
  valign=c
]{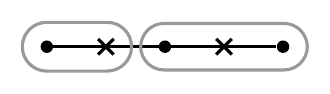}
\hspace{0.8em}
\includegraphics[
  scale=0.3,
  valign=c
]{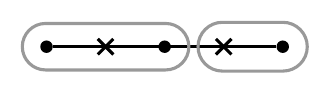}
\hspace{0.8em}
\includegraphics[
  scale=0.3,
  valign=c
]{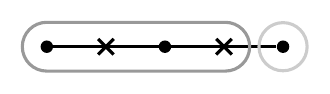}
\end{equation}
Since these four functions can only be obtained through the splitting of $Z$, their total differentials depend exclusively on themselves and $Z$. Moreover, we observe that the splitting processes to $g$ and $\tilde g$ produce no tube containing only a single site, i.e.\includegraphics[scale=0.3,valign=c]{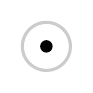}, which implies they correspond to the inverse of absorptions. Therefore, in the differential equations for $q_2$ and $g$ eqs.~\eqref{q2diffeq} and \eqref{gdiffeq2}, the contributions from $Z$ differ by a relative minus sign.

As another nontrivial example of tubing splitting, we consider the function $q_2$ derived from $Z$. Since one splitting has already been performed, further splittings can only take place at the three remaining locations
\begin{equation} \label{q2splitting}
\includegraphics[scale=0.7,valign=c]{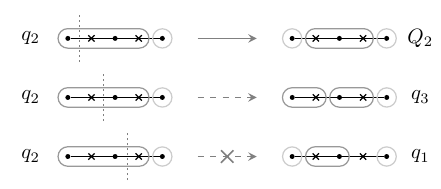}
\end{equation}
Remarkably, this splitting process involves all three scenarios outlined in our rules. Following our previous discussion, we can deduce that $q_2$ appears in the differential equations for both $Q_2$ and $q_3$, but differing by a relative minus sign. This can be readily verified in eqs.~\eqref{diffeqQ2} and \eqref{diffeqq3}. 
Furthermore, the third line in this process eq.~\eqref{q2splitting} is forbidden, as the splitting yields a tube containing solely a cross, i.e.\includegraphics[scale=0.3,valign=c]{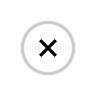}.
This implies that the total differential of the function $q_1$ is independent of $q_2$, as explicitly shown in eq.~\eqref{diffeqq1}.

Finally, let us turn to a slightly different example, where we investigate which basis functions can directly split into the tubing of the wavefunction coefficient $\psi$. 
Given that generating $\psi$ requires exactly four splitting steps, the level immediately preceding it consists of four basis functions. The splitting process is given by
\begin{equation}
\includegraphics[scale=0.7,valign=c]{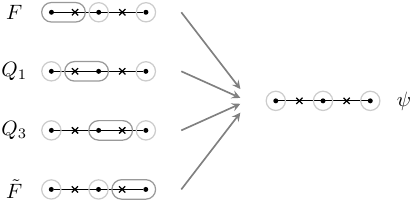}
\end{equation}
For the four functions $\{F,Q_1,Q_3,\tilde F\}$, their splitting processes all correspond to a parent tube with a site and a cross yielding a single-vertex tube, and are thus allowable. 
Since all four channels are denoted by solid lines, their contributions to the differential equations for $\psi$ take a universal form: $(\psi - \{F,Q_1,Q_3,\tilde F\})$, which can be examined in eq.~\eqref{diffeqpsi}. 
Based on these rules, any tree-level basis function can be analyzed in the same manner.

However, the splitting rules outlined above are insufficient to fully determine the differential equation for an arbitrary function. On the one hand, an activated function branch in the kinematic flow does not undergo merely a single growth or absorption step. Therefore, when analyzing the splitting of tubings, we must trace back to functions beyond the immediately preceding level. On the other hand, distinct letters also appear in the differential equations and is accompanied by specific function combinations. Thus, it is essential to establish the relations among the letters directly from the perspective of splittings.
To achieve this, a key insight is that these letters inherently satisfy specific linear relations, which are known as \textit{the three-letter relations} \cite{Arkani-Hamed:2023kig,Baumann:2025qjx}.
As we shall see, these identities emerge precisely as a consequence of analyzing the tubing structures from the perspective of splittings. This enables us to formulate a complete set of rules, which is strictly equivalent to the kinematic flow.
\subsection{Three-Letter Relations and Splittings of Letters}
\label{sec:generation-rule}
Once the relations among these basis functions have been reconstructed from tubing splittings, their differential equations can readily be derived through the kinematic flow.
However, it will be more illuminating to construct the rules for the differential equations from function generation via splittings.
Since each linear combination of the functions (e.g. $(q_2 -Z)$ in eq.~\eqref{q2diffeq}) in the differential equations is accompanied by a specific letter, we need to unravel the relations among these letters.
For the case of the three-site chain, the letters entering the differential equations are given by 
\begin{equation}
\begin{alignedat}{2}
\hat B_1&\equiv X_1+Y\,,
&\quad \hat B_2 &\equiv X_2+Y+Y'\,,
\\[-1pt] \hat B_3 &\equiv X_3+Y'\,,
&\quad \hat B_4 &\equiv X_1+X_2+X_3\,,
\\[-1pt] \hat B_5 &\equiv X_1+X_2+Y'\,,
&\quad \hat B_6 &\equiv X_2+X_3+Y\,,
\\[-1pt] \hat B_7 &\equiv X_1-Y\,,
&\quad \hat B_8 &\equiv X_2-Y+Y'\,,
\\[-1pt] \hat B_9 &\equiv X_2+Y-Y'\,,
&\quad \hat B_{10} &\equiv X_2-Y-Y'\,,
\\[-1pt] \hat B_{11} &\equiv X_3-Y'\,,
&\quad \hat B_{12} &\equiv X_1+X_2-Y'\,,
\\[-1pt] \hat B_{13} &\equiv X_2+X_3-Y\,.
\end{alignedat}
\label{three-site-letters}
\end{equation}
A quick glance at these expressions reveals that they are not all linearly independent.
In fact, this is a consequence of the local encoding of kinematic data by the marked graphs, and thus the letters associated to certain tubings must be related \cite{Arkani-Hamed:2023kig}. 
Since the minimal set of linearly dependent letters consists of  three elements, such an identity is referred to as a three-letter relation.
More specifically, if a tubing associated to a given letter admits a splitting into two smaller ones, then the three corresponding letters will satisfy a linear relation. For example, in the three-site chain, the tubing containing all sites and crosses admits four distinct splittings, which are identical to those of the generating function. This leads to \cite{Arkani-Hamed:2023kig}
\begin{align}
1:\quad \hat B_4
&=\hat B_5+\hat B_{11} \notag\\[-2pt]
&\Leftrightarrow\; \includegraphics[scale=0.3,valign=c]{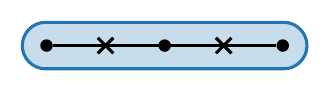}
= \includegraphics[scale=0.3,valign=c]{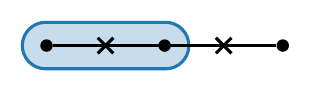}
+\includegraphics[scale=0.3,valign=c]{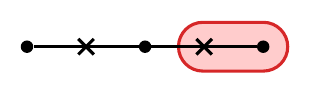}
\,,\label{three-letter1}
\\[2pt] 2:\quad \hat B_4
&=\hat B_{12}+\hat B_3 \notag\\[-2pt]
&\Leftrightarrow\; \includegraphics[scale=0.3,valign=c]{kinematic_flow/Figures/three-site_case/colored/Zcolored.pdf}
=\includegraphics[scale=0.3,valign=c]{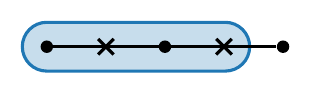}
+\includegraphics[scale=0.3,valign=c]{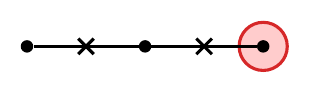}
\,,\\[2pt]
3:\quad \hat B_4 &=\hat B_6+\hat B_7
\notag\\[-2pt] &\Leftrightarrow\;
\includegraphics[scale=0.3,valign=c]{kinematic_flow/Figures/three-site_case/colored/Zcolored.pdf}
= \includegraphics[scale=0.3,valign=c]{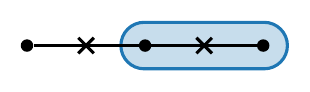}
+ \includegraphics[scale=0.3,valign=c]{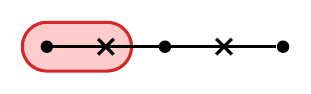}
\,, \\[2pt]
4:\quad \hat B_4 &=\hat B_{13}+\hat B_1
\notag\\[-2pt] &\Leftrightarrow\;
\includegraphics[scale=0.3,valign=c]{kinematic_flow/Figures/three-site_case/colored/Zcolored.pdf}
=\includegraphics[scale=0.3,valign=c]{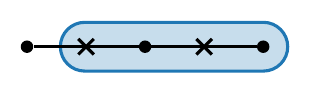}
+\includegraphics[scale=0.3,valign=c]{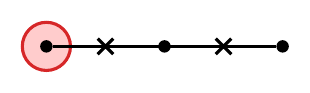}
\,.\label{three-letter4}
\end{align}
Note that, unlike in the differential equations, the tubings here represent the letters themselves but not their dlog forms. 
Analogously, any tubing containing more than one site
\footnote{In this work, the range of splittings is slightly broader than that discussed in \cite{Arkani-Hamed:2023kig}. Specifically, establishing the equivalence between the kinematic flow and the splitting rules requires us to include splittings of the following form
\begin{equation}
\includegraphics[scale=0.3,valign=c]{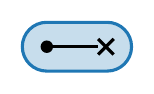} = \includegraphics[scale=0.3,valign=c]{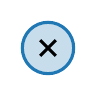} + \includegraphics[scale=0.3,valign=c]{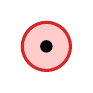}\,,
\end{equation}
although the tubing on the right containing only a cross is forbidden (and thus omitted). If we interpret this as a letter representing $2Y$, where $Y$ is the internal energy intersecting the tube, the above equation can also be regarded as a three-letter relation.}
can split in a similar fashion, generating a set of corresponding three-letter relations.
From the geometric viewpoint, each letter is associated with a hyperplane. Thus, the three-letter relations can also be recognized as the incidences of three of these hyperplanes along a codimension-two surface. 
These relations are utilized in \cite{Arkani-Hamed:2023kig,Baumann:2025qjx} to establish the integrability condition for the basis functions, i.e. $\dd^2 = 0$.

It is obvious that the tubings associated with these letters follow the same splitting rules as those of the basis functions. 
In other words, given a tubing associated with a basis function, it can be regarded as consisting of different sub-tubings corresponding to distinct letters. 
Consequently, all allowed splittings of these letters give rise to the complete set of next-level functions for this given function. This implies that every splitting of a basis function also encodes a three-letter relation. 
Motivated by these observations, we can collect all generated letters on the right-hand side of eqs.~\eqref{three-letter1} to \eqref{three-letter4} into a single marked graph and apply the splitting rules defined in section~\ref{sec:splitting-rules}.
Then, we can obtain
\begin{align}
\includegraphics[scale=0.7,valign=c]{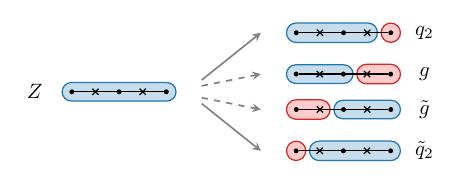}
\end{align}
These are precisely the four splitting channels of the generating function $Z$.
A comparison with eqs.~\eqref{q2diffeq} to \eqref{tilderq2diffeq} reveals that all letters involved in the splitting processes appear in the differential equations for the descendant functions. 
Moreover, the generated letters can be  classified into two categories by color, as they play different roles in the differential equations: 
The blue letters appear in the differential equations exclusively as factors multiplying the basis functions, whereas the red letters multiply the difference between the basis functions and their parent functions.
We refer to the former as \textit{passive letters} and the latter as \textit{active letters}. Strikingly, these simple extra definitions are enough to determine all the differential equations. Taking one of the descendants $q_2$ as an example, we can deduce that
\begin{equation}
    \dd q_2 \supset  \eps \Big[\,2q_2\includegraphics[scale=0.3,valign=c]{kinematic_flow/Figures/three-site_case/letters/X12-blue.pdf} +(q_2-Z) \includegraphics[scale=0.3,valign=c]{kinematic_flow/Figures/three-site_case/letters/X3+red.pdf} +Z \includegraphics[scale=0.3,valign=c]{kinematic_flow/Figures/three-site_case/colored/Zcolored.pdf} \Big]\,.
\end{equation}
Fortunately, this completes the differential equation for $q_2$. However, for a generic basis function, its differential equation may involve some ancestor functions beyond the immediate parent. Furthermore, a systematic criterion is also required  to assign the active and passive roles to the descendant letters.
With these ingredients in place, we now summarize a complete set of rules valid for arbitrary tree-level graphs. We begin with the splitting of the letters. Representing the passive and active letters with blue and red tubings, we define the following splitting rules
\begin{itemize}
    \item \textbf{Splitting the passive tubings.} A passive (blue) tube can split into two descendants following the rules in section~\ref{sec:splitting-rules}. The descendant tube containing a cross inherits the passive property, while the other becomes active. If both the descendants contain crosses, the tube involving the site adjacent to the splitting position remains passive, and the other becomes active.
    \item \textbf{Splitting the active tubings.} Similarly, an active (red) tube can also undergo a splitting process. In this case, the passive part of the original basis function's tubing remains unchanged, while the descendant tube without involving a cross inherits the active property. If both the descendants contain crosses, then the tube
    lacking the site adjacent to the splitting position becomes active.
\end{itemize}
Furthermore, if the tubing of a basis function cannot be obtained via splitting, we denote all its sub-tubing(s) as passive.

We now illustrate these two splitting rules through two examples. We first consider the splitting of the following basis function in the four-site chain case
\begin{align}
\nonumber
\includegraphics[scale=0.7,valign=c]{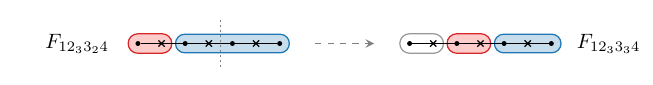}
\end{align}
This corresponds to a passive tube splitting into two smaller tubes. Since both resulting tubes contain crosses, the tube involving the third site from the left remains passive, and the other becomes active. The property of the remaining uncolored tube should be determined from the splitting of other parent functions of $F_{12_33_34}$. 
From this, we can deduce specific terms in the differential equation for the descendant basis function
\begin{equation}
\begin{aligned}
\dd F_{12,33,34} \supset\eps\Bigl[\,
&2F_{12,33,34}\,\includegraphics[scale=0.3,valign=c]{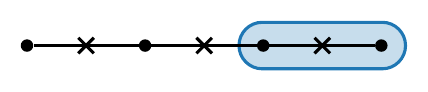}
\\[-2pt]
&+(F_{12,33,34}+F_{12,33,24})\,\includegraphics[scale=0.3,valign=c]{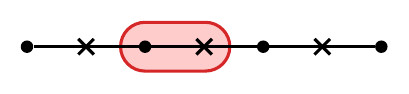}
\\[-2pt]
&-F_{12,33,24}\,\includegraphics[scale=0.3,valign=c]{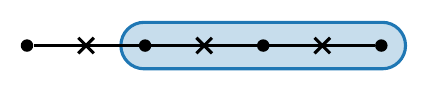}\Bigr]\,,\end{aligned}
\end{equation}
which can be easily verified in the supplementary Mathematica notebook of \cite{Arkani-Hamed:2023kig}. Here, the positive sign in the second term arises from the fact that this splitting is denoted by a dashed line, whose inverse process corresponds to an absorption. 

Next, we consider an example involving the splitting of an active tube 
\begin{align}
\nonumber
\includegraphics[scale=0.7,valign=c]{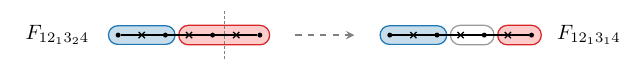}
\end{align}
In this example, the passive tube within the tubing of the parent function remains unchanged, whereas the tube lacking the third site from the left becomes active.
From this process, we can deduce that the differential equation for the basis function $F_{12_13_14}$ involves
\begin{equation}
\begin{aligned}
\dd F_{12,13,14} \supset\eps\Bigl[\,
&2F_{12,13,14}\,\includegraphics[scale=0.3,valign=c]{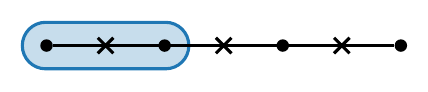}
\\[-2pt]
&+(F_{12,13,14}+F_{12,13,24})\,
\includegraphics[scale=0.3,valign=c]{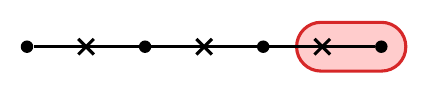}
\\[-2pt]
&-F_{12,13,24}\,\includegraphics[scale=0.3,valign=c]{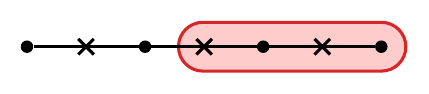}\Bigr]\,.
\end{aligned}
\end{equation}

Given these splitting rules, one can systematically derive the colored tubings corresponding to all basis functions starting from the generating function $Z$. 
\footnote{In practice, if we are only concerned with the colored tubing configuration of a specific function, it suffices to analyze all of its parent functions. We relegate the detailed discussion of this procedure to appendix~\ref{app:more-example}.} 
Now it remains only to determine how the ancestor functions appear in the differential equation of a given basis function. From a careful examination of when growth and absorption are permitted within the kinematic flow, we extract a pair of straightforward rules: 
(1) Given the tubing of a basis function, if the passive tube adjacent to an active one does not terminate at a site on the opposite side, the immediate ancestor function connected by a dashed line in this direction also contributes to its differential equation.
(2) If two (or more) parents of a given function trace back to the same ancestors via dashed lines (including forbidden ones), we also treat these ancestor functions as its parent. 
Thus, the complete differential equation is obtained by multiplying the combination of basis functions by the letters of corresponding properties, with each letter used exactly once. 
Note that we have implicitly adopted certain conventions identical to those in the kinematic flow, such as the introduction of the $\eps$ factor and the multiplication by the number of sites within the letters.

Before proceeding to some explicit examples, let us first elucidate how these splitting rules are equivalent to the kinematic flow. 
\subsection{Comparison with Kinematic Flow}
The kinematic flow and the splitting rules evolve the tubings in opposite directions. 
The former starts from the tubing of the wavefunction coefficient,
describing how an activated tube continuously expands via absorbing crosses or tubings with crosses, and finally reaches its maximal configuration. Conversely, the latter first defines the maximal tubing of the marked graph, and iteratively splits it into smaller ones.
Therefore, a necessary condition for the splitting rules to hold is the invertibility of the kinematic flow. Within the basis employed in \cite{Arkani-Hamed:2023kig} and this section, the kinematic flow rules involve several subtle properties. This subsection is devoted to illustrating the one-to-one correspondence between these two sets of rules.

We first consider the three possible types of splitting. For the splitting processes indicated by solid lines, their inverses correspond to growths or mergers. This is because such a single-site tube obtained from the splittings is just an activated tube of the kinematic flow.
Accordingly, we generally prohibit the emergence of single-cross tubes during splitting, unless the parent tube contains only one site and one cross. The latter precisely corresponds to the scenario where an activated tube grows by enclosing a cross. 
As for the processes denoted by  dashed lines, their inverses correspond to absorptions. Since both tubes resulting from such a splitting are activated in the kinematic flow rules, the differential equation of the descendant function must involve this parent. 
Finally, for the processes denoted by crossed dashed lines, their inverses are strictly forbidden, implying that the two involved functions are irrelevant.

Next, we turn to the splitting of the tubings associated to letters. It is evident that the splitting of either a passive or an active tube consistently generates a new active tube.
Under the kinematic flow rules, this active tube becomes activated and absorbs the other part through any allowed process, yielding the corresponding parent function.
Notably, the splitting behavior of the active tubes is even more special. 
This is because, in order to generate an active tube that can continue to split, the inverse process must correspond to an absorption. Thus, when a descendant function is generated via the splitting of an active tube, its differential equation inevitably involves the ancestor function along this direction. Readers can explicitly verify this point in the subsequent examples.
Furthermore, we emphasize that while the three-letter relations enlightened the splitting of tubings, they do not directly dictate the explicit form of the differential equations.

Finally, we give some explanations of the two rules for determining how the ancestor functions contribute to the differential equation of a given function.
Within the kinematic flow, absorption processes are strictly directional; they are permitted only when the cross of the absorbing tube is oriented toward the absorbed tube. Therefore, given a basis function, one cannot deduce the presence of its ancestor functions in the differential equation merely by relying on extended dashed lines.
The introduction of Rule (1) is specifically designed to address this shortcoming: after an active tube absorbs its adjacent component, if the resulting tube still possesses an outward-oriented cross, then further absorption is valid.
As for Rule (2), it stems from a simple fact within the kinematic flow: A tube without any crosses can grow by enclosing more than one adjacent cross. Should it absorb more than one cross, the inverse tracking process will admit multiple trajectories. 

Thus, we demonstrated that every splitting rule formulated in this work exhibits a precise one-to-one correspondence with the rules governing the kinematic flow. From this, it is evident that these splitting rules are equivalent to the kinematic flow at tree level.
\subsection{Selected Examples}
\label{sec:examples}
In this subsection, we illustrate with a few concrete examples how these splitting rules can be used to yield the correct differential equations.  We will begin with the simplest two-site chain case, and subsequently turn to the three-site chain to demonstrate how these rules yield the correct differential equations. By construction, the domain of validity for the splitting rules should coincide precisely with that of the kinematic flow (at least at tree level).  We have also successfully checked the more intricate four-site cases, with certain nontrivial steps detailed in appendix~\ref{app:more-example}.
In fact, we believe that the existence of such rules itself is more fundamental than their explicit formulations. Readers not interested in the explicit examples can skip this subsection and move directly to section.~\ref{sec:phy-implications} to see the physical implications of these rules.
\subsubsection{Two-site chain}
\label{sec:two-site}
For the two-site chain case, the system of differential equations for the wavefunction coefficient $\psi$ involves four basis functions. We first consider the generating function $Z$, which is associate with the tubing enclosing all the sites and crosses. Since this function cannot be generated via splittings, the corresponding differential equation is the simplest
\begin{equation}
    \dd Z = 2  \eps\, Z  \includegraphics[scale=0.3,valign=c]{kinematic_flow/Figures/two-site_chain/colored/ZZcolored.pdf} \,.
\end{equation}
As a warmup for the splitting rules, we next consider the splitting of the function $Z$.  Since the tubing of $Z$ allow for two distinct splitting positions, it leads to the following two channels
\begin{align}
\label{twosite-splitZ}
\includegraphics[scale=0.7,valign=c]{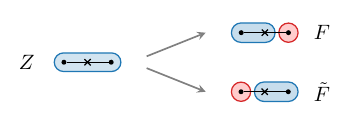}
\end{align}
A direct comparison with eq.~\eqref{kinematic-flow-F} reveals that each active tube generated by the splitting processes precisely corresponds to the growing tube in the kinematic flow. Notably, these two channels are mapped into each other under the interchange $X_1 \leftrightarrow X_2$, which reflects the exchange symmetry between the two equivalent vertices.
The differential equations for the two resulting functions can be straightforwardly read off
\begin{align}
    \dd F& = \eps \Big[\,F \includegraphics[scale=0.3,valign=c]{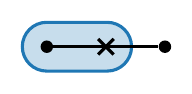} + (F-Z) \includegraphics[scale=0.3,valign=c]{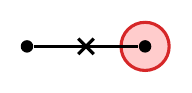}  +Z \includegraphics[scale=0.3,valign=c]{kinematic_flow/Figures/two-site_chain/colored/ZZcolored.pdf}\Big]\,,  \\
    \dd \tilde F &= \eps \Big[\,\tilde F \includegraphics[scale=0.3,valign=c]{kinematic_flow/Figures/two-site_chain/letters/X2-blue.pdf} +(\tilde F - Z) \includegraphics[scale=0.3,valign=c]{kinematic_flow/Figures/two-site_chain/letters/X1+red.pdf}  +Z \includegraphics[scale=0.3,valign=c]{kinematic_flow/Figures/two-site_chain/colored/ZZcolored.pdf}\Big] \, .
\end{align}
Taking $F$ and $\tilde F$ as the parent functions and implementing the splitting rules, it is evident that they both yield the same descendant function $\psi$
\begin{align}
\includegraphics[scale=0.9,valign=c]{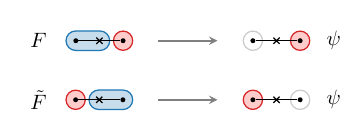}
\end{align}
Here, the two channels generate distinct active tubes, implying that the two parent functions will multiply different letters in the differential equation of $\psi$. Moreover, no passive tubes are generated in these two splitting processes. Since the splittings of $Z$ into $F$ and $\tilde F$ are both represented by solid lines, the differential equation for $\psi$ is naturally free of its ancestor functions. More concretely, we could obtain
\begin{equation}
\begin{aligned}
\dd\psi=\eps\Bigl[\,&(\psi-F)\,\includegraphics[scale=0.3,valign=c]{kinematic_flow/Figures/two-site_chain/letters/X1+red.pdf}
\\[-2pt]
&+(\psi-\tilde F)\,\includegraphics[scale=0.3,valign=c]{kinematic_flow/Figures/two-site_chain/letters/X2+red.pdf}
\\[-2pt]
&+F\,\includegraphics[scale=0.3,valign=c]{kinematic_flow/Figures/two-site_chain/letters/X1-blue.pdf}
+\tilde F\,\includegraphics[scale=0.3,valign=c]{kinematic_flow/Figures/two-site_chain/letters/X2-blue.pdf}\Bigr]\,.
\end{aligned}
\end{equation}
These differential equations are consistent with the results derived in section~\ref{sec:kinematicflow}. We are now ready to investigate some more complicated examples.
\subsubsection{Three-site chain}
We then turn to the three-site chain case, where the system of differential equations for $\psi$ involves 16 basis functions. The relevant graph tubings for these basis functions are present in eq.~\eqref{three-site source}, while the letters appearing in the differential equations can be found in eq.~\eqref{three-site-letters}. At this time, the generating function continues to satisfy the simplest differential equation, which only depends on the dlog of the total energy
\begin{equation}
    \dd Z = 3\eps \, Z \includegraphics[scale=0.35,valign=c]{kinematic_flow/Figures/three-site_case/colored/Zcolored.pdf}\, .
\end{equation}
In what follows, we will demonstrate how this generating function $Z$ iteratively splits into other basis functions, and ultimately yields the wavefunction coefficient $\psi$. We will also show that these splitting rules lead to the correct differential equations.
\begin{itemize}
\item We first observe that the tubing of $Z$ contains four regions between a site and a cross, each of which corresponds to a distinct splitting channel
\begin{align}
\includegraphics[scale=0.7,valign=4]{kinematic_flow/Figures/three-site_case/rules/Z_split.pdf}
\end{align}
Note that for the channels denoted by dashed lines, the corresponding parent functions acquire a relative minus sign in the differential equations. The differential equations for $q_2$ and $g$ are thus given by
\begin{align}
\dd q_2 &=\eps\Bigl[\,2q_2\,
\includegraphics[scale=0.3,valign=c
]{kinematic_flow/Figures/three-site_case/letters/X12-blue.pdf}
 \notag\\[-2pt]
&\quad{}+(q_2-Z)\,
 \includegraphics[scale=0.3,valign=c
]{kinematic_flow/Figures/three-site_case/letters/X3+red.pdf}
+Z\,\includegraphics[scale=0.3,valign=c
]{kinematic_flow/Figures/three-site_case/colored/Zcolored.pdf}
\Bigr]\,,
\\[2pt]
\dd g&=\eps\Bigl[\,2g\,\includegraphics[scale=0.3,valign=c]{kinematic_flow/Figures/three-site_case/letters/X12+blue.pdf}\notag\\[-2pt]
&\quad{}+(g+Z)\,\includegraphics[scale=0.3,valign=c]{kinematic_flow/Figures/three-site_case/letters/X3-red.pdf}-Z\,
 \includegraphics[scale=0.3,valign=c]{kinematic_flow/Figures/three-site_case/colored/Zcolored.pdf}\Bigr]\,.\end{align}
and the differential equations for $\tilde g$ and $\tilde q_2$ can be obtained by interchanging the two vertices $X_1 \leftrightarrow X_3$.
\end{itemize}

We then study the functions generated by two successive splittings of $Z$, whose corresponding function trees are more complicated. Since there are four admissible splitting locations in the tubing of $Z$, this level naturally contains $\begin{pmatrix}
    4\\ 2
\end{pmatrix} =6$ basis functions. We will explicitly present the differential equations for four of them, while the remaining two can be readily derived from the symmetry between $X_1$ and $X_3$.
\begin{itemize}
\item The first example is the basis function $q_1$, which can be generated from two splitting channels
\begin{align}
\nonumber
\includegraphics[scale=0.7,valign=c]{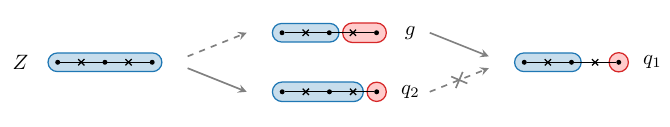}
\end{align}
In the tubing of $q_1$, the active tube (corresponding to $\dd \log \, (X_3 +Y')$) is not adjacent to a passive one. Thus, the ancestor function $Z$ will contribute to this differential equation via $g$. Additionally, since the splitting channel $q_2 \rightarrow q_1$ is forbidden, the total differential of $q_1$ receives no contribution from $q_2$. Recalling that dashed line imparts a minus sign to the parent function, we can obtain the differential equation corresponding to this function tree
\begin{align}
    \dd q_1 = \eps \Big[\,2q_1 \includegraphics[scale=0.3,valign=c]{kinematic_flow/Figures/three-site_case/letters/X12+blue.pdf} + (q_1-g) &\includegraphics[scale=0.3,valign=c]{kinematic_flow/Figures/three-site_case/letters/X3+red.pdf} \nonumber \\ +(g+Z) &\includegraphics[scale=0.3,valign=c]{kinematic_flow/Figures/three-site_case/letters/X3-red.pdf} \nonumber \\ 
    -Z &\includegraphics[scale=0.3,valign=c]{kinematic_flow/Figures/three-site_case/colored/Zcolored.pdf}
    \Big] \, .
\end{align}
It is evident that all letters appearing in the permitted channels contribute to this differential equation, multiplying different function combinations as dictated by their colors. 
\item The next example is the function $q_3$, whose function tree is given by
\begin{align}
\includegraphics[scale=0.7,valign=c]{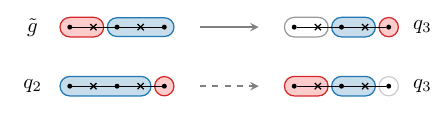}
\end{align}
In the top $\tilde g \rightarrow q_3$ channel, the active tube within the $q_3$ tubing is adjacent to a passive one that terminates at a site on the opposite side.
Thus, this direction requires no further tracing to the ancestor functions. As for the bottom channel $q_2 \rightarrow q_3$, although the tracing condition is satisfied, the parent function $q_2$ is connected to the ancestor by solid lines. This is also excluded from consideration. Therefore, the differential equation for $q_3$ is simply
\begin{align}
\dd q_3 &=\eps\Bigl[\,q_3\,
\includegraphics[scale=0.3,valign=c]{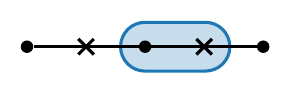}
\notag\\[-2pt]
&\quad{}+(q_3-\tilde g)\,
 \includegraphics[scale=0.3,valign=c]{kinematic_flow/Figures/three-site_case/letters/X3+red.pdf}
 +(q_3+q_2)\,\includegraphics[scale=0.3,valign=c]{kinematic_flow/Figures/three-site_case/letters/X1-red.pdf}
 \notag\\[-2pt]
&\quad{}+\tilde g\,
 \includegraphics[scale=0.3,valign=c]{kinematic_flow/Figures/three-site_case/letters/X23+blue.pdf}
 -q_2\,\includegraphics[scale=0.3,valign=c]{kinematic_flow/Figures/three-site_case/letters/X12-blue.pdf}\Bigr]\,.
\end{align}
\item Next, we consider another special function $f$, with the associated function tree given by
\begin{align}
\label{three-sitefsplit}
\includegraphics[scale=0.7,valign=c]{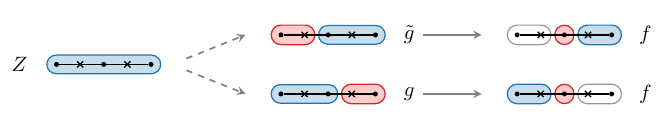}
\end{align}
In contrast to the previous cases, the tubing of $f$ now contains one active tube and two passive tubes. Furthermore, we can observe that $f$ originates from two splitting channels $\tilde g \rightarrow f$ and $g \rightarrow f$, both of which can be traced back to the same ancestor function $Z$. Following the established rules, $Z$ should now be treated as another parent function of $f$ on an equal footing with $\{g,\tilde g\}$. In such a case, we can effectively treat the $Z\rightarrow f$ splitting as a solid-line process, since the active tube in $f$ contains only a single site. The differential equation for $f$ is then
\begin{align}
\dd f &=\eps\Bigl[\,f\,
\includegraphics[scale=0.3,valign=c]{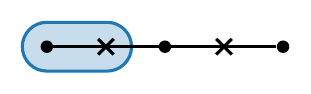}
+f\,\includegraphics[scale=0.3,valign=c]{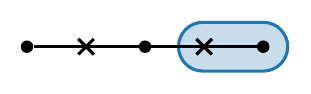}
\notag\\[-2pt]
&\quad{}+(f-g-\tilde g-Z)\,\includegraphics[scale=0.3,valign=c]{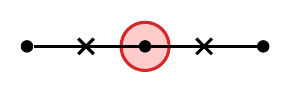}
+\tilde g\,\includegraphics[scale=0.3,valign=c]{kinematic_flow/Figures/three-site_case/letters/X23+blue.pdf}
\notag\\[-2pt]
&\quad{}+g\,\includegraphics[scale=0.3,valign=c]{kinematic_flow/Figures/three-site_case/letters/X12+blue.pdf}
+Z\,\includegraphics[scale=0.3,valign=c]{kinematic_flow/Figures/three-site_case/colored/Zcolored.pdf}\Bigr]\,.
\end{align}
This feature also provides an alternative formulation of the Rule (2) introduced in section~\ref{sec:generation-rule}: If different channels yielding a given function generate the same active tube with a single site, then the ancestor function along this direction should be treated as another parent. One can readily verify this statement through the following examples, as well as the more complicated cases in appendix~\ref{app:more-example}.
\item Our final example at this level is the function $Q_2$, whose function tree exhibits no surprising features
\begin{align}
\includegraphics[scale=0.7,valign=c]{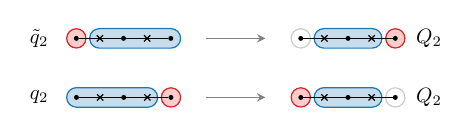}
\end{align}
Evidently, no ancestor functions will appear in the differential equation for $Q_2$, which takes the following form
\begin{align}
\dd Q_2&=\eps\Bigl[\,Q_2\,
\includegraphics[scale=0.3,valign=c]{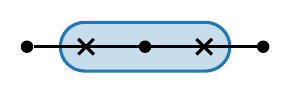}
\notag\\[-2pt]
&\quad{}+(Q_2-q_2)\,\includegraphics[scale=0.3,valign=c]{kinematic_flow/Figures/three-site_case/letters/X1+red.pdf}+(Q_2-\tilde q_2)\,\includegraphics[scale=0.3,valign=c]{kinematic_flow/Figures/three-site_case/letters/X3+red.pdf}
 \notag\\[-2pt]
&\quad{}+q_2\,\includegraphics[scale=0.3,valign=c]{kinematic_flow/Figures/three-site_case/letters/X12-blue.pdf}
 +\tilde q_2\,\includegraphics[scale=0.3,valign=c]{kinematic_flow/Figures/three-site_case/letters/X23-blue.pdf}\Bigr]\,.
\end{align}
\item The tubings of the two remaining functions at this level, $\tilde q_1$ and $\tilde q_3$, are related to those of $q_1$ and $q_3$ via the exchange $X_1\leftrightarrow X_3$. Writing the differential equations for them is thus a trivial exercise.
\end{itemize}

We now proceed to the functions at the next level. Splitting the generating function $Z$ three times yields $\begin{pmatrix}
    4 \\ 3
\end{pmatrix} =4$ basis functions. Since the functions at this level have at least 3 parent functions, their differential equations will be more complicated.
\begin{itemize}
\item We first consider the function tree of the basis function $F$
\begin{equation}
\includegraphics[scale=0.7,valign=c]{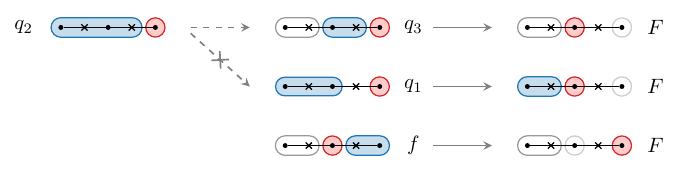}
\end{equation}
We can observe that $F$ now admits three parent functions, two of which split into the same active tube (corresponding to $\dd \log \, (X_2+Y+Y')$ ). Thus, the same ancestor function $q_2$ along these two directions should also be treated as a parent of $F$. The third channel remains conventional, since the parent $f$ does not trace back to other functions through dashed lines. Then the corresponding differential equation is given by
\begin{align}
\dd F &=\eps\Bigl[\,F\,
 \includegraphics[scale=0.3,valign=c]{kinematic_flow/Figures/three-site_case/letters/X1-blue.pdf}
 +(F-f)\,\includegraphics[scale=0.3,valign=c
]{kinematic_flow/Figures/three-site_case/letters/X3+red.pdf}\notag\\[-2pt]
&\quad{}+(F-q_1-q_2-q_3)\,\includegraphics[scale=0.3,valign=c]{kinematic_flow/Figures/three-site_case/letters/X2++red.pdf}
\notag\\[-2pt]
&\quad{}+f\,\includegraphics[scale=0.3,valign=c]{kinematic_flow/Figures/three-site_case/letters/X3-blue.pdf}
+q_1\,\includegraphics[scale=0.3,valign=c]{kinematic_flow/Figures/three-site_case/letters/X12+blue.pdf}
 \notag\\[-2pt]
&\quad{}+q_2\,\includegraphics[scale=0.3,valign=c]{kinematic_flow/Figures/three-site_case/letters/X12-blue.pdf}
+q_3\,\includegraphics[scale=0.3,valign=c]{kinematic_flow/Figures/three-site_case/letters/X2+-blue.pdf}\Bigr]\,.
\end{align}
\item The next example is the basis function $Q_1$, whose function tree is given by
\begin{align}
\includegraphics[scale=0.7,valign=c]{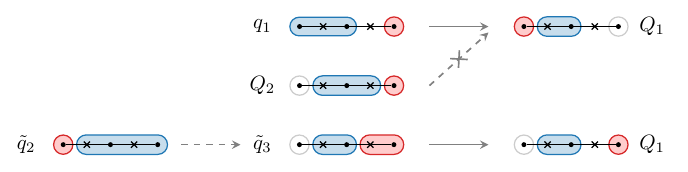}
\end{align}
At this stage, the function $Q_1$ still has three parent functions, corresponding to three distinct splitting channels. For the top channel, the corresponding active tube is adjacent to a passive one that terminates at a site, hence there are no contributions from ancestor functions along this direction. Moreover, the middle channel is forbidden. The bottom channel corresponds to the splitting of an active tube, which can be traced back one further level. Note that the tubings of $Q_1$ and $\tilde q_3$ contain the same passive tube. However, since each letter can be used only once in the differential equation, terms proportional to this tube multiplied by $\tilde q_3$ are absent. The differential equation for the function $Q_1$ is then
\begin{align}
\dd Q_1&=\eps\Bigl[\,Q_1\,
 \includegraphics[scale=0.3,valign=c]{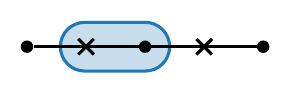}
 +(Q_1-q_1)\,\includegraphics[scale=0.3,valign=c]{kinematic_flow/Figures/three-site_case/letters/X1+red.pdf}\notag\\[-2pt]
&\quad{}+(Q_1-\tilde q_3)\,
\includegraphics[scale=0.3,valign=c]{kinematic_flow/Figures/three-site_case/letters/X3+red.pdf}
+(\tilde q_3+\tilde q_2)\,
 \includegraphics[scale=0.3,valign=c]{kinematic_flow/Figures/three-site_case/letters/X3-red.pdf}
 \notag\\[-2pt]
&\quad{}+q_1\,\includegraphics[scale=0.3,valign=c]{kinematic_flow/Figures/three-site_case/letters/X12+blue.pdf}
 -\tilde q_2\,\includegraphics[scale=0.3,valign=c]{kinematic_flow/Figures/three-site_case/letters/X23-blue.pdf}\Bigr]\,.
\end{align}
\item The function trees and differential equations for the functions $\tilde F$ and $Q_3$ can be obtained analogously. We refrain from elaborating on this point further.
\item At the final level, there is a unique basis function, which corresponds to the physically relevant wavefunction coefficient $\psi$. Emerging as the endpoint for all splitting paths, $\psi$ is associated with the following function tree
\begin{equation}
\includegraphics[scale=0.7,valign=c]{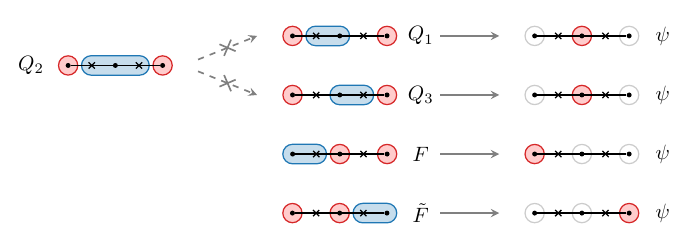}
\end{equation}
Despite admitting more parent functions, the function tree for $\psi$ is otherwise straightforward. After properly applying the splitting rules, we can directly read off the corresponding differential equation
\begin{align}
\dd\psi &=\eps\Bigl[\,(\psi-F)\,
\includegraphics[scale=0.3,valign=c]{kinematic_flow/Figures/three-site_case/letters/X1+red.pdf}
 +(\psi-\tilde F)\,\includegraphics[scale=0.3,valign=c]{kinematic_flow/Figures/three-site_case/letters/X3+red.pdf}
\notag\\[-2pt]
&\quad{}+(\psi-Q_1-Q_2-Q_3)\,
\includegraphics[scale=0.3,valign=c]{kinematic_flow/Figures/three-site_case/letters/X2++red.pdf}
\notag\\[-2pt]
&\quad{}+F\,\includegraphics[scale=0.3,valign=c]{kinematic_flow/Figures/three-site_case/letters/X1-blue.pdf}
 +\tilde F\,\includegraphics[scale=0.3,valign=c]{kinematic_flow/Figures/three-site_case/letters/X3-blue.pdf}
+Q_1\,\includegraphics[scale=0.3,valign=c]{kinematic_flow/Figures/three-site_case/letters/X2-+blue.pdf}\notag\\[-2pt]
&\quad{}+Q_2\,\includegraphics[scale=0.3,valign=c]{kinematic_flow/Figures/three-site_case/letters/X2--blue.pdf}
+Q_3\,\includegraphics[scale=0.3,valign=c]{kinematic_flow/Figures/three-site_case/letters/X2+-blue.pdf}\Bigr]\,.
\end{align}
\end{itemize}

By comparing with the results in \cite{Arkani-Hamed:2023kig} or appendix~\ref{app:diffeq3site}, we confirm that the splitting rules precisely reproduce the differential equations. In this case, all basis functions are organized into distinct levels based on the number of splitting operations performed on the generating function, with their total number given by
\begin{equation}
    1 + \begin{pmatrix}
        4 \\ 1
    \end{pmatrix} + \begin{pmatrix}
        4\\ 2  
    \end{pmatrix} + \begin{pmatrix}
        4 \\ 3
    \end{pmatrix} +\begin{pmatrix}
        4\\ 4
    \end{pmatrix} = 1 + \,4 +\,6 +\, 4 +\,1 = 16\,,
\end{equation}
which is the consistent result. Having explicitly verified the more involved four-site chain and four-site star cases in appendix~\ref{app:more-example}, we confirm that the splitting rules are completely equivalent to the kinematic flow at tree level. 
Therefore, these rules share a drawback similar to that of the kinematic flow: as the number of vertices $n$ increases, the number of basis functions grows rapidly,  making the function trees exceedingly intricate. Even compared to the rules in \cite{Arkani-Hamed:2023kig}, our rules are more complicated, because deriving the differential equation for a specific function requires tracing its splitting history back to its ancestors.

Nevertheless, this method provides a fresh perspective for understanding the relationships among the basis functions. We start from certain objects that depend purely on the geometric structure of the graphs, with the wavefunction coefficients emerging as the final outputs. This may suggest that it is possible to construct the evolution rules for the basis functions without involving the bulk-time integral. Moreover, we observe that the generating functions universally entail the minimal number of time integrals, as they correspond to the collapse of all time orderings. As the splitting occurs, specific splitting channels will introduce additional time integrals, eventually leading to the most complicated structure of $\psi$. Thus, this may offer an alternative physical interpretation for the emergence of time \cite{Arkani-Hamed:2023bsv}. This pattern is particularly manifest within a recently proposed basis of functions, which we will discuss in detail in section~\ref{sec:another-basis}.
\section{Discussions and Physical Implications}
\label{sec:phy-implications}
The splitting rules we constructed in the previous section are equivalent to the kinematic flow at tree level. However, their true significance is that they motivate us to ask whether graph tubings encode richer physical information. This section aims to address this very issue. We first introduce the role of the splitting rules, and then explain how they clearly show the singularity structure of the wavefunction coefficients being built from the differential equations. Neither of these aspects is manifest in the kinematic flow formulation. Even more importantly, this perspective provides a graphical representation of local evolution. That is, a tube enclosing several vertices treats them as an effective contact substructure, meaning that the corresponding basis function depends only on the total external energy flowing into it. Thus, these vertices can be considered non-local relative to one another, while a splitting process restores this locality by separating vertices and introducing the corresponding internal energy. As we will see in the next section, this perspective provides deep insights into the emergence of time integrals in kinematic space.

\subsection{The Role of Splitting Rules}
Let us first recall the motivation for constructing the splitting rules: we wish to determine which other functions contain a given basis function in their total differentials. Therefore, once the tubing of a basis function is obtained, this becomes the most straightforward question to answer. We just need to identify all splitting channels for this tubing, and extend the dashed line processes by an additional level. Then the splitting rules are just what we need.

As a concrete example, consider a relatively complicated function $F_{12_33_24}$ in the four-site chain case, whose corresponding tubing is \includegraphics[scale=0.2,valign=c]{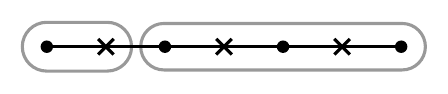}. Addressing this problem via the kinematic flow is challenging, since one must first derive the differential equations for all of its higher-level functions. On the other hand, determining all splitting channels for this function is simple, from which we obtain
\begin{widetext}
\begin{equation}
\includegraphics[scale=1,valign=c
]{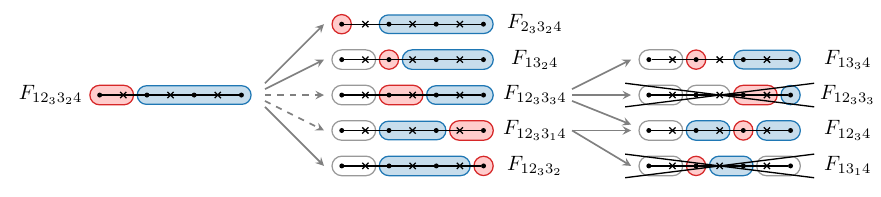}
\label{eq:splitF12-33-24}
\end{equation}
\end{widetext}
Note that at the next level of $F_{12_33_24}$, we need not consider the splitting of uncolored tubes. We observe that the channels for the two functions $F_{12_33_3}$ and $F_{13_14}$ are forbidden, because the passive tube adjacent to the active one terminates at a site on the other side. The remaining seven channels are all admissible. Thus, we deduce that $F_{12_33_24}$ contributes to the differential equations for the following seven basis functions $\{F_{23_24}, F_{13_24}, F_{12_33_34},F_{12_33_14}, F_{12_33_2},F_{13_34},F_{12_34}\}$. Furthermore, following our established rules, writing down the contribution of $F_{12_33_24}$ to the differential equations for these functions becomes a trivial task.

This approach can be systematically employed to analyze the source structure of arbitrary basis functions.
\subsection{Singularities}
Singularities play a crucial role in the study of cosmological correlations.
In the present model, the alphabet of letters associated to a marked graph corresponds to two possible classes of singularity configurations. The first class consists of kinematic configurations where the total energy flowing into a subgraph vanishes. Although such limits are physically inaccessible, these partial energy singularities (as well as total energy singularities) remain of crucial importance, since the wavefunction coefficients here are fixed in terms of simpler building blocks \cite{Arkani-Hamed:2017fdk,Baumann:2020dch,Baumann:2021fxj,Goodhew:2021oqg,Jazayeri:2021fvk}. The second class, known as the collinear singularities, corresponds to configurations where two (or more) external momenta become collinear. Such singularities are typically forbidden by the Bunch-Davies vacuum, and the regular behavior of the wavefunction coefficient in these limits provides an essential boundary condition for solving the differential equations \cite{Arkani-Hamed:2018kmz,Arkani-Hamed:2023kig}.

Although the singularity structure of $\psi$ can be deduced from its properties as a wavefunction coefficient, one might wonder how singularities are distributed for an arbitrary basis function. A further natural question is how the complicated singularity structure of $\psi$ emerges from the interplay of other basis functions. The answer can be obtained by explicitly solving the differential equations. Once we substitute some known source functions into the differential equation for another basis function, the singularities of these source functions and those of the differential equation both contribute to this new basis function. While the boundary conditions of the differential equations eliminate all collinear singularities in $\psi$, they are nevertheless present in the solutions of other basis functions. This can be directly verified from the explicit expressions for the solutions given in section~3.3 (FRW) or appendix~D.1 (de Sitter) of \cite{Arkani-Hamed:2023kig}.

These singularity structures naturally coincide with the rationale behind our splitting rules, whereas they remain less transparent through the kinematic flow. First, the generating function contains a total energy singularity, which is represented by the corresponding graph tubing. Subsequently, when a splitting occurs, every letter involved in the process appears in the associated differential equation. Thus, the generated basis function will inherit the singularities encoded by all participating letters. 

As a quick check, consider the splitting of the function $Z$ in the two-site chain case in eq.~\eqref{twosite-splitZ}, which allows for two channels. For the channel associated to $F$, the letters involved in the process represent three singularities $\{X_1+X_2, X_1-Y, X_2+Y\} = 0$, one of which is a collinear singularity. After imposing the appropriate boundary conditions, the solution for the basis function $F$ takes the following form \cite{Arkani-Hamed:2023kig}
\begin{equation}
\begin{aligned}
F &=c_F\,(X_1-Y)^\eps(X_2+Y)^\eps
\\[-1pt] &\quad +\frac{c_Z}{2}
 \left(\frac{X_1-Y}{X_2+Y}\right)^\eps
(X_1+X_2)^{2\eps}
\\[-1pt]
&\qquad\times {}_2F_1\!\left[
 \begin{matrix} \eps,\,2\eps\\
 1+2\eps\end{matrix}
 \,\middle|\,\frac{X_1+X_2}{X_2+Y}
 \right]\,.
\end{aligned}
\end{equation}
where ${}_{2}F_1$ is the Gauss hypergeometric function and the normalization coefficients are given by
\begin{align}
c_F &=-\pi^2\, \csc(\pi\eps)\, \csc(2\pi\eps)\,,
\\[-1pt]
c_Z &=4^{-\eps}\sqrt{\pi}\, \csc(\pi\eps)\, \Gamma(\eps)\,
\Gamma\left(\frac{1}{2}-\eps\right)\,.
\end{align}
Since the universe under consideration typically corresponds to $\eps \leq 0$, the solution becomes singular when any of $\{X_1+X_2, X_1-Y, X_2+Y\}$ vanishes.\footnote{In the general case, the singularities for the basis functions can be rather complicated, for example, appearing as higher-order poles or branch points \cite{Baumann:2021fxj}.} 
Hence, the collinear singularities in the basis functions are not entirely removed by the boundary conditions, consistent with our prediction. In practice, the boundary conditions of these differential equations can only remove the collinear singularities in the homogeneous solutions of each basis function (except for $\psi$).

From this, we can summarize how to determine the loci of the singularities for an arbitrary basis function. We should start from the generating function and identify all splitting channels that lead to the target function tubing. In this way, this function will inherit the singularities represented by all letters involved in these channels. Note that we should be careful with the forbidden channels denoted by crossed dashed lines. Unless they satisfy the Rule (2) in section~\ref{sec:generation-rule}, no further backward tracing of basis functions should be performed. As a concrete example, consider the basis function $f$ in the three-site chain case, whose splitting channels can be traced back to $Z$ via eq.~\eqref{three-sitefsplit}. Therefore, the solution for $f$ contains all of the following singularities
\begin{equation}
\left\{
\begin{gathered} X_1+X_2+X_3,\; X_1-Y,\;
X_2+X_3+Y,\\[-2pt] X_1+X_2+Y',\; X_3+Y',\; X_1-Y,
\\[-2pt]
X_2+Y+Y',\; X_3-Y'
\end{gathered} \right\} =0\,.
\end{equation}
Next, we turn to another function $F$, whose tubing is generated from those of $Z$ through three splitting processes. Although the splitting channels seem intricate at this stage, they can be readily derived as follows
\begin{equation}
\includegraphics[scale=1,valign=c]{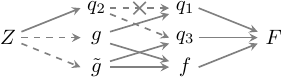}
\end{equation}
Thus, the solution for $F$ incorporates all singularities represented by the letters participating in these channels. We leave it to the reader to write them out explicitly. In conclusion, the splitting rules reveal how the singularities of the basis functions naturally accumulate through the splitting of tubings.
\subsection{Internal Energies and Local Evolution}
\label{local-evolution}
The splitting rules also make transparent the distinct roles played by the kinematic variables, and therefore provide new insight into how local evolution is encoded by graph tubings. For a marked graph, these kinematic variables can be divided into two classes: the external energies flowing into individual vertices, and the internal energies flowing through the diagram. These two classes naturally play distinct physical roles in the graph-tubing representation, yet this distinction is not manifest in the conventional kinematic-flow formulation.

This issue is clarified from the perspective of tubing splittings. Taking the three-site chain case as an example, recall that the generating function $Z$ depends only on the total external energy and is independent of internal energies (up to a normalization factor). Since each splitting process corresponds to a three-letter relation, the splitting of the $Z$ tubing can be interpreted as dividing the total energy into two parts. To observe this, consider one of the splitting channels $Z \rightarrow g$, the letters before and after the splitting are given by
\begin{align}
\underset{X_1+X_2+X_3}{\includegraphics[scale=0.3,valign=c]{kinematic_flow/Figures/three-site_case/colored/Zcolored.pdf}}  &\rightarrow  \underset{X_1+X_2+Y'}{\includegraphics[scale=0.3,valign=c]{kinematic_flow/Figures/three-site_case/letters/X12+blue.pdf}} + \underset{X_3-Y'}{\includegraphics[scale=0.3,valign=c]{kinematic_flow/Figures/three-site_case/letters/X3-red.pdf}} \,.
\end{align}
Given that the splitting takes place between a site and a cross, the sign of $Y'$ is reversed in the induced letters. 
Evidently, this procedure is equivalent to severing the connection between $X_1+X_2$ and $X_3$ through the introduction of an internal energy $Y'$, i.e. we should express the function $g$ in the form of $g(X_{12},X_3,Y')$ rather than $g(X_{123})$, where $X_{i\cdots j}$ represents $X_i+\cdots + X_j$. 
From this perspective, given the tubing of a basis function, vertices enclosed by the same tube are treated as an effective contact substructure, meaning that the corresponding function depends only on the total energy flowing into the tube. We can refer to such vertices as being mutually non-local.
In the differential equations, such vertices can be treated as a single vertex carrying an appropriate total energy. Conversely, vertices belonging to different tubes are considered as local with respect to each other.

As the splitting processes successively decompose the tubings, the resulting complete tubing retains all preceding information. Therefore, the sequence from the generating function to the wavefunction coefficient can be understood as progressively introducing internal energies, making some vertices or some combinations of vertices local. Since the introduction of each new internal energy localizes two components in the graph tubing, we propose that the internal energies are fundamentally linked to the locality structure of the basis functions.

In \cite{Arkani-Hamed:2023kig,Arkani-Hamed:2023bsv}, locality is argued to manifest as a reduction in the order of differential equations satisfied by the wavefunction coefficients. Specifically, in the two-site chain case, there are four basis functions in the family of master integrals, indicating that the wavefunction coefficient should generally satisfy a fourth-order ordinary homogeneous differential equation in each of $X_1$ and $X_2$. However, it turns out to satisfy only a second-order differential equation with respect to $X_1$ (or $X_2$),
\begin{equation}
\begin{aligned}
\Bigl[\,&(X_1^2-Y^2)\partial_{X_1}^{\,2}
 +2(1-\eps)X_1\partial_{X_1}\\[-2pt]
 &{}-\eps(1-\eps)\Bigr]\psi= g\left(\frac{1}{X_1+X_2}\right)^{1-2\eps}\,.
\end{aligned}
\end{equation}
This is because the internal line corresponding to $Y$ satisfies a Green's function equation, and the vertices associated to $X_1$ and $X_2$ are local. Compared to other basis functions, it is precisely the locality at each vertex that dictates the wavefunction coefficient to satisfy the most complicated differential equation. In the context of kinematic flow, each merger and absorption process induces non-locality between a pair of components.

Returning to the three-site chain case, while the family of master integrals contains 25 elements, only 16 of them are involved in the system of differential equations for $\psi$. This reduction may also be explained from the viewpoint of locality. 
\footnote{In \cite{Arkani-Hamed:2023kig}, this pattern is interpreted as a consequence of the integrands associated with the wavefunction coefficients are non-generic twisted integrals, meaning that only the coordinate lines are twisted.} 
Note that the splitting of the $Z$ tubing decomposes the three vertices into either $\{X_{12},X_3\}$ or $\{X_1,X_{23}\}$, but never into $\{X_{13},X_2\}$. This is because the vertices for $X_1$ and $X_3$ are not connected by an internal line, implying that they should be local except in the generating function. We argue that the basis functions absent from the system of differential equations for $\psi$ are associated to the non-local structures between $X_1$ and $X_3$. More concretely, if we introduce an auxiliary graphical operation in which the two non-adjacent vertices $X_1$ and $X_3$ are contacted, the original marked graph will be transformed to that of the one-loop bubble diagram investigated in \cite{Baumann:2024mvm}

\begin{equation}
\begin{tikzpicture}[
  baseline=(current bounding box.center)
]\node[inner sep=0pt] (chain) {\includegraphics[scale=0.5]
{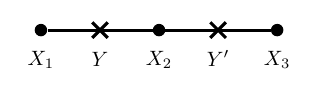}
};\node[inner sep=0pt,right=2.9cm of chain,yshift=0.35cm
] (bubble) {\includegraphics[scale=0.5]
{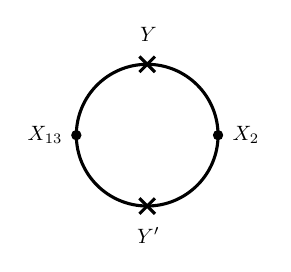}
};\draw[ gray,thick,-stealth,shorten <=2pt,shorten >=2pt]
([yshift=0.35cm]chain.east)
--node[midway,above=3pt,text=black,font=\small]{contact $X_1$ and $X_3$}(bubble.west);
\end{tikzpicture}
\end{equation}
For the one-loop bubble diagram, although there are 16 corresponding complete tubings, some of them vanish trivially, since they can be expressed as the difference of functions obtained by two identical replacement operations \cite{Baumann:2024mvm}. Consequently, only the following 10 non-vanishing basis functions survive \footnote{The following picture is reproduced from \cite{Baumann:2024mvm} under the terms of the Creative Commons CC-BY 4.0 license.}
\begin{align}
\includegraphics[scale=1,valign=c]{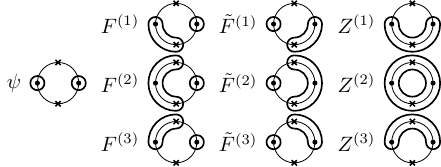}
\end{align}
Among these basis functions, the function $Z^{(2)}$, after the identification $X_{13} \equiv X_1 +X_3$, satisfies the same differential equation as the total-energy generating function in the three-site chain case. 
We therefore regard $Z^{(2)}$ as the image of an already-present function, rather than as an additional missing tubing. The remaining 9 graph tubings can then be interpreted as the basis functions that do not appear in the system of differential equations for $\psi_{(3)}$. 

Therefore, the reason why these basis functions do not contribute to the wavefunction coefficient is physically transparent: they reflect the non-local evolution processes that are forbidden by the graph geometry.
It will be interesting to investigate whether more complicated cases exhibit analogous properties.
\section{Another Basis}
\label{sec:another-basis}
Recently, an alternative physics-motivated function basis has been investigated in \cite{Baumann:2025qjx,He:2024olr,Glew:2025ugf,Glew:2025ypb} (and \cite{Pimentel:2026kqc} for loop integrands), under which the differential equations and kinematic flow rules are substantially simplified. The motivation of this basis is that each bulk-to-bulk propagator in eq.~\eqref{bulk-to-bulk-propagator} can be decomposed into three pieces according to the possible time orderings. Since there is one internal line in the two-site chain, the wavefunction coefficient $\psi_{(2)}$ is built from three pieces
\begin{align}
    \psi_{(2)} = \psi_{\includegraphics[scale=0.25,valign=c]{kinematic_flow/Figures/two-site_chain/functions/tildeF.pdf}} + \psi_{\includegraphics[scale=0.25,valign=c]{kinematic_flow/Figures/two-site_chain/functions/F.pdf}} +\psi_{\includegraphics[scale=0.25,valign=c]{kinematic_flow/Figures/two-site_chain/functions/psi.pdf}}\,.
\end{align}
From the expression in eq.~\eqref{psi-twosite}, the integrand associated to each piece is defined as
\begin{align}
\label{psi+}
\psi_{\includegraphics[scale=0.25,valign=c]{kinematic_flow/Figures/two-site_chain/functions/tildeF.pdf}} & \equiv N \int \frac{\dd \eta_1\, \dd \eta_2}{(\eta_1 \eta_2)^{1+\eps}}e^{i(X_1+Y)\eta_1} e^{i(X_2-Y)\eta_2} \theta (\eta_2-\eta_1)\,,\\
\label{psi-}
\psi_{\includegraphics[scale=0.25,valign=c]{kinematic_flow/Figures/two-site_chain/functions/F.pdf}} & \equiv N \int \frac{\dd \eta_1\, \dd \eta_2}{(\eta_1 \eta_2)^{1+\eps}} e^{i(X_1-Y)\eta_1}e^{i(X_2+Y)\eta_2}\theta (\eta_1-\eta_2)\,, \\
\psi_{\includegraphics[scale=0.25,valign=c]{kinematic_flow/Figures/two-site_chain/functions/psi.pdf}} & \equiv N \int \frac{\dd \eta_1\, \dd \eta_2}{(\eta_1 \eta_2)^{1+\eps}}  e^{i(X_1+Y)\eta_1}e^{i(X_2+Y)\eta_2}\,.
\end{align}
Similarly, the tree-level wavefunction coefficient with $N$ internal lines can be decomposed into $3^N$ components. These constituent pieces of the wavefunction coefficients are elegantly encoded in terms of the complete tubings defined before. Namely, any complete tubing where all vertices are local (i.e. each tube encloses only one site) contributes to the wavefunction coefficient, and other tubings correspond to the auxiliary functions appearing in the differential equations.

At this stage, the derivatives with respect to kinematic variables $X_1,X_2,Y$ can be naturally converted into time derivatives, which then act on the bulk-to-bulk propagators via integration by parts. Therefore, the total differential of each component of the wavefunction coefficient is expressed in terms of itself, along with the terms generated by the collapse of the corresponding time orderings. By iteratively differentiating the new functions until the system of equations closes, we obtain the system of differential equations for $\psi$. The differential equations in this basis can also be derived via a distinct kinematic flow, whose rules are much simpler. Now, there are only two steps \cite{Baumann:2025qjx}:
\begin{itemize}
    \item[1.] \textbf{Activation.}  This step is the same as that in section~\ref{sec:kinematicflow}. Each tube in the complete tubing can get activated and become a letter in the differential equation. This activated tube appears simply multiplying the original basis function.
    \item[2.] \textbf{Merger.} If two tubes in the graph tubing are adjacent to each other, they can merge and become a larger tube. The function associated to the new tubing is the source function, which will multiply the difference between the two merging tubes. The tube containing the cross on the merged edge carries a plus sign. 
\end{itemize}
In this basis, the enlargement of any tube in the complete tubing occurs at most once. This simplification obviates the need to involve any ancestor functions when constructing the inverse splitting rules. Moreover, since the evolution of graph tubings (basis functions) is defined via the collapse of time orderings, the associated splitting rules provide valuable insights into the emergence of time integral from kinematic space. We are now ready to explain this in detail.
\subsection{Splitting Rules}
To construct the appropriate splitting rules, we need to analyze the properties of mergers in this basis. First, a merger can only occur between two adjacent tubes, and each differentiation induces only one possible merger. Moreover, all mergers can be described by the same rule in building the differential equations. This motivates us to simply classify the splittings as either allowed or forbidden, with the former denoted by solid lines while the latter are omitted. As before, the forbidden processes are defined as the splittings which lead to a tube with a single cross \includegraphics[scale=0.2,valign=c]{kinematic_flow/Figures/three-site_case/splitting/a_cross.pdf}.
As for the letters, we now only need to assign a simple set of splitting rules
\begin{itemize}
    \item \textbf{Splitting Rules.} Any tube containing more than two sites can split into two smaller ones. The resulting tube containing the cross adjacent to the splitting location becomes passive, while the remaining one becomes active.
\end{itemize}
In this way, given the tubing of a basis function, we can draw all its generating channels through splittings. Then, the differential equation for this function includes all tubes in its graph tubing multiplied by the function itself, plus the difference between the tubes generated from all splitting channels multiplied by the corresponding parent functions. Each passive tube generated from splitting carries a plus sign in the differential equation. Finally, we assign a factor of $\eps$ multiplied by the number of vertices in the letters to the former part of the differential equation. It is straightforward to verify that these rules are equivalent to the previous kinematic flow.

\textbf{Two-site chain.} To illustrate these splitting rules, we again begin with the two-site chain case, where the four graph tubings are given by eq.~\eqref{two-site function}. Evidently, the generating function in this case satisfies the same differential equation as the one in the basis of section~\ref{sec:two-site}
\begin{equation}
\label{timeless}
    \dd F_{\includegraphics[scale=0.2,valign=c]{kinematic_flow/Figures/two-site_chain/functions/ZZ.pdf}} = 2 \eps \includegraphics[scale=0.3,valign=c]{kinematic_flow/Figures/two-site_chain/colored/ZZcolored.pdf} F_{\includegraphics[scale=0.2,valign=c]{kinematic_flow/Figures/two-site_chain/functions/ZZ.pdf}}\,.
\end{equation}
Thus, they can be identified as the same function. In addition, there is also a tubing that cannot be derived through splitting in this case, i.e. \includegraphics[scale=0.3,valign=c]{kinematic_flow/Figures/two-site_chain/functions/psi.pdf}, whose differential equation simply reads
\begin{equation}    \psi_{\includegraphics[scale=0.2,valign=c]{kinematic_flow/Figures/two-site_chain/functions/psi.pdf}} = \eps \,\Big(\includegraphics[scale=0.3,valign=c]{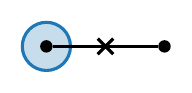} + \includegraphics[scale=0.3,valign=c]{kinematic_flow/Figures/two-site_chain/letters/X2+blue.pdf}\Big)\,\psi_{\includegraphics[scale=0.2,valign=c]{kinematic_flow/Figures/two-site_chain/functions/psi.pdf}} \, .
\end{equation}
Next, we consider the two splitting channels for the tubing \includegraphics[scale=0.3,valign=c]{kinematic_flow/Figures/two-site_chain/functions/ZZ.pdf}, which are given as follows
\begin{align}
\label{anotherbasis-channel}
\includegraphics[scale=0.7,valign=c]{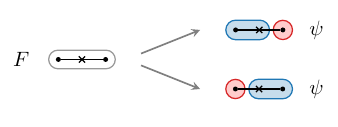}
\end{align}
It is evident that the total differential of the basis function $\psi_{\includegraphics[scale=0.2,valign=c]{kinematic_flow/Figures/two-site_chain/functions/F.pdf}}$ depends on itself and the parent function $F_{\includegraphics[scale=0.2,valign=c]{kinematic_flow/Figures/two-site_chain/functions/ZZ.pdf}}$. Following the present splitting rules, the corresponding differential equation can be written as
\begin{equation}
\begin{aligned}
\dd\psi_{\includegraphics[scale=0.2,valign=c]{kinematic_flow/Figures/two-site_chain/functions/F.pdf}}
&=\eps\Bigl(\includegraphics[scale=0.3,valign=c]{kinematic_flow/Figures/two-site_chain/letters/X1-blue.pdf}
+\includegraphics[scale=0.3,valign=c]{kinematic_flow/Figures/two-site_chain/letters/X2+red.pdf}
\Bigr)\psi_{\includegraphics[scale=0.2,valign=c]{kinematic_flow/Figures/two-site_chain/functions/F.pdf}
}\\[-2pt] &\quad{} +\Bigl(
\includegraphics[scale=0.3,valign=c]{kinematic_flow/Figures/two-site_chain/letters/X1-blue.pdf}
 - \includegraphics[scale=0.3,valign=c]{kinematic_flow/Figures/two-site_chain/letters/X2+red.pdf}\Bigr)
F_{\includegraphics[scale=0.2,valign=c]{kinematic_flow/Figures/two-site_chain/functions/ZZ.pdf}}\,.
\end{aligned}
\end{equation}
and the differential equation for the other channel can be obtained by the $X_1 \leftrightarrow X_2$ interchange. These results coincide precisely with those in \cite{Baumann:2025qjx}.

\textbf{More examples.} Since the splitting rules are quite simple in this basis, we proceed directly to the four-site chain case. The generating function in this basis satisfies the same differential equation as the function $F_{12_23_24}$ defined in eq.~\eqref{generating-functions}
\begin{equation}
    \dd H_{\includegraphics[scale=0.2,valign=c]{kinematic_flow/Figures/four-site_chain/functions/F12_23_24.pdf}} = 4 \eps \includegraphics[scale=0.3,valign=c]{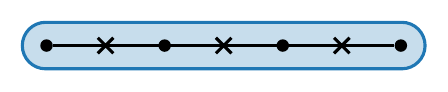} \,H_{\includegraphics[scale=0.2,valign=c]{kinematic_flow/Figures/four-site_chain/functions/F12_23_24.pdf}}\,.
\end{equation}
Next, we consider one of the splitting channels for the tubing \includegraphics[scale=0.2,valign=c]{kinematic_flow/Figures/four-site_chain/functions/F12_23_24.pdf},
\begin{align}
\includegraphics[scale=0.7,valign=c]{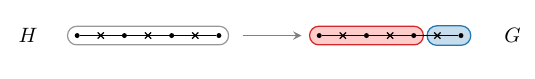}
\end{align}
Note that the passive tube containing the cross adjacent to the splitting location is indicated in blue, contributing to the differential equation with a plus sign. The contribution from the other tube is instead subtracted. \footnote{There may be some minor sign typos in the corresponding differential equations in \cite{Baumann:2025qjx}.}
The differential equation for the new function is then given by 
\begin{equation}
\begin{aligned}
\dd G_{\includegraphics[scale=0.2,valign=c
]{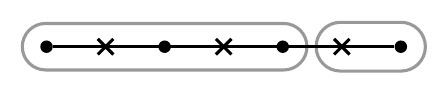}
} &= \eps\Bigl( 3\,\includegraphics[scale=0.3,valign=c]{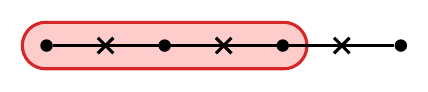}
+ \includegraphics[scale=0.3,valign=c]{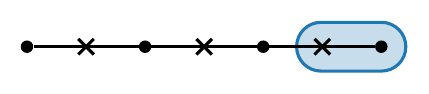}
\Bigr) \\[-2pt] &\qquad\times
G_{ \includegraphics[scale=0.2,valign=c]{kinematic_flow/Figures/four-site_chain/functions/F12_23_14.pdf}
} \\[-2pt] &\quad{} +
\Bigl( \includegraphics[scale=0.3,valign=c]{kinematic_flow/Figures/four-site_chain/letters/X4-blue.pdf}
-\includegraphics[scale=0.3,valign=c]{kinematic_flow/Figures/four-site_chain/letters/X123+red.pdf}
\Bigr)\\[-2pt]
&\qquad\times H_{\includegraphics[scale=0.2,valign=c]{kinematic_flow/Figures/four-site_chain/functions/F12_23_24.pdf}
}\,.
\end{aligned}
\end{equation}
We observe that each splitting channel will introduce two letters in the differential equations.

We then consider a new tubing at the next level, \includegraphics[scale=0.2,valign=c]{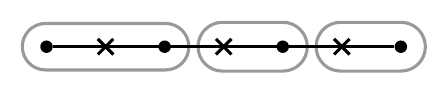}. Since this tubing possesses two disjoint locations, it should be derived through two successive splittings of the generating function. Therefore, there are two parent functions in this case. The function tree is as follows 
\begin{align}
\includegraphics[scale=0.7,valign=c]{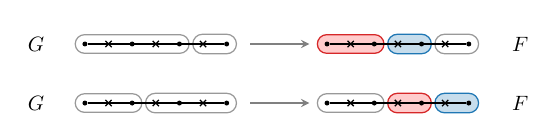}
\end{align}
In this basis, the same tube within a complete tubing may exhibit different characteristics in different splitting channels. For instance, the letter corresponding to $X_3^{-+}$ is passive in the upper channel, but active in the lower channel. Following the rules, the two channels contribute independently to the differential equation, which takes the following form
\begin{equation}
\begin{aligned}
\dd F_{ \includegraphics[scale=0.2,valign=c]{kinematic_flow/Figures/four-site_chain/functions/F12_13_14.pdf}
}&= \eps\Bigl( 2\,\includegraphics[scale=0.3,valign=c]{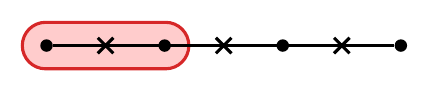}
+\includegraphics[scale=0.3,valign=c]{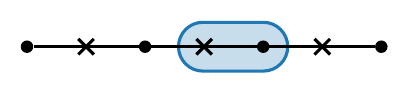}
\\[-2pt] &\qquad{} + \includegraphics[scale=0.3,valign=c]{kinematic_flow/Figures/four-site_chain/letters/X4-blue.pdf}
\Bigr) \times
F_{ \includegraphics[scale=0.2,valign=c]{kinematic_flow/Figures/four-site_chain/functions/F12_13_14.pdf}}
\\[-2pt] &\quad{} + \Bigl(
\includegraphics[scale=0.3,valign=c]{kinematic_flow/Figures/four-site_chain/letters/X3-+blue.pdf}
- \includegraphics[scale=0.3,valign=c]{kinematic_flow/Figures/four-site_chain/letters/X12+red.pdf}\Bigr)
\\[-2pt]
&\qquad \times
G_{\includegraphics[scale=0.2,valign=c]{kinematic_flow/Figures/four-site_chain/functions/F12_23_14.pdf}}
\\[-2pt] &\quad{} +\Bigl(
\includegraphics[scale=0.3,valign=c]{kinematic_flow/Figures/four-site_chain/letters/X4-blue.pdf}
- \includegraphics[scale=0.3,valign=c]{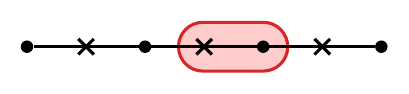} \Bigr)
\\[-2pt] &\qquad\times
G_{\includegraphics[scale=0.2,valign=c]{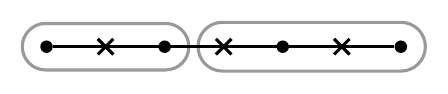}}\,.
\end{aligned}
\end{equation}
Finally, we consider another non-trivial example, the tubing \includegraphics[scale=0.2,valign=c]{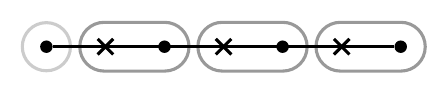}. Note that each tube in this tubing encloses only one site. Therefore, the corresponding basis function involves four time integrals and constitutes a component of the wavefunction coefficients. The function tree now contains three parent functions
\begin{align}
\includegraphics[scale=0.7,valign=c]{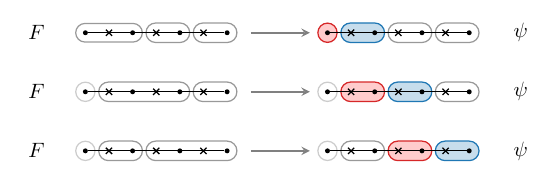}
\end{align}
Generally, each pair of adjacent tubes contributes to the differential equation, where the tube enclosing the cross at their boundary plays a passive role. While the differential equation in this case exhibits increased complexity, its derivation is still straightforward
\begin{equation}
\begin{aligned}
\dd\psi_{\includegraphics[scale=0.2,valign=c]{kinematic_flow/Figures/four-site_chain/functions/F2_13_14.pdf}}
&= \eps\Bigl(\includegraphics[scale=0.3,valign=c]{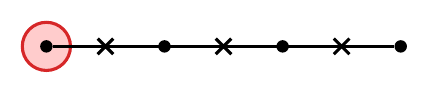}
+\includegraphics[scale=0.3,valign=c]{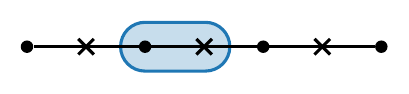}
\\[-2pt] &\qquad{} +
\includegraphics[scale=0.3,valign=c]{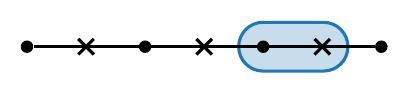}
+ \includegraphics[scale=0.3,valign=c]{kinematic_flow/Figures/four-site_chain/letters/X4-blue.pdf}\Bigr)
\\[-2pt] &\qquad\times
\psi_{\includegraphics[scale=0.2,valign=c]{kinematic_flow/Figures/four-site_chain/functions/F2_13_14.pdf}}
\\[-2pt] &\quad{} +\Bigl(
\includegraphics[scale=0.3,valign=c]{kinematic_flow/Figures/four-site_chain/letters/X2+-blue.pdf}-
\includegraphics[scale=0.3,valign=c]{kinematic_flow/Figures/four-site_chain/letters/X1+red.pdf}\Bigr)
\\[-2pt] &\qquad\times
F_{\includegraphics[scale=0.2,valign=c]{kinematic_flow/Figures/four-site_chain/functions/F12_13_14.pdf}}
\\[-2pt] &\quad{} + \Bigl(
\includegraphics[scale=0.3,valign=c]{kinematic_flow/Figures/four-site_chain/letters/X3-+blue.pdf} -
\includegraphics[scale=0.3,valign=c]{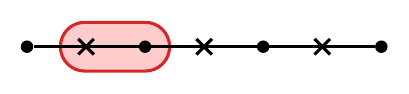}\Bigr)
\\[-2pt] &\qquad\times
F_{\includegraphics[scale=0.2,valign=c]{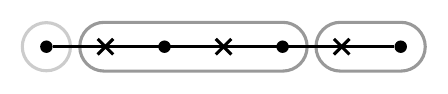}}
\\[-2pt] &\quad{} +\Bigl(\includegraphics[scale=0.3,valign=c]{kinematic_flow/Figures/four-site_chain/letters/X4-blue.pdf}-
\includegraphics[scale=0.3,valign=c]{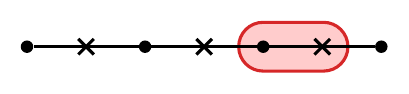}
\Bigr)\\[-2pt] &\qquad\times
F_{\includegraphics[scale=0.2,valign=c]{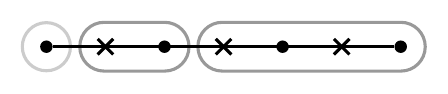}}\,.
\end{aligned}
\end{equation}
The differential equations for other basis functions can be easily derived using the same approach, without any surprise. It is evident that the splitting rules in this basis are more concise.
\subsection{Emergent Time}
In the conventional state-evolution picture, all observables on the future boundary (cosmological correlators) are described in terms of integrals over time \cite{Weinberg:2005vy,Maldacena:2002vr,Chen:2010xka,Wang:2013zva,Arkani-Hamed:2015bza,Zaldarriaga:2003my}. However, in curved spacetime, evaluating the time integrals even for tree-level correlators is quite challenging, which obscures the exploration of their analytic properties.
Mediated by graph tubings, the recently developed kinematic flow method circumvents this difficulty by investigating how a cosmological correlator changes when we vary the kinematic data, facilitating a description rooted purely in boundary kinematics. Therefore, understanding how time naturally emerges from kinematic space has become an important theme.

To make this question more explicit, let us examine how the time integrals are introduced within the state-evolution picture: they are introduced by each vertex in a Feynman graph (rather than the propagators). In the kinematic flow formulation, these vertices become the constituents of the marked graph, and we define a tube by circling a collection of vertices and crosses. The evolution of these tubes exhibits a clear directionality --- they grow by continuously absorbing crosses and other tubes, which corresponds precisely to taking derivatives with respect to the kinematic variables. If time is an emergent concept in kinematic space, then the kinematic process most likely to encode this emergence should involve an alteration in the relations between two (or more) vertices. 

Although the preceding discussion is in principle independent of the choice of function basis, this relation is most manifest in the current one, where all basis functions are organized into building blocks associated with the time-integral representation. For this reason, the evolution of each tubing is endowed with a physical interpretation related to time. In this model, the derivative with respect to each kinematic variable is related to the time derivative, which then acts on a bulk-to-bulk propagator after integration by parts. The activation process in the kinematic flow exactly corresponds to the time derivative acting on the non-time-ordered part of the bulk-to-bulk propagator. This yields the propagator itself multiplied by the corresponding letter. The merger process arises from the time derivative of the step function, which will collapse a time-ordering and eliminate a time integral. The explicit structure of derivatives in this basis allows the relationships between the tubings to be represented by a geometric pattern \cite{Baumann:2025qjx}.

It is then interesting to ask what will happen if we reverse the direction of tubing evolution. Since all splitting processes we have constructed are the inverse of specific mergers, it is expected that each splitting is accompanied by the emergence of a time integral. To observe this, consider the basis functions in the two-site chain case. The generating function can be derived by selecting an arbitrary time ordering in the wavefunction coefficient and subsequently collapsing it 
\begin{align}
\label{expressF}
F_{\includegraphics[scale=0.25,valign=c]{kinematic_flow/Figures/two-site_chain/functions/ZZ.pdf}} &= N \int \frac{\dd \eta_1\, \dd\eta_2}{(\eta_1 \eta_2)^{1+\eps}} \,e^{i(X_1+Y)\eta_1} e^{i(X_2-Y)\eta_2}\, \eta_1\delta(\eta_2-\eta_1) \nonumber\\
&= N \int\frac{\dd \eta}{(-\eta)^{1+2\eps}}\, e^{i(X_1+X_2)\eta} \,.
\end{align}
Here, the collapsing process corresponds to replacing the step function $\theta(\eta_2-\eta_1)$ with its time derivative $\eta_1 \delta(\eta_2-\eta_1)$. In the timeless description, this function is determined by the differential equation eq.~\eqref{timeless}, with the solution as $c_1(X_1 +X_2)^{2\eps}$, where $c_1$ is a constant. Therefore, the time integrand in eq.~\eqref{expressF} can be viewed as a consequence of Schwinger parametrization \cite{Schwinger:1951nm} 
\begin{equation}
F_{\includegraphics[scale=0.25,valign=c]{kinematic_flow/Figures/two-site_chain/functions/ZZ.pdf}} \propto(X_1 +X_2)^{2\eps} = \frac{1}{i\, \Gamma(-2 \eps)}\int\frac{\dd \eta}{(-\eta)^{1+2\eps}} \,e^{i(X_1+X_2)\eta}\,.   
\end{equation}
The emergence of this time integral stems from the fact that all vertices in the corresponding graph tubing are enclosed by one tube. Next, we consider the two splitting channels for this function in eq.~\eqref{anotherbasis-channel}, both of which localize two components of this tubing. By comparing eq.~\eqref{expressF} with the expressions for the two descendants eqs.~\eqref{psi+} and \eqref{psi-}, we can observe that each splitting is equivalent to introducing a time integral, and the two splitting channels just correspond to the two time orderings in the bulk-to-bulk propagator. Similarly, a tree-level wavefunction coefficient with $N$ vertices contains $N-1$ bulk-to-bulk propagators, yielding $2(N-1)$ ways to introduce time integrals for the generating function. This pattern can be easily generalized to arbitrary basis functions: Each viable splitting within this function basis corresponds to the introduction of a new time integral accompanied by a time-ordering. As for the tubings that cannot split further, every vertex is localized, meaning that they carry the maximal number of time integrals. This perspective provides a natural mechanism for the emergence of time.

It is worth noting that the preceding discussion is not necessarily restricted to this particular basis. In section~\ref{local-evolution}, we established the connection between splittings and the introduction of internal energies in another basis, and stated that the introduction of each new internal energy localizes two previous non-local parts. For example, in the case of the three-site chain, the basis function $g$ should depend on $X_{12}$ and $X_3$ rather than $X_{123}$. But what kind of time integral leads to such a dependence for a basis function? Recall that the wavefunction coefficient is given by  
\begin{equation}
\begin{aligned}
\psi_{(3)} &= \int_{-\infty}^{0}
\frac{\dd\eta_1\,\dd\eta_2\,\dd\eta_3
}{(-\eta_1\eta_2\eta_3)^{1+\eps}}
\,e^{iX_1\eta_1}\, G(Y,\eta_1,\eta_2)
\\[-2pt] &\quad\times e^{iX_2\eta_2}\,
G(Y',\eta_2,\eta_3)\, e^{iX_3\eta_3}\,.
\end{aligned}
\end{equation}
and all basis functions can only be related to each other through certain derivatives with respect to kinematic variables. Thus, the only remaining possibility is that the external energies flowing into the two non-local vertices are multiplied by the same time coordinate and subsequently integrated, i.e. $\propto \int \dd \eta\, e^{i(X_1+X_2)\eta}$. This implies that, even in the basis constructed from projective simplices, the number of tubes in a complete tubing still matches the number of time integrals in the corresponding function. Although this is not a rigorous proof, we can confirm its validity through some concrete examples. While the transformation relations between the two bases were given in \cite{Arkani-Hamed:2023kig}, we utilize their inverse in the present discussion. For the two-site chain case, we have
\begin{align}
    Z &= F_{\includegraphics[scale=0.2,valign=c]{kinematic_flow/Figures/two-site_chain/functions/ZZ.pdf}}\,,\\[1ex]
\hline \rule{0pt}{4ex}
    F & = \psi_{\includegraphics[scale=0.2,valign=c]{kinematic_flow/Figures/two-site_chain/functions/F.pdf}} +\frac{1}{2}\,F_{\includegraphics[scale=0.2,valign=c]{kinematic_flow/Figures/two-site_chain/functions/ZZ.pdf}}\,,\\
    \tilde F & = \psi_{\includegraphics[scale=0.2,valign=c]{kinematic_flow/Figures/two-site_chain/functions/tildeF.pdf}} +\frac{1}{2}\,F_{\includegraphics[scale=0.2,valign=c]{kinematic_flow/Figures/two-site_chain/functions/ZZ.pdf}}\,,\\[1ex]
\hline \rule{0pt}{4ex}
    \psi &= \psi_{\includegraphics[scale=0.2,valign=c]{kinematic_flow/Figures/two-site_chain/functions/F.pdf}} +\psi_{\includegraphics[scale=0.2,valign=c]{kinematic_flow/Figures/two-site_chain/functions/tildeF.pdf}} -\psi_{\includegraphics[scale=0.2,valign=c]{kinematic_flow/Figures/two-site_chain/functions/psi.pdf}}\,.
\end{align}
Clearly, excluding the generating function $Z$, all other basis functions can be expressed as two-fold integrals over time, as their expressions include the $\psi$ functions within the new basis based on time ordering. This is consistent with the prediction dictated by the splitting rules.

We can also verify this property in the case of the three-site chain. Now, there are 16 basis functions in both bases, with the transformation relations given by
\begin{align}
Z &= G_{\includegraphics[scale=0.2,valign=c]{kinematic_flow/Figures/three-site_case/functions/Z.pdf}} \,, \\[1ex]\hline \rule{0pt}{4ex} g &= F_{\includegraphics[scale=0.2,valign=c]{kinematic_flow/Figures/three-site_case/functions/g.pdf}} -\frac{1}{3}\, G_{\includegraphics[scale=0.2,valign=c]{kinematic_flow/Figures/three-site_case/functions/Z.pdf}}\,, \\ \tilde g & = F_{\includegraphics[scale=0.2,valign=c]{kinematic_flow/Figures/three-site_case/functions/tildeg.pdf}} -\frac{1}{3}\,G_{\includegraphics[scale=0.2,valign=c]{kinematic_flow/Figures/three-site_case/functions/Z.pdf}}\,, \\ q_2 &= F_{\includegraphics[scale=0.2,valign=c]{kinematic_flow/Figures/three-site_case/functions/q2.pdf}} +\frac{1}{3}\, G_{\includegraphics[scale=0.2,valign=c]{kinematic_flow/Figures/three-site_case/functions/Z.pdf}}\,, \\ \tilde q_2 &= F_{\includegraphics[scale=0.2,valign=c]{kinematic_flow/Figures/three-site_case/functions/tildeq2.pdf}} +\frac{1}{3} \,G_{\includegraphics[scale=0.2,valign=c]{kinematic_flow/Figures/three-site_case/functions/Z.pdf}}\,, \\ q_1 &= F_{\includegraphics[scale=0.2,valign=c]{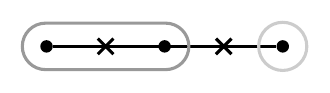}} +F_{\includegraphics[scale=0.2,valign=c]{kinematic_flow/Figures/three-site_case/functions/g.pdf}}-\frac{1}{3}\, G_{\includegraphics[scale=0.2,valign=c]{kinematic_flow/Figures/three-site_case/functions/Z.pdf}}\,, \\ \tilde q_1 &= F_{\includegraphics[scale=0.2,valign=c]{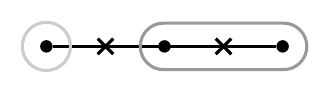}} + F_{\includegraphics[scale=0.2,valign=c]{kinematic_flow/Figures/three-site_case/functions/tildeg.pdf}} -\frac{1}{3}\,G_{\includegraphics[scale=0.2,valign=c]{kinematic_flow/Figures/three-site_case/functions/Z.pdf}}\,, \\[1ex]\hline \rule{0pt}{4ex} q_3 &= \psi_{\includegraphics[scale=0.2,valign=c]{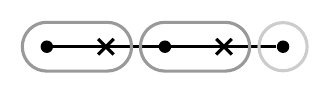}} -\frac{1}{2}\,F_{\includegraphics[scale=0.2,valign=c]{kinematic_flow/Figures/three-site_case/functions/q2.pdf}} +\frac{1}{2}\,F_{\includegraphics[scale=0.2,valign=c]{kinematic_flow/Figures/three-site_case/functions/tildeg.pdf}}\,,\\ \tilde q_3 &= \psi_{\includegraphics[scale=0.2,valign=c]{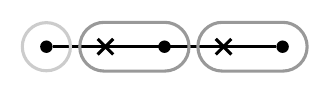}}-\frac{1}{2}\,F_{\includegraphics[scale=0.2,valign=c]{kinematic_flow/Figures/three-site_case/functions/tildeq2.pdf}} +\frac{1}{2}\,F_{\includegraphics[scale=0.2,valign=c]{kinematic_flow/Figures/three-site_case/functions/g.pdf}} \,,\\
f &= \psi_{\includegraphics[scale=0.2,valign=c]{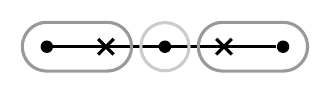}}
+\tfrac{1}{2} F_{\includegraphics[scale=0.2,valign=c]{kinematic_flow/Figures/three-site_case/functions/g.pdf}} \notag\\[-2pt] &\quad{} +\tfrac{1}{2}
F_{\includegraphics[scale=0.2,valign=c]{kinematic_flow/Figures/three-site_case/functions/tildeg.pdf}} +\tfrac{1}{3} G_{\includegraphics[scale=0.2,valign=c]{kinematic_flow/Figures/three-site_case/functions/Z.pdf}}
\,, \\[1pt] Q_2&=
\psi_{\includegraphics[scale=0.2,valign=c]{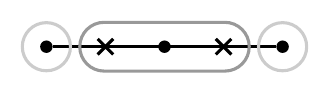}}
\notag\\[-2pt] &\quad{} +\tfrac{1}{2}
F_{\includegraphics[scale=0.2,valign=c]{kinematic_flow/Figures/three-site_case/functions/q2.pdf}} +\tfrac{1}{2} F_{\includegraphics[scale=0.2,valign=c]{kinematic_flow/Figures/three-site_case/functions/tildeq2.pdf}}
\,, \\[1pt] Q_1&=
\psi_{\includegraphics[scale=0.2,valign=c]{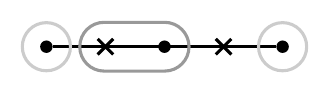}}
\notag\\[-2pt] &\quad{}
+\psi_{\includegraphics[scale=0.2,valign=c]{kinematic_flow/Figures/three-site_case/functions/tildeq3.pdf}}
+\tfrac{1}{2}
F_{\includegraphics[scale=0.2,valign=c]{kinematic_flow/Figures/three-site_case/functions/q1.pdf}}
\notag\\[-2pt] &\quad{} -\tfrac{1}{2}
F_{\includegraphics[scale=0.2,valign=c]{kinematic_flow/Figures/three-site_case/functions/tildeq2.pdf}}
+\tfrac{1}{2}
F_{\includegraphics[scale=0.2,valign=c]{kinematic_flow/Figures/three-site_case/functions/g.pdf}}
\,, \\[1pt] Q_3 &=
\psi_{\includegraphics[scale=0.2,valign=c]{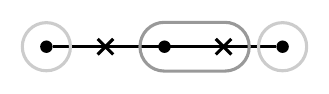}}
\notag\\[-2pt] &\quad{}
+\psi_{\includegraphics[scale=0.2,valign=c]{kinematic_flow/Figures/three-site_case/functions/q3.pdf}}
+\tfrac{1}{2}
F_{\includegraphics[scale=0.2,valign=c]{kinematic_flow/Figures/three-site_case/functions/tildeq1.pdf}}
\notag\\[-2pt] &\quad{} -\tfrac{1}{2}
F_{\includegraphics[scale=0.2,valign=c]{kinematic_flow/Figures/three-site_case/functions/q2.pdf}}
+\tfrac{1}{2}
F_{\includegraphics[scale=0.2,valign=c]{kinematic_flow/Figures/three-site_case/functions/tildeg.pdf}}
\,, \\[1pt] F &=
\psi_{\includegraphics[scale=0.2,valign=c]{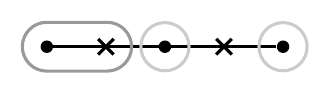}}
+\psi_{\includegraphics[scale=0.2,valign=c]{kinematic_flow/Figures/three-site_case/functions/f.pdf}}
\notag\\[-2pt] &\quad{}
+\psi_{\includegraphics[scale=0.2,valign=c]{kinematic_flow/Figures/three-site_case/functions/q3.pdf}} +\tfrac{1}{2} F_{\includegraphics[scale=0.2,valign=c]{kinematic_flow/Figures/three-site_case/functions/g.pdf}}
\notag\\[-2pt] &\quad{} +\tfrac{1}{2}
F_{\includegraphics[scale=0.2,valign=c]{kinematic_flow/Figures/three-site_case/functions/q1.pdf}} +\tfrac{1}{2}
F_{\includegraphics[scale=0.2,valign=c]{kinematic_flow/Figures/three-site_case/functions/q2.pdf}} \,,
\\[1pt] \tilde F &=
\psi_{\includegraphics[scale=0.2,valign=c]{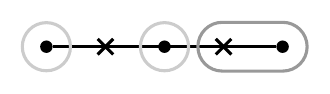}}
+\psi_{\includegraphics[scale=0.2,valign=c]{kinematic_flow/Figures/three-site_case/functions/f.pdf}}
\notag\\[-2pt] &\quad{}
+\psi_{\includegraphics[scale=0.2,valign=c]{kinematic_flow/Figures/three-site_case/functions/tildeq3.pdf}}
+\tfrac{1}{2}
F_{\includegraphics[scale=0.2,valign=c]{kinematic_flow/Figures/three-site_case/functions/tildeg.pdf}}
\notag\\[-2pt] &\quad{} +\tfrac{1}{2}
F_{\includegraphics[scale=0.2,valign=c]{kinematic_flow/Figures/three-site_case/functions/tildeq1.pdf}}
+\tfrac{1}{2}
F_{\includegraphics[scale=0.2,valign=c]{kinematic_flow/Figures/three-site_case/functions/tildeq2.pdf}}
\,, \\[1pt] \psi &=
\psi_{\includegraphics[scale=0.2,valign=c]{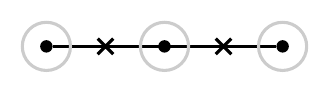}}
+\psi_{\includegraphics[scale=0.2,valign=c]{kinematic_flow/Figures/three-site_case/functions/tildeq3.pdf}}
+\psi_{\includegraphics[scale=0.2,valign=c]{kinematic_flow/Figures/three-site_case/functions/Q2.pdf}}
\notag\\[-2pt] &\quad{}
+\psi_{\includegraphics[scale=0.2,valign=c]{kinematic_flow/Figures/three-site_case/functions/q3.pdf}}
+\psi_{\includegraphics[scale=0.2,valign=c]{kinematic_flow/Figures/three-site_case/functions/f.pdf}}
\notag\\[-2pt] &\quad{} -\psi_{\includegraphics[scale=0.2,valign=c]{kinematic_flow/Figures/three-site_case/functions/Q1.pdf}}
-\psi_{\includegraphics[scale=0.2,valign=c]{kinematic_flow/Figures/three-site_case/functions/F.pdf}}
\notag\\[-2pt] &\quad{} -\psi_{\includegraphics[scale=0.2,valign=c]{kinematic_flow/Figures/three-site_case/functions/tildeF.pdf}}
-\psi_{\includegraphics[scale=0.2,valign=c]{kinematic_flow/Figures/three-site_case/functions/Q3.pdf}}
\,.\end{align}
Here, these equations have been classified according to the maximal number of time integrals present in the right-hand side. By comparing these results with the tubings associated with the basis functions defined in eq.~\eqref{three-site source}, it is believed that the maximal number of tubes in a complete tubing corresponds to the number of time integrals present in this basis function. We conclude that under both bases, the splitting process which introduces a new internal energy leads to the emergence of a time integral.   

We always expect the wavefunction coefficients which are associated with physical observables to occupy a privileged status. Although it is not apparent in the original literature \cite{Arkani-Hamed:2023kig}, under this new basis, a basis function contributes to the wavefunction coefficient only if every vertex in its tubing is local. In contrast, in the remaining tubings, at least two vertices are enclosed by a tube, losing the information of the internal energy connecting them. Exploring the additional insights yielded by this feature is an interesting avenue for future research.
\section{Beyond Single Graphs}
\label{sec:beyond}
While the preceding discussion was restricted to individual Feynman diagrams, as in the context of scattering amplitudes, physical observables require summing over all channels contributing to a given process \cite{Parke:1986gb,Benincasa:2007xk,Hodges:2012ym,Cheung:2014dqa,Arkani-Hamed:2017mur}. It has been demonstrated in \cite{Arkani-Hamed:2023kig,Arkani-Hamed:2023bsv,Baumann:2025qjx} that the kinematic flow can be systematically generalized to $\text{tr}\,\phi^3$ theory of colored scalars, leading to the correct differential equations. Since the three steps of activation, growth (merger), and absorption remain reversible in this case, we can accordingly construct appropriate splitting rules. In this section, we briefly outline the splitting rules after summing over graphs in this model, and then use them to analyze the relations among the basis functions.

In cosmology, an $n$-point correlation function depends on $n$ associated momentum vectors, which naturally form an $n$-gon due to the translation invariance. Hence, varying the kinematic data is equivalent to altering the side lengths and the overall shape of this polygon. The wavefunction now can be expressed as flavor-ordered partial wavefunctions multiplied by flavor factors. At tree level, the flavor-ordered wavefunction receives contributions exclusively from planar polygons formed by the kinematic variables \cite{Arkani-Hamed:2023lbd}. Each triangulation of the polygon corresponds to an exchange process within the flavor-ordered wavefunction, and these triangulations establish a systematic mathematical relationship with the flat-space wavefunction. For example, in the four-point function case, the following two triangulations of the momentum quadrilateral correspond to the $s$- and $t$-channel exchanges of the flavor-ordered wavefunction
\begin{align}
\nonumber\includegraphics[scale=0.8,valign=c]{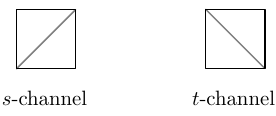}
\end{align}
Similarly, for the five-point function, the associated pentagon can be triangulated in five distinct ways
\begin{align}
\nonumber\includegraphics[scale=0.5,valign=c]{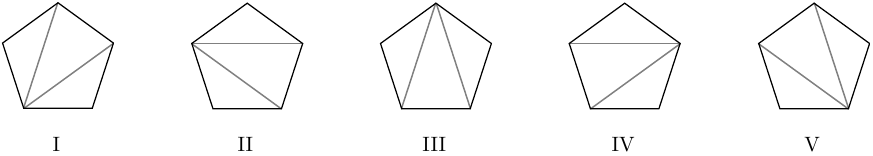}
\end{align}
The explicit relationship between these triangulations and the flat-space wavefunction is detailed in \cite{Arkani-Hamed:2023kig}. We now turn to the differential equations satisfied by these functions. First, the letters appearing in the differential equations correspond to the shaded sub-polygons of the kinematic polygon. Next, the basis functions are associated to some (maybe disconnected) shaded sub-polygons, wherein each constituent shaded triangle possesses at least one dashed internal edge. The kinematic flow starts from the triangulations of a given polygon and evolves these configurations by either shading specific sub-polygon(s) or converting solid internal edge(s) into dashed ones. The final result of this evolution reduces to a complete shaded polygon without any internal edges. The formulation of the kinematic flow, along with some concrete examples of its application, is comprehensively illustrated in \cite{Arkani-Hamed:2023kig}.

From this, we can construct appropriate splitting rules. We only need to note that,
compared to the previous tubing representation, each sub-polygon now corresponds to a tube, where the solid or dashed lines indicate whether or not the tube contains the internal energy associated with that line. Following the philosophy of the preceding discussion, we state the splitting rules below without further elaboration:
\begin{itemize}
    \item[1.] The first kind of splitting corresponds to introducing a new internal energy. Specifically, we connect any two sites on the polygon to form an internal line that does not intersect any existing ones. This procedure yields a triangle and another sub-polygon, with the latter becoming passive (denoted in blue). For the new internal line, the side facing the triangle should be dashed, while the opposite side can be either dashed or solid. If the latter is solid, this triangle becomes active (denoted in red).
    \item[2.] The second kind of splitting converts a dashed line into a solid one, leaving the triangle on the dashed side uncolored. For an internal line that also has a solid edge, the splitting can only be performed when both of its sides correspond to triangles. 
\end{itemize}
These simple rules suffice for analyzing the relationships between the basis functions. If we wish to derive the complete differential equations, we only need to further distinguish different types of splittings. For example, for the  basis based on time ordering introduced in section~\ref{sec:another-basis}, we should drop the second splitting rule.

\noindent\textbf{Four-point function.} The splitting rules are best illustrated through some concrete examples. Let us begin with the four-point function, where the four wavenumbers form a quadrilateral. The corresponding generating function is fully shaded, which we consider to be passive in this context
\begin{align}
\includegraphics[scale=0.3,valign=c]{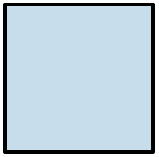}
\end{align}
Following the splitting rules, it yields the following splitting channels
\begin{align}
\includegraphics[scale=1,valign=c]{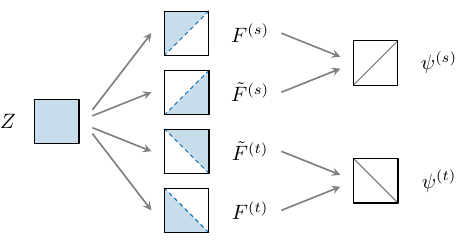}
\end{align}
In the first step, there are two distinct ways to connect two sites to form an internal line. Since the original quadrilateral is divided into two triangles, the introduced internal line should be dashed. Moreover, each splitting configuration admits two choices for the passive sub-polygon. We can deduce that $Z$ will appear in the differential equations for the four descendant functions. In the second step, only one kind of splitting is permitted, which converts the introduced dashed line into a solid line. The final result consists of the two triangulations of the quadrilateral, which will contribute to the flavor-ordered wavefunction. These are precisely the inverse processes in the case of the kinematic flow introduced in \cite{Arkani-Hamed:2023kig}.

\noindent\textbf{Five-point function.} For the five-point function case, the splitting configurations exhibit a much richer variety. At this stage, the generating function comprises five sites, with each site allowing four distinct ways to introduce an internal energy. Taking one such site as an example, this leads to
\begin{align}
\includegraphics[scale=1,valign=c]{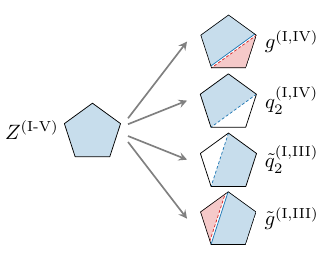}
\end{align}
Here, the introduction of each internal line introduces a triangle and a sub-polygon (a quadrilateral), and thus the edge on the sub-polygon side can be either dashed or solid. Since each site admits four splitting patterns and every two triangulations share a common basis function, there are 10 descendants for the generating function $Z^{(\text{I-V})}$.

Taking one of the descendant functions $g^{(\text{I,IV})}$ as an example, the following splitting channels are given by
\begin{align}
\includegraphics[scale=1,valign=c]{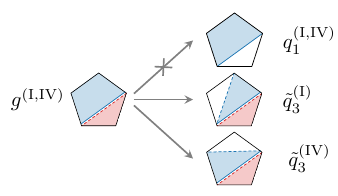}   
\end{align}
As the two sides of the internal line in the sub-polygon of $g^{(\text{I,IV})}$ do not both correspond to triangles, this line cannot be converted to a solid one, making the first channel forbidden. As for introducing a new internal line, there are two distinct splitting channels, and this internal line should be dashed. We note that $g^{(\text{I,IV})}$ can now split into functions associated with different triangulations, which is the direct result of summing over individual graphs.

Finally, we investigate the splitting of the function $\tilde q_3^{(\text{I})}$. Now, we can only convert some dashed lines to solid ones, leading to the following processes
\begin{align}
\includegraphics[scale=1,valign=c]{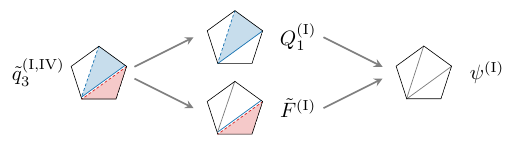}
\end{align}
Similarly, the final outcome of the splittings will contribute to the flavor-ordered wavefunction.

Compared to the case of individual Feynman diagrams, after summing over different channels, a basis function may appear in multiple full triangulations of the polygon. As shown in \cite{Arkani-Hamed:2023kig}, the compatibility among different channels is captured by a geometric structure called the dressed associahedron \cite{Arkani-Hamed:2017mur}. Here, we briefly analyze how the relationship between the basis functions and the associahedron can be understood from the perspective of kinematic flow and splitting rules. The evolution of the kinematic flow starts from the triangulations of the momentum polygon, which correspond to all vertices of the associahedron. Subsequently, if two triangulations share $N-1$ identical internal line(s), where $N$ is the number of bulk-to-bulk propagators in each channel,
the corresponding two vertices can be connected to form an edge. This edge represents a partial triangulation that serves as the descendant of the two vertex functions, with the triangulation obtained precisely by removing the unshared internal lines between the triangulations of the two vertices. Then, the edges sharing $N-2$ identical internal lines may further form faces, continuing the evolution until reaching the central untriangulated polygon (shaded). Therefore, the geometric evolution from the vertices through the edges to the center is exactly the procedure of systematically stripping away internal energies (collapsing time integrals). \footnote{In the basis based on time ordering proposed in \cite{Baumann:2025qjx}, the basis functions are represented in an alternative way, such that each vertex of the associahedron is generalized to $2^N$ ones, where $N$ is the number of the bulk-to-bulk propagators of the graph. However, these time-orderings can still be encoded in the same sub-polygons within this basis, as illustrated in section~\ref{sec:another-basis}.}

It is more interesting to view the evolution of these sub-polygons via splittings. This time we begin with the generating function at the center, and each splitting that introduces a new internal energy yields a geometric object whose codimension increases by one. These objects subsequently intersect to form higher-codimensional ones, and finally terminate at the vertices, which are the full triangulations of the polygon. For example, faces intersect to form edges, and edges intersect to yield vertices. Since summing over graphs only involves the mixing of functions from different channels rather than levels, the previous conclusion remains robust: Introducing a new internal energy leads to the emergence of a time integral. Thus, the intersection of the sub-elements within the associahedron may also encode the physical mechanism for the emergence of time. The kinematic flow and splitting rules also provide a flowing character to the elements of the associahedron.

Recently, for the flat-space $\text{tr}\,\phi^3$ theory, the associahedron has been generalized into a richer geometric structure known as the cosmohedron \cite{Arkani-Hamed:2024jbp,Glew:2025otn}. It will then be interesting to explore whether deep connections between these geometric structures can be established through appropriate rules.
\section{Conclusions and Outlook}
\label{sec:conclusion}
Conformally coupled scalars in a power-law cosmology serve as the simplest model for our study of cosmological correlations, yet they may capture the most general physical principles. By encoding the essential information in the differential equations into graph tubings, we can construct a set of graphical rules describing a flow in kinematic space. In this paper, we propose a novel perspective on the kinematic flow by inverting the direction in which the tubings flow.

The motivation for this approach may appear strange, as it lacks a  direct mathematical correspondence analogous to the kinematic flow, where the flow of tubings represents the total differentials of basis functions. However, if these graph tubings are viewed as higher-level objects that transcend the apparent differential equations, the splitting rules can then be regarded as the natural principles governing their kinematics. Rather than merely yielding the differential equations, the most crucial contribution of these rules is the recasting of the relationships between the basis functions. In this manner, deeper physical structures such as singularities and locality are directly manifested. Moreover, this framework provides some hints of the emergence of time from kinematic space.

Regardless of whether the tubings constitute more fundamental building blocks, the splitting rules provide some directions for future research: 
\begin{itemize}
    \item In this work, the splitting rules we constructed are  restricted to tree-level correlators. It will then be interesting to explore whether these rules can be generalized to cosmological loop integrands. Although the graph tubings for the loop graphs are similar to those in the tree-level case, some basis functions vanish because they are equal to the difference of functions obtained via two identical replacement operations \cite{Baumann:2024mvm}. While these functions have to be dropped by hand in the kinematic flow, they would be naturally absent if suitable splitting rules are formulated. If one does not aim to yield the full differential equations, the splitting rules may serve as a simpler alternative for explaining the relationships between these functions. We will return to this topic in our future work. 
    
    \item So far, our discussion is restricted to the specific model of conformally coupled scalars. The simplest extension of this is to allow the exchanged particles to have a generic mass. This leads to the bulk-to-bulk propagators taking a form that involves Hankel functions. By employing the integral representation of the Hankel functions, the authors of \cite{Gasparotto:2024bku,Baumann:2026atn} obtained a system of differential equations for the two-point tree-level wavefunction coefficient in a de Sitter background. It is therefore interesting to investigate whether this system can be recast into a form described by the kinematic flow and splitting rules. In particular, the splitting rules help us to identify the simplest basis function by collapsing all time orderings. It would be rewarding to explore whether further physical insights can be extracted from this generating function.  
   
    \item Both the kinematic flow and splitting rules in this paper depend crucially on the choice of function basis. Then, a natural question to ask is how much information encoded in the graph tubings is basis-independent. Motivated by the splitting rules, we conjecture that for every complete tubing, the number of tubes exactly matches the number of time integrals in the associated basis function. Moreover, we observe that the manner in which these tubes enclose the vertices is closely related to physical concepts such as locality and the emergence of time. If we can fully exploit the information encoded within the tubings, we might be able to bypass certain mathematical complexities and generalize the kinematic flow to more general cases using physical arguments. 
\end{itemize}
We believe that the graph tubings and the kinematic flow are not merely artifacts of the differential equations, but rather possess a life of their own.

\textbf{Note added.} As this work was being completed, the kinematic flow rules for massive correlators \cite{Baumann:2026atn}, as well as those for banana loops and unparticles \cite{Westerdijk:2026msm} were proposed in recent literature. While the shrinking and cutting procedures defined in \cite{Baumann:2026atn} and \cite{Westerdijk:2026msm} share similarities with the splitting of tubes in this work, they are motivated by entirely different physical objectives. 
\vspace{0.2cm}
\appendix
\section{Differential Equations for Three-Site Chain Case}
\label{app:diffeq3site}
In this appendix, we present the differential equations satisfied by the basis functions in the three-site chain case, categorized according to a slightly different level classification scheme. For convenience, we define the following shorthand notations
\begin{equation}
\begin{aligned}
    X_1^\pm &\equiv X_1 \pm Y \,, &\qquad  X_{12}^\pm &\equiv X_1 +X_2 \pm Y'\,, \\ X_2^{\pm \pm} &\equiv X_2 \pm Y \pm Y' \,, &\qquad X_{23}^\pm &\equiv X_2 +X_3 \pm Y\,,\\ 
    X_3 ^\pm & \equiv X_3 \pm Y'\,, &\qquad X_{123} & \equiv X_1 +X_2 +X_3\,.
\end{aligned}
\end{equation}
\noindent\textbf{Level 0.}  We first observe the existence of a basis function $Z$ whose total differential depends only on itself,
\begin{equation}
  \dd Z = 3 \eps \, Z\, \dd \log   X_{123}\, .
\end{equation}

\noindent\textbf{Level 1.} Next, using only $Z$ as the source function, one can generate the differential equations for the other four basis functions,
\begin{align}
\dd q_2
&=\eps\Bigl[2q_2\,\dd\log X_{12}^{-}
 +(q_2-Z)\,\dd\log X_3^{+} +Z\,\dd\log X_{123} \Bigr]\,,\\[-1pt]
\dd g &=\eps\Bigl[2g\,\dd\log X_{12}^{+}
 +(g+Z)\,\dd\log X_3^{-}-Z\,\dd\log X_{123}\Bigr]\,,
\label{gdiffeq}
\\[-1pt]
\dd\tilde g &=\eps\Bigl[ 2\tilde g\,\dd\log X_{23}^{+} +(\tilde g+Z)\,\dd\log X_1^{-} -Z\,\dd\log X_{123} \Bigr]\,,
\label{tildeg1}
\\[-1pt]
\dd\tilde q_2 &=\eps\Bigl[2\tilde q_2\,\dd\log X_{23}^{-} +(\tilde q_2-Z)\,\dd\log X_1^{+} +Z\,\dd\log X_{123}
\Bigr]\,.
\end{align}
Note that the basis functions with and without tildes are related by the exchange of sites $1 \leftrightarrow 3$.
The analogous equation structures satisfied by $q_2$ and $g$ (as well as $\tilde q_2$ and $\tilde g$) motivate their assignment to the same function level, which is different from \cite{Arkani-Hamed:2023kig}. 

\noindent\textbf{Level 2.} We then consider three functions that are sourced by $g$, $\tilde g$ and $Z$,
\begin{align}
\dd q_1 &=\eps\Bigl[
 2q_1\,\dd\log X_{12}^{+} +(q_1-g)\,\dd\log X_3^{+} \notag\\[-2pt]
&\quad{} +(g+Z)\,\dd\log X_3^{-} -Z\,\dd\log X_{123} \Bigr]\,,
\label{diffeqq1}
\\[1pt]
\dd f &=\eps\Bigl[ f\bigl(
\dd\log X_1^{-}+\dd\log X_3^{-}
\bigr)\notag\\[-2pt]
&\quad{}
+(f-g-\tilde g-Z)\,\dd\log X_2^{++} +g\,\dd\log X_{12}^{+}
\notag\\[-2pt]
&\quad{} +\tilde g\,\dd\log X_{23}^{+}
+Z\,\dd\log X_{123}\Bigr]\,,
\label{diffeqf}
\\[1pt]
\dd\tilde q_1&=\eps\Bigl[2\tilde q_1\,\dd\log X_{23}^{+}+(\tilde q_1-\tilde g)\,\dd\log X_1^{+}\notag\\[-2pt]
&\quad{}+(\tilde g+Z)\,\dd\log X_1^{-}
 -Z\,\dd\log X_{123}\Bigr]\,.
\label{diffeqtildeq1}
\end{align}
The total differentials of these three functions depend not only on the basis functions at the preceding level, but also at higher levels. This feature is a direct consequence of permitting absorptions among graph tubings within the kinematic flow. Meanwhile, there are three basis functions generated solely by the functions at level 1, whose total differentials are
\begin{align}
\dd q_3 &=\eps\Bigl[ q_3\,\dd\log X_2^{+-} \notag\\[-2pt]
&\quad{} +(q_3-\tilde g)\,\dd\log X_3^{+} +(q_3+q_2)\,\dd\log X_1^{-} \notag\\[-2pt]
&\quad{} +\tilde g\,\dd\log X_{23}^{+}
 -q_2\,\dd\log X_{12}^{-} \Bigr]\,,
\label{diffeqq3}
\\[1pt]
\dd Q_2 &=\eps\Bigl[Q_2\,\dd\log X_2^{--} \notag\\[-2pt] &\quad{} +(Q_2-q_2)\,\dd\log X_1^{+} +(Q_2-\tilde q_2)\,\dd\log X_3^{+} \notag\\[-2pt]
&\quad{} +q_2\,\dd\log X_{12}^{-}
+\tilde q_2\,\dd\log X_{23}^{-}\Bigr]\,,
\label{diffeqQ2}
\\[1pt]
\dd\tilde q_3 &=\eps\Bigl[\tilde q_3\,\dd\log X_2^{-+} \notag\\[-2pt]
&\quad{}+(\tilde q_3-g)\,\dd\log X_1^{+}
+(\tilde q_3+\tilde q_2)\,\dd\log X_3^{-} \notag\\[-2pt]
&\quad{} +g\,\dd\log X_{12}^{+} -\tilde q_2\,\dd\log X_{23}^{-}\Bigr]\,.
\end{align}
These three functions also share a similar differential structure.

\noindent\textbf{Level 3.} There are four basis functions at this level. Although the differential equations at this stage become more complicated, one can still identify two functions that are analogous to $q_1$ and $q_3$ ($\tilde q_1$ and $\tilde q_3$) 
\begin{align}
\dd Q_1&=\eps\Bigl[Q_1\,\dd\log X_2^{-+}
 +(Q_1-q_1)\,\dd\log X_1^{+}\notag\\[-2pt]&\quad{}+(Q_1-\tilde q_3)\,\dd\log X_3^{+}+(\tilde q_3+\tilde q_2)\,\dd\log X_3^{-}\notag\\[-2pt]
&\quad{}+q_1\,\dd\log X_{12}^{+}-\tilde q_2\,\dd\log X_{23}^{-}\Bigr]\,,
\\[1pt]
\dd Q_3&=\eps\Bigl[Q_3\,\dd\log X_2^{+-}
 +(Q_3-\tilde q_1)\,\dd\log X_3^{+}
\notag\\[-2pt]
&\quad{}+(Q_3-q_3)\,\dd\log X_1^{+}
 +(q_3+q_2)\,\dd\log X_1^{-}\notag\\[-2pt]
&\quad{}\tilde q_1\,\dd\log X_{23}^{+}
 -q_2\,\dd\log X_{12}^{-}\Bigr]\,.
\end{align}
The remaining two basis functions are analogous to $f$:
\begin{align}
\dd F&=\eps\Bigl[F\,\dd\log X_1^{-}
 +(F-f)\,\dd\log X_3^{+}\notag\\[-2pt]
&\quad{}+(F-q_1-q_2-q_3)\,\dd\log X_2^{++}\notag\\[-2pt]
&\quad{}+f\,\dd\log X_3^{-}+q_1\,\dd\log X_{12}^{+}\notag\\[-2pt]
&\quad{}+q_3\,\dd\log X_2^{+-}+q_2\,\dd\log X_{12}^{-}\Bigr]\,,
\\[1pt]
\dd\tilde F&=\eps\Bigl[\tilde F\,\dd\log X_3^{-}+(\tilde F-f)\,\dd\log X_1^{+}\notag\\[-2pt]
&\quad{}+(\tilde F-\tilde q_1-\tilde q_2-\tilde q_3)\,\dd\log X_2^{++}
\notag\\[-2pt]
&\quad{}+f\,\dd\log X_1^{-}+\tilde q_1\,\dd\log X_{23}^{+}\notag\\[-2pt]
&\quad{}+\tilde q_3\,\dd\log X_2^{-+}+\tilde q_2\,\dd\log X_{23}^{-}\Bigr]\,.
\end{align}

\noindent\textbf{Level 4.} Finally, we arrive at the wavefunction coefficient $\psi$, which is of primary physical interest and serves as the starting point for the construction of these differential equations in \cite{Arkani-Hamed:2023kig},
\begin{align}
\dd\psi&=\eps\Bigl[
 (\psi-F)\,\dd\log X_1^{+} +(\psi-\tilde F)\,\dd\log X_3^{+} \notag\\[-2pt]
&\quad{} +(\psi-Q_1-Q_2-Q_3)\,\dd\log X_2^{++} \notag\\[-2pt]
&\quad{} +F\,\dd\log X_1^{-} +\tilde F\,\dd\log X_3^{-} \notag\\[-2pt]
&\quad{} +Q_1\,\dd\log X_2^{-+} +Q_3\,\dd\log X_2^{+-} +Q_2\,\dd\log X_2^{--}\Bigr]\,.
\label{diffeqpsi}
\end{align}
\section{More Examples}
\label{app:more-example}
In this appendix, we will show how to implement the splitting rules in more complicated cases to derive the correct differential equations. We will explicitly derive some selected differential equations for the four-site chain and four-site star cases, while the remaining ones can be obtained following the identical procedure.

To achieve this, we first elucidate how to derive the colored tubing for a given basis function by only analyzing all of its parent functions. Recall that for the basis introduced in section~\ref{sec:alternative}, no tube within a complete tubing exhibits different properties (passive or active) across different splitting channels. Moreover, the splitting of either an active or passive tube generates an active tube by the same rule. Thus, given the complete tubing of a basis function, we only need to identify all its parent functions to locate the active tubes, leaving the rest naturally classified as passive. Taking one of the complete tubing \includegraphics[scale=0.3,valign=c]{kinematic_flow/Figures/four-site_chain/functions/F2_13_14.pdf} as an example, we can deduce that
\begin{equation}
\begin{aligned}
\includegraphics[scale=0.7,valign=c]{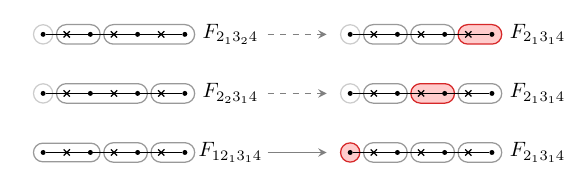}
\end{aligned}
\end{equation}
From this we obtain three active tubes, and the last one should be assigned to be passive. Consequently, we arrive at the colored complete tubing \includegraphics[scale=0.2,valign=c]{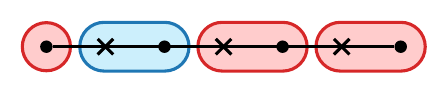}, which is precisely the correct result. Analogously, the colored complete tubing for arbitrary basis functions at tree-level can be obtained via this approach. We are now ready to study some more complicated examples. 
\subsection{Four-Site Chain}
In the case of the four-site chain, the generating function $F_{12_23_24}$ (corresponding to the tubing \includegraphics[scale=0.2,valign=c]{kinematic_flow/Figures/four-site_chain/functions/F12_23_24.pdf}) possesses six splitting locations, and the total number of basis functions is
\begin{equation}
    1 + \begin{pmatrix}
        6 \\ 1
    \end{pmatrix} + \begin{pmatrix}
        6\\ 2  
    \end{pmatrix} + \begin{pmatrix}
        6 \\ 3
    \end{pmatrix} +\begin{pmatrix}
        6\\ 4
    \end{pmatrix} +\begin{pmatrix}
        6\\ 5
    \end{pmatrix}+\begin{pmatrix}
        6\\ 6
    \end{pmatrix}= 1 + \,6 +\,15 +\,20 +\, 15 +\,6 +\,1 = 64\,.
\end{equation}
Here, each term in this equation represents the number of basis functions from level 0 to level 6, respectively. In this context, the generating function continues to satisfy the simplest differential equation
\begin{equation}
    \dd F_{12_23_24} = 4\eps \, F_{12_23_24} \,\includegraphics[scale=0.3,valign=c]{kinematic_flow/Figures/four-site_chain/letters/X1234blue.pdf} \, .
\end{equation}
Furthermore, we note the existence of an additional basis function that cannot be generated via splitting, $F_{12_13_34}$, i.e. \includegraphics[scale=0.3,valign=c]{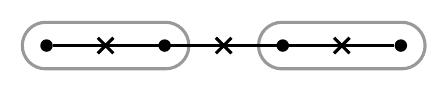}, since all channels that could yield it are forbidden. This can also be verified through the kinematic flow, which reveals that neither growth nor absorption can be applied to this function. Treating all tubes in this complete tubing as passive, we can read off the corresponding differential equation
\begin{equation}
    \dd F_{12_13_34} = 2  \eps  \,F_{12_13_34} \,\Big(\includegraphics[scale=0.3,valign=c]{kinematic_flow/Figures/four-site_chain/letters/X12+blue.pdf} + \includegraphics[scale=0.3,valign=c]{kinematic_flow/Figures/four-site_chain/letters/X34+blue.pdf} \,\Big)\, .
\end{equation}
Now we will present the differential equations for some concrete examples.

\noindent \textbf{Level 1.} We first consider two of the splitting channels for the generating function
\begin{align}
\includegraphics[scale=0.7,valign=c]{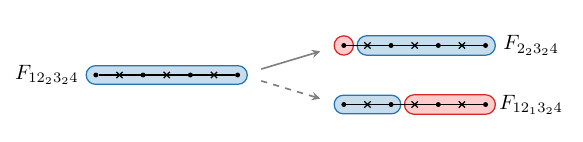}
\end{align}
Note that the parent functions associated with the splittings denoted by solid and dashed lines enter the differential equation with a relative minus sign. The differential equations now are 
\begin{align}
\dd F_{2_23_24} &=\eps\Bigl[\,3F_{2_23_24}\,
\includegraphics[scale=0.3,valign=c]{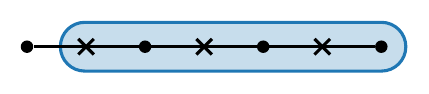}
 \notag\\[-2pt] &\quad{}
 +(F_{2_23_24}-F_{12_23_24})\,
\includegraphics[scale=0.3,valign=c]{kinematic_flow/Figures/four-site_chain/letters/X1+red.pdf}
 \notag\\[-2pt]&\quad{}
 +F_{12_23_24}\,
\includegraphics[scale=0.3,valign=c]{kinematic_flow/Figures/four-site_chain/letters/X1234blue.pdf}
 \Bigr]\,,
\\[2pt]
\dd F_{12_13_24} &=\eps\Bigl[\,2F_{12_13_24}\,\includegraphics[scale=0.3,valign=c]{kinematic_flow/Figures/four-site_chain/letters/X12+blue.pdf}\notag\\[-2pt]
&\quad{}+2(F_{12_13_24}+F_{12_23_24})\,
\includegraphics[scale=0.3,valign=c]{kinematic_flow/Figures/four-site_chain/letters/X34-red.pdf}
\notag\\[-2pt] &\quad{}-F_{12_23_24}\,
\includegraphics[scale=0.3,valign=c]{kinematic_flow/Figures/four-site_chain/letters/X1234blue.pdf}
 \Bigr]\,.
\end{align}

\noindent \textbf{Level 2.} The functions at this level can be obtained by splitting the generating function twice, and thus they generally have two parents. Taking one of the basis function $F_{2_23_34}$ associated with the tubing \includegraphics[scale=0.3,valign=c]{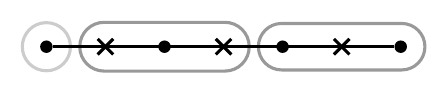} as an example, the function tree is given by
\begin{widetext}
\begin{align}
\includegraphics[scale=1,valign=c]{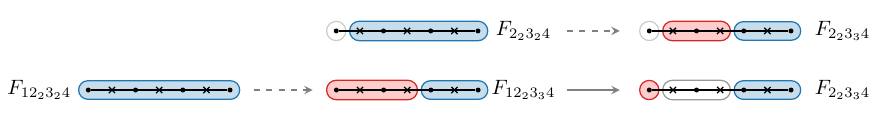} 
\end{align}
\end{widetext}
In the bottom channel, since the active tube and the passive tube are not adjacent in the tubing of $F_{2_23_34}$, we could trace back to an ancestor function in this direction. Thus, we can obtain the corresponding differential equation
\begin{align}
\dd F_{2,23,34}
&=\eps\Bigl[\, 2F_{2_23_34}\,\includegraphics[scale=0.3,valign=c]{kinematic_flow/Figures/four-site_chain/letters/X34+blue.pdf}\notag\\[-2pt] &\quad{}+(F_{2_23_34}+F_{2_23_24})\,
\includegraphics[scale=0.3,valign=c]{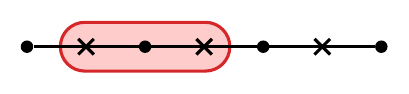}
 \notag\\[-2pt]&\quad{}+(F_{2_23_34}-F_{12_23_34})\,
\includegraphics[scale=0.3,valign=c]{kinematic_flow/Figures/four-site_chain/letters/X1+red.pdf}
 \notag\\[-2pt] &\quad{}-F_{2_23_24}\,
\includegraphics[scale=0.3,valign=c]{kinematic_flow/Figures/four-site_chain/letters/X234-blue.pdf}
\notag\\[-2pt]&\quad{}+(F_{12_23_34}+F_{12_23_24})\,
\includegraphics[scale=0.3,valign=c]{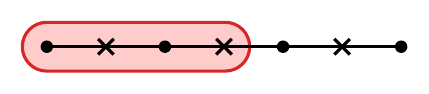}\notag\\[-2pt] &\quad{}-F_{12_23_24}\,
\includegraphics[scale=0.3,valign=c
 ]{kinematic_flow/Figures/four-site_chain/letters/X1234blue.pdf}
 \Bigr]\,.
\end{align}

\noindent\textbf{Level 3.} Next, we consider the most nontrivial example at level 3, which is the function $F_{2_13_14}$ associated to the tubing \includegraphics[scale=0.3,valign=c]{kinematic_flow/Figures/four-site_chain/functions/F2_13_14.pdf}. The corresponding function tree is
\begin{widetext}
\begin{align}
    \includegraphics[scale=1,valign=c]{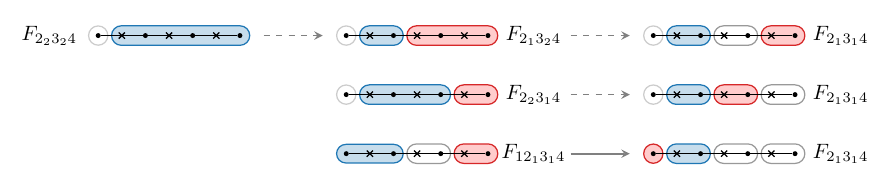}
\end{align}
\end{widetext}
Although the situation is more involved at this stage, we can simply write down the differential equation following the splitting rules
\begin{align}
\dd F_{2_13_14} &=\eps\Bigl[\,(F_{2_13_14}-F_{12_13_14})\,\includegraphics[scale=0.3,valign=c]{kinematic_flow/Figures/four-site_chain/letters/X1+red.pdf}\notag\\[-2pt]&\quad{} +(F_{2_13_14}+F_{2_23_14})\,\includegraphics[scale=0.3,valign=c]{kinematic_flow/Figures/four-site_chain/letters/X3-+red.pdf}\notag\\[-2pt] &\quad{} +F_{12_13_14}\,
\includegraphics[scale=0.3,valign=c]{kinematic_flow/Figures/four-site_chain/letters/X12+blue.pdf}\notag\\[-2pt] &\quad{} -F_{2_23_14}\,\includegraphics[scale=0.3,valign=c]{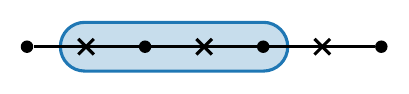}\notag\\[-2pt] &\quad{} +F_{2_13_14}\, \includegraphics[scale=0.3,valign=c]{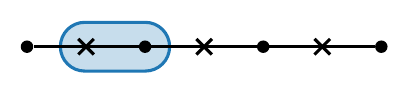}\notag\\[-2pt] &\quad{} +(F_{2_13_14}+F_{2_13_24})\,\includegraphics[scale=0.3,valign=c]{kinematic_flow/Figures/four-site_chain/letters/X4-red.pdf}\notag\\[-2pt] &\quad{} -(F_{2_13_24}+F_{2_23_24})\,\includegraphics[scale=0.3,valign=c]{kinematic_flow/Figures/four-site_chain/letters/X34-red.pdf}\notag\\[-2pt] &\quad{} +F_{2_23_24}\,\includegraphics[scale=0.3,valign=c]{kinematic_flow/Figures/four-site_chain/letters/X234-blue.pdf}\Bigr]\,.
\end{align}

\noindent \textbf{Level 4.} This level contains the largest number of basis functions, and their differential equations are more complicated. Among these, the tubing associated with the most intricate differential equation is \includegraphics[scale=0.3,valign=c]{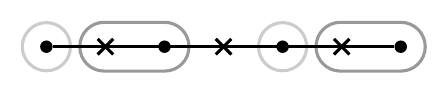}, corresponding to the function $F_{2_14}$. The function tree now involves four levels
\begin{widetext}
\begin{align}
\includegraphics[scale=1,valign=c]{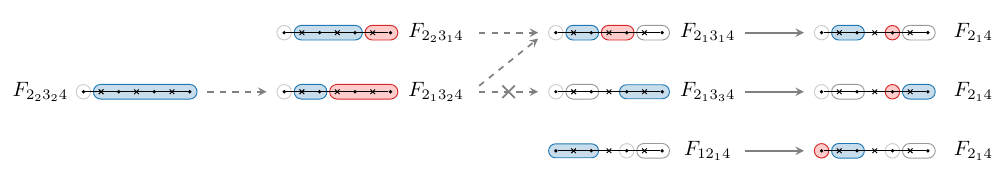}
\end{align}
\end{widetext}
Note that the two parents $F_{2_13_14}$ and $F_{2_13_34}$ of this function yield the same active tube. Thus, the function $F_{2_13_24}$ along the two paths should also be treated as a parent. The differential equation is then given by
\begin{align}
\dd F_{2_14} &=\eps\Bigl[\, F_{2_14} \Bigl(\includegraphics[scale=0.3,valign=c]{kinematic_flow/Figures/four-site_chain/letters/X2-+blue.pdf}+\includegraphics[scale=0.3,valign=c]{kinematic_flow/Figures/four-site_chain/letters/X4-blue.pdf}\Bigr)\notag\\[-2pt] &\quad{}+(F_{2_14}-F_{12_14})\,\includegraphics[ scale=0.3,valign=c]{kinematic_flow/Figures/four-site_chain/letters/X1+red.pdf} \notag\\[-2pt] &\quad{} +\left( F_{2_14}-\sum F_{2,13_i4}\right)\includegraphics[scale=0.3,valign=c]{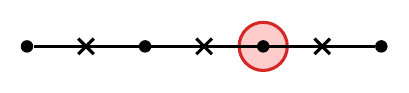}\notag\\[-2pt] &\quad{} +F_{12_14}\,\includegraphics[scale=0.3,valign=c]{kinematic_flow/Figures/four-site_chain/letters/X12+blue.pdf} +F_{2_13_34}\, \includegraphics[scale=0.3,valign=c]{kinematic_flow/Figures/four-site_chain/letters/X34+blue.pdf}\notag\\[-2pt] &\quad{} +(F_{2_13_14}+F_{2_23_14})\,\includegraphics[scale=0.3,valign=c]{kinematic_flow/Figures/four-site_chain/letters/X3-+red.pdf}\notag\\[-2pt] &\quad{} -F_{2_23_14}\,\includegraphics[scale=0.3,valign=c]{kinematic_flow/Figures/four-site_chain/letters/X23-+blue.pdf}\notag\\[-2pt] &\quad{} +(F_{2_13_24}+F_{2_23_24})\, \includegraphics[scale=0.3,valign=c]{kinematic_flow/Figures/four-site_chain/letters/X34-red.pdf}\notag\\[-2pt] &\quad{} -F_{2_23_24}\,\includegraphics[scale=0.3,valign=c]{kinematic_flow/Figures/four-site_chain/letters/X234-blue.pdf}\Bigr]\,.
\end{align}
Similarly, the function $F_{2_13_2}$ should also be treated as a parent of $F_{2_1}$, since it serves as an ancestor function that can be traced back to through multiple parents. The differential for this basis function is
\begin{align}
\dd F_{2_1} &=\eps\Bigl[\, F_{2_1}\,\includegraphics[scale=0.3,valign=c]{kinematic_flow/Figures/four-site_chain/letters/X2-+blue.pdf} \notag\\[-2pt] &\quad{} +(F_{2_1}-F_{12_1})\, \includegraphics[scale=0.3,valign=c]{kinematic_flow/Figures/four-site_chain/letters/X1+red.pdf} \notag\\[-2pt] &\quad{} +\left( F_{2_1}-\sum F_{2_13_i} \right) \includegraphics[scale=0.3,valign=c]{kinematic_flow/Figures/four-site_chain/letters/X3++red.pdf} \notag\\[-2pt] &\quad{} +F_{2_1}\, \includegraphics[scale=0.3,valign=c]{kinematic_flow/Figures/four-site_chain/letters/X12+blue.pdf} +F_{2_13_3}\, \includegraphics[scale=0.3,valign=c]{kinematic_flow/Figures/four-site_chain/letters/X3+-blue.pdf} \notag\\[-2pt] &\quad{} +(F_{2_1}-F_{2_14})\,\includegraphics[scale=0.3,valign=c]{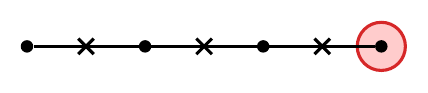} \notag\\[-2pt] &\quad{} +(F_{2_13_1}+F_{2_23_1})\, \includegraphics[scale=0.3,valign=c]{kinematic_flow/Figures/four-site_chain/letters/X3-+red.pdf} \notag\\[-2pt] &\quad{} +F_{2_14}\, \includegraphics[scale=0.3,valign=c]{kinematic_flow/Figures/four-site_chain/letters/X4-blue.pdf} -F_{2_23_1}\, \includegraphics[scale=0.3,valign=c]{kinematic_flow/Figures/four-site_chain/letters/X23-+blue.pdf} \notag\\[-2pt] &\quad{} +(F_{2_13_2}+F_{2_23_2})\, \includegraphics[scale=0.3,valign=c]{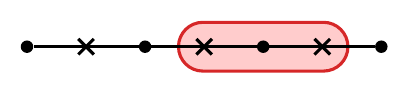} \notag\\[-2pt] &\quad{} +F_{2_23_2}\, \includegraphics[scale=0.3,valign=c]{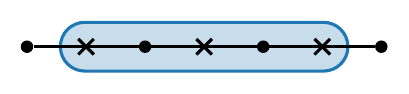}\Bigr]\,.
\end{align}
 \textbf{Level 6.} At this level, the unique basis function  corresponds to the wavefunction coefficient $\psi$, whose function tree is as follows
\begin{widetext}
\begin{align}
    \includegraphics[scale=1,valign=c]{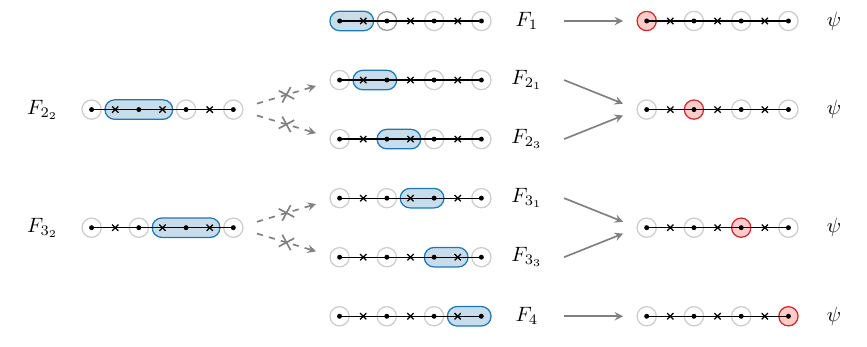}
\end{align}
\end{widetext}
Thus, the differential equation for $\psi$ is easy to derive
\begin{align}
\dd\psi &=\eps\Bigl[\, (\psi-F_1)\,\includegraphics[scale=0.3,valign=c]{kinematic_flow/Figures/four-site_chain/letters/X1+red.pdf}\notag\\[-2pt] &\quad{} +\left(\psi-{\textstyle\sum}F_{2_i}\right) \includegraphics[scale=0.3,valign=c]{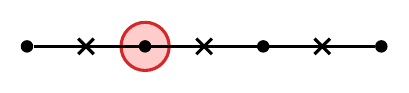}\notag\\[-2pt] &\quad{} +\left(\psi-{\textstyle\sum}F_{3_i}\right)\includegraphics[scale=0.3,valign=c]{kinematic_flow/Figures/four-site_chain/letters/X3++red.pdf}\notag\\[-2pt] &\quad{} +(\psi-F_4)\,\includegraphics[scale=0.3,valign=c]{kinematic_flow/Figures/four-site_chain/letters/X4+red.pdf} \notag\\[-2pt] &\quad{} +F_1\,\includegraphics[scale=0.3,valign=c]{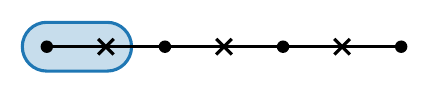}+F_4\,\includegraphics[scale=0.3,valign=c]{kinematic_flow/Figures/four-site_chain/letters/X4-blue.pdf}\notag\\[-2pt] &\quad{} +F_{2_1}\,\includegraphics[scale=0.3,valign=c]{kinematic_flow/Figures/four-site_chain/letters/X2-+blue.pdf} +F_{3_1}\, \includegraphics[scale=0.3,valign=c] {kinematic_flow/Figures/four-site_chain/letters/X3-+blue.pdf} \notag\\[-2pt] &\quad{} +F_{2_2}\, \includegraphics[scale=0.3,valign=c]{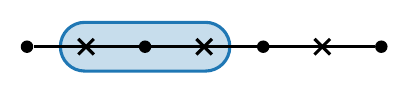} +F_{3_2}\, \includegraphics[scale=0.3,valign=c] {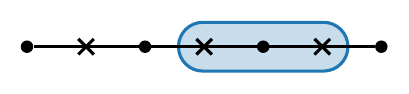} \notag\\[-2pt] &\quad{} +F_{2_3}\, \includegraphics[scale=0.3,valign=c]{kinematic_flow/Figures/four-site_chain/letters/X2+-blue.pdf} +F_{3_3}\, \includegraphics[scale=0.3,valign=c]{kinematic_flow/Figures/four-site_chain/letters/X3+-blue.pdf}\Bigr]\,.
\end{align}
\subsection{Four-Site Star}
Compared to the four-site chain case, the graph geometry of the four-site star exhibits a higher degree of permutation symmetry, $X_1 \leftrightarrow X_2 \leftrightarrow X_3$. Consequently, the corresponding differential equations for the basis functions are somewhat simpler, despite the system still containing 64 basis functions. The generating function $F_{1234_7}$, which corresponds to the tubing \includegraphics[scale=0.3,valign=c]{kinematic_flow/Figures/four-site_star/functions/F1234_7.pdf}, satisfies the same differential equation as $F_{12_23_24}$ in the four-site chain case
\begin{align}
    \dd F_{1234_7} = 4 \eps \,\includegraphics[scale=0.3,valign=c]{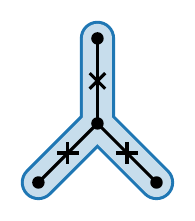}\,.
\end{align}
We can interpret this phenomenon as the result of all points being non-local in the tubing of these two functions. 

\noindent \textbf{Level 1.} The generating function can split into six other basis functions. However, due to the symmetry of the three vertices corresponding to $X_1,X_2$ and $X_3$, only two of them are independent. We choose the following two splitting channels
\begin{align}
\includegraphics[scale=0.7,valign=c]{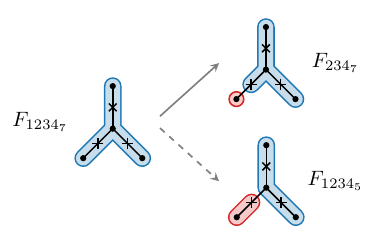}
\end{align}
This will lead to the following two differential equations
\begin{align}
\dd F_{234_7} &=\eps\Bigl[\, 3F_{234_7}\, \includegraphics[scale=0.3,valign=c]{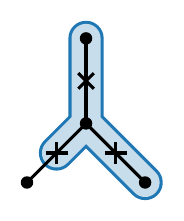} +F_{1234_7}\, \includegraphics[scale=0.3,valign=c]{kinematic_flow/Figures/four-site_star/letters/X1234.pdf}\notag\\[-2pt] &\quad{} +(F_{234_7}-F_{1234_7})\, \includegraphics[scale=0.3,valign=c]{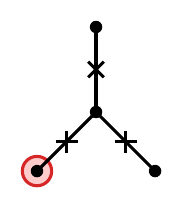} \Bigr]\,, \\[3pt] \dd F_{1234_5} &=\eps\Bigl[\,  3F_{1234_5}\, \includegraphics[scale=0.3,valign=c]{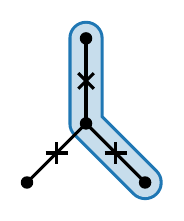}-F_{1234_7}\, \includegraphics[scale=0.3,valign=c]{kinematic_flow/Figures/four-site_star/letters/X1234.pdf}\notag\\[-2pt] &\quad{} +(F_{1234_5}+F_{1234_7})\, \includegraphics[scale=0.3,valign=c]{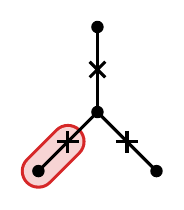}\Bigr]\,.
\end{align}

\noindent \textbf{Level 2.} Next, we consider an example at level 2, \includegraphics[scale=0.3,valign=c]{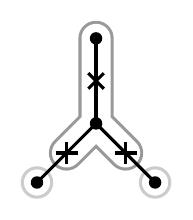}, corresponding to the function $F_{34_7}$. The function tree is
\begin{align}
\includegraphics[scale=0.7,valign=c]{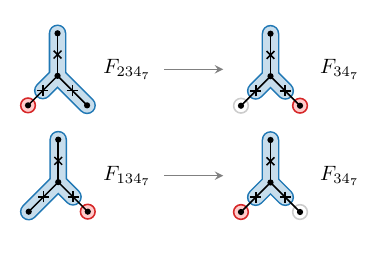}
\end{align}
The differential equation for this basis function is easy to derive
\begin{align}
    \dd F_{34_7} = \eps \,\Big[\,2 \,F_{34_7} &\, \includegraphics[scale=0.,valign=c]{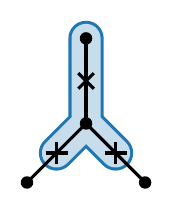} +\, (F_{34_7}-F_{234_7})\,\includegraphics[scale=0.3,valign=c]{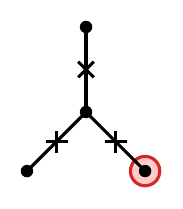} +\,F_{234_7}\,\includegraphics[scale=0.3,valign=c]{kinematic_flow/Figures/four-site_star/letters/X234-blue.pdf} \nonumber \\
    +\,(F_{34_7}-F_{134_7})\, &\includegraphics[scale=0.3,valign=c]{kinematic_flow/Figures/four-site_star/letters/X1+.pdf} +\,F_{134_7}\,\includegraphics[scale=0.3,valign=c]{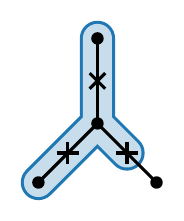} \Big]\,.
\end{align}
It is straightforward to see that the differential equations possess a high degree of symmetry. 

\noindent \textbf{Level 3.} Next, we consider the most nontrivial example at level 3, which is the function $F_{134_3}$ associated to the tubing \includegraphics[scale=0.3,valign=c]{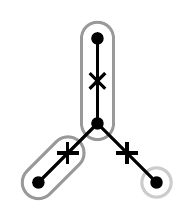}. The function tree is then
\begin{align}
\includegraphics[scale=0.7,valign=c]{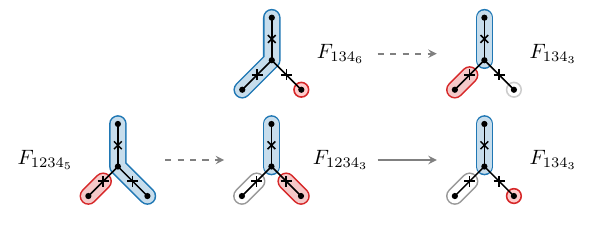}
\end{align}
It is evident that this basis function has only two parents. This is not an isolated case, since many splittings are forbidden in the four-site star tubings. The differential equation is given by
\begin{align}
\dd F_{134_3} &=\eps\Bigl[\, 2F_{134_3}\, \includegraphics[scale=0.3,valign=c]{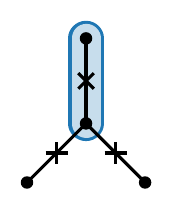} +(F_{134_3}+F_{134_6})\, \includegraphics[scale=0.3,valign=c]{kinematic_flow/Figures/four-site_star/letters/X1-red.pdf} \notag\\[-2pt] &\quad{} +(F_{134_3}-F_{1234_3})\, \includegraphics[scale=0.3,valign=c]{kinematic_flow/Figures/four-site_star/letters/X2+.pdf} -F_{134_6}\, \includegraphics[scale=0.3,valign=c] {kinematic_flow/Figures/four-site_star/letters/X134-blue.pdf} \notag\\[-2pt] &\quad{} +(F_{1234_3}+F_{1234_5})\, \includegraphics[scale=0.3,valign=c]{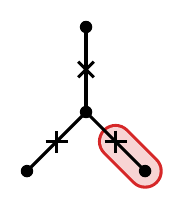} -F_{1234_5}\, \includegraphics[scale=0.3,valign=c] {kinematic_flow/Figures/four-site_star/letters/X234-blue.pdf} \Bigr]\,.
\end{align}

\noindent \textbf{Level 4.} Nothing surprising happens at this level either. Take as an example one of the tubings  \includegraphics[scale=0.2,valign=c]{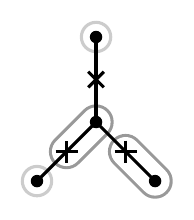}, which corresponds to the function $F_{24_1}$. We can obtain the following function tree
\begin{align}
\includegraphics[scale=0.7,valign=c]{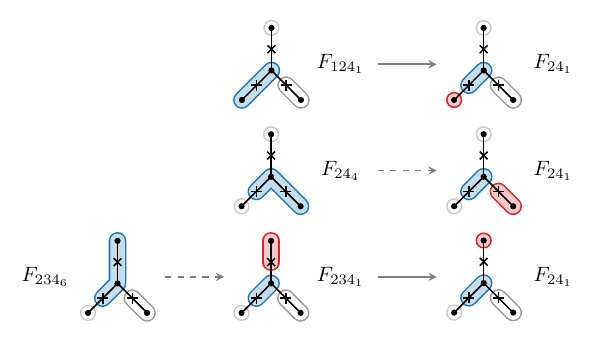}
\end{align}
Also, although this function is derived from the generating function through four splittings, it has only three parents. The differential equation is given by
\begin{align}
\dd F_{24_1} &=\eps\Bigl[\, F_{24_1}\, \includegraphics[scale=0.3,valign=c]{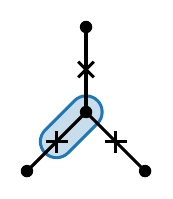} +(F_{24_1}-F_{124_1})\, \includegraphics[scale=0.3,valign=c] {kinematic_flow/Figures/four-site_star/letters/X1+.pdf} \notag\\[-3pt] &\quad{} +F_{124_1}\, \includegraphics[scale=0.3,valign=c] {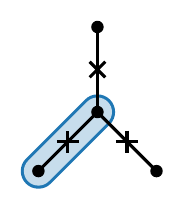} +(F_{24_1}+F_{24_4})\, \includegraphics[scale=0.3,valign=c]{kinematic_flow/Figures/four-site_star/letters/X2-red.pdf} \notag\\[-3pt] &\quad{} +(F_{24_1}-F_{234_1})\, \includegraphics[scale=0.3,valign=c] {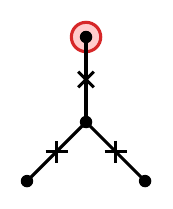} -F_{24_4}\, \includegraphics[scale=0.3,valign=c] {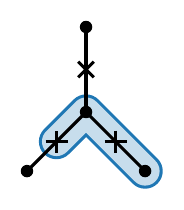} \notag\\[-3pt] &\quad{} +(F_{234_1}+F_{234_6})\, \includegraphics[scale=0.3,valign=c] {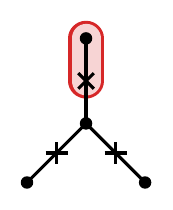} -F_{234_6}\, \includegraphics[scale=0.3,valign=c]{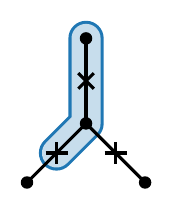}\Bigr]\,.
\end{align}

\noindent \textbf{Level 5.} The most nontrivial example at this level is the tubing \includegraphics[scale=0.2,valign=c]{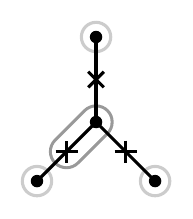}. The corresponding function tree is 
\begin{align}
\includegraphics[scale=0.7,valign=c]{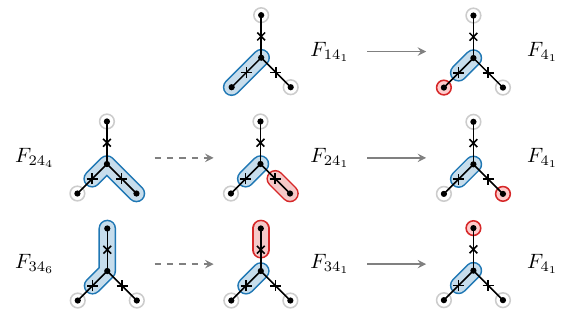}
\end{align}
There are also only three parents for $F_{4_1}$ at this time. The differential for this function is
\begin{align}
\dd F_{4_1} &=\eps\Bigl[\, F_{4_1}\, \includegraphics[scale=0.3,valign=c]{kinematic_flow/Figures/four-site_star/letters/X4-++blue.pdf} +(F_{4_1}-F_{14_1})\, \includegraphics[scale=0.3,valign=c]{kinematic_flow/Figures/four-site_star/letters/X1+.pdf} \notag\\[-3pt] &\quad{} +F_{14_1}\, \includegraphics[scale=0.3,valign=c] {kinematic_flow/Figures/four-site_star/letters/X14++blue.pdf} +(F_{4_1}-F_{24_1})\, \includegraphics[scale=0.3,valign=c]{kinematic_flow/Figures/four-site_star/letters/X2+.pdf} \notag\\[-3pt] &\quad{} +(F_{4_1}-F_{34_1})\, \includegraphics[scale=0.3,valign=c]{kinematic_flow/Figures/four-site_star/letters/X3+.pdf}\notag\\[-3pt] &\quad{} +(F_{24_1}+F_{24_4})\, \includegraphics[scale=0.3,valign=c] {kinematic_flow/Figures/four-site_star/letters/X2-red.pdf} +(F_{34_1}+F_{34_6})\, \includegraphics[scale=0.3,valign=c] {kinematic_flow/Figures/four-site_star/letters/X3-red.pdf} \notag\\[-3pt] &\quad{} -F_{24_4}\, \includegraphics[scale=0.3,valign=c] {kinematic_flow/Figures/four-site_star/letters/X24-+blue.pdf} -F_{34_6}\, \includegraphics[scale=0.3,valign=c] {kinematic_flow/Figures/four-site_star/letters/X34-+blue.pdf}\Bigr]\,.
\end{align}

\noindent \textbf{Level 6.} The basis function at the final level is the wavefunction coefficient $\psi$, which has the most complicated function tree
\begin{align}
\includegraphics[scale=0.7,valign=c]{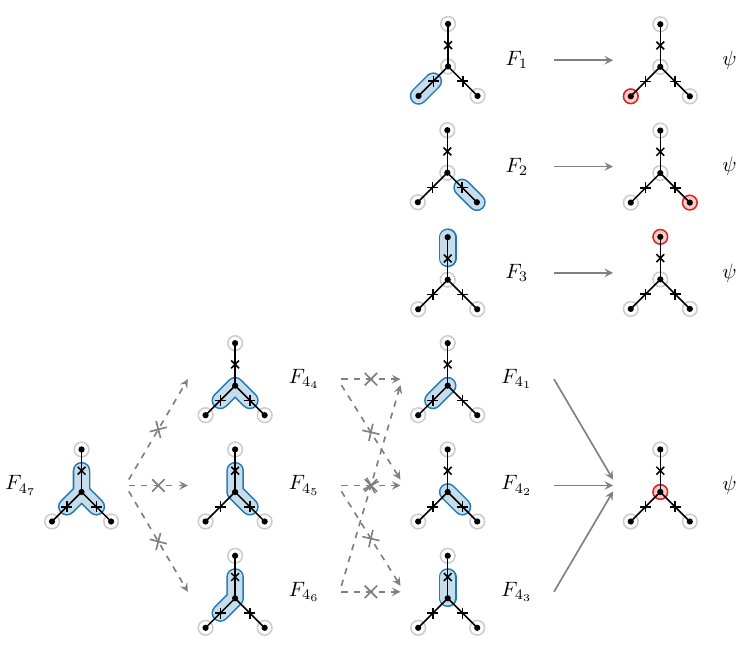}
\end{align}
In this case, the four basis functions $F_{4_4}, F_{4_5},F_{4_6},F_{4_7}$ should all be considered as the parents of $\psi$, since they can split into $\psi$ through distinct paths (including forbidden ones). Then we can write down the following differential equation 
\begin{equation}
\begin{aligned}
\dd\psi &=\eps\Bigl[\, (\psi-F_1)\,\includegraphics[scale=0.3,valign=c]{kinematic_flow/Figures/four-site_star/letters/X1+.pdf}+F_1\, \includegraphics[scale=0.3,valign=c]{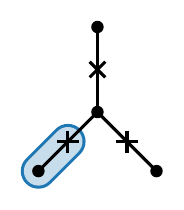}\\[-3pt] &\quad{} +(\psi-F_2)\, \includegraphics[scale=0.3,valign=c] {kinematic_flow/Figures/four-site_star/letters/X2+.pdf} +F_2\, \includegraphics[scale=0.3,valign=c] {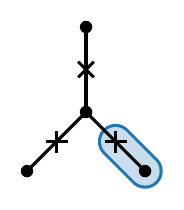} \\[-3pt] &\quad{} +(\psi-F_3)\, \includegraphics[scale=0.3,valign=c] {kinematic_flow/Figures/four-site_star/letters/X3+.pdf}+F_3\, \includegraphics[scale=0.3,valign=c]{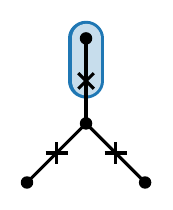}\\[-3pt] &\quad{} -\bigl(\psi- \sum F_{4_i}\bigr)\, \includegraphics[scale=0.3,valign=c]{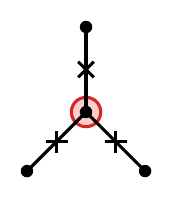}\\[-3pt] &\quad{} +F_{4_1}\,\includegraphics[scale=0.3,valign=c]{kinematic_flow/Figures/four-site_star/letters/X4-++blue.pdf} +F_{4_2}\, \includegraphics[scale=0.3,valign=c]{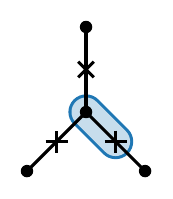} +F_{4_3}\, \includegraphics[scale=0.3,valign=c]{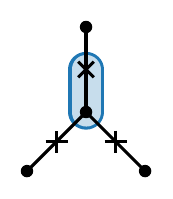} \\[-3pt] &\quad{} +F_{4_4}\, \includegraphics[scale=0.3,valign=c]{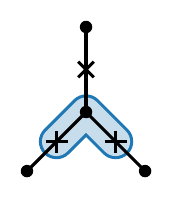} +F_{4_5}\, \includegraphics[scale=0.3,valign=c]{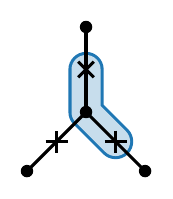} +F_{4_6}\, \includegraphics[scale=0.3,valign=c] {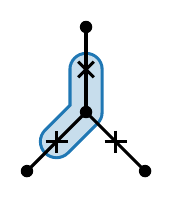} \\[-3pt] &\quad{} +F_{4_7}\, \includegraphics[scale=0.3,valign=c] {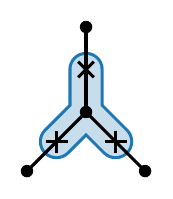} \Bigr]\,.
\end{aligned}
\end{equation}
\clearpage
\bibliographystyle{utphys}
\linespread{1.075}\selectfont
\bibliography{Flow-Refs}

\end{document}